\documentclass[onefignum,onetabnum]{siamart171218}

\usepackage{lipsum}
\usepackage{amsfonts}
\usepackage{graphicx}
\usepackage{epstopdf}
\usepackage{algorithmic}
\ifpdf
  \DeclareGraphicsExtensions{.eps,.pdf,.png,.jpg}
\else
  \DeclareGraphicsExtensions{.eps}
\fi

\newsiamremark{remark}{Remark}
\newsiamremark{hypothesis}{Hypothesis}
\crefname{hypothesis}{Hypothesis}{Hypotheses}
\newsiamthm{claim}{Claim}

\headers{Discrete breathers of nonlinear dimer lattices}{A. Hofstrand, H. Li, and M. I. Weinstein}

\title{Discrete breathers of nonlinear dimer lattices: bridging the anti-continuous and continuous limits}
\author{Andrew Hofstrand\thanks{New York Institute of Technology, New York, NY (\email{ahofstra@nyit.edu}).}
\and Huaiyu Li\thanks{Department of Applied Physics and Applied Mathematics, Columbia University, New York, NY (\email{hl3002@columbia.edu}).} \and Michael I. Weinstein\thanks{Department of Applied Physics and Applied Mathematics and Department of Mathematics, Columbia University, New York, NY (\email{miw2103@columbia.edu}).}
}

\usepackage{amsopn}

\makeatletter
\newcommand*{\addFileDependency}[1]{% argument=file name and extension
  \typeout{(#1)}% latexmk will find this if $recorder=0 (however, in that case, it will ignore #1 if it is a .aux or .pdf file etc and it exists! if it doesn't exist, it will appear in the list of dependents regardless)
  \@addtofilelist{#1}% if you want it to appear in \listfiles, not really necessary and latexmk doesn't use this
  \IfFileExists{#1}{}{\typeout{No file #1.}}% latexmk will find this message if #1 doesn't exist (yet)
}
\makeatother

\newcommand*{\myexternaldocument}[1]{%
    \externaldocument{#1}%
    \addFileDependency{#1.tex}%
    \addFileDependency{#1.aux}%
}
\myexternaldocument{ex_supplement}
\usepackage{comment}
\begin{document}

\maketitle

% REQUIRED
\begin{abstract} 
 In this work, we study the dynamics of an infinite array of nonlinear dimer oscillators which are linearly coupled as in the classical model of Su, Schrieffer and Heeger (SSH). The ratio of in-cell and out-of-cell couplings of the SSH model defines distinct {\it phases}: topologically trivial and topologically non-trivial. We first consider the case of weak out-of-cell coupling, corresponding to the topologically trivial regime for linear SSH; for any prescribed isolated dimer frequency, $\omega_b$, which satisfies non-resonance and non-degeneracy assumptions,
we prove that there are discrete breather solutions for sufficiently small values of the out-of-cell coupling parameter. These states are $2\pi/\omega_b$- periodic in time and  exponentially localized in space. We then study the global continuation with respect to this coupling parameter.
We first consider the case where $\omega_b$, the seeding discrete breather frequency, is in the (coupling dependent)  phonon gap of the underlying linear infinite array. As the coupling is increased, the phonon gap decreases in width and tends to a point (at which the topological transition for linear SSH occurs).  In this limit, the spatial scale of the discrete breather grows and its amplitude decreases, indicating the weakly nonlinear long wave regime. Asymptotic analysis shows that in this regime the discrete breather envelope is determined by a vector gap soliton of the limiting envelope equations. We use the envelope theory to describe discrete breathers for SSH- coupling parameters corresponding to topologically trivial and, by exploiting an emergent symmetry, topologically nontrivial regimes, when the spectral gap is small. Our asymptotic theory shows excellent agreement with extensive numerical simulations over a wide range of parameters. Analogous asymptotic and numerical results are obtained for the continuations from the anti-continuous regime for frequencies, $\omega_b$, below the acoustic or above the optical phonon bands.
\end{abstract}

% REQUIRED

% REQUIRED

\section{Introduction}
\label{sec:intro}

\subsection{Motivation and background} 
There is great current interest in the study of wave propagation through discrete and continuous periodic media, which exhibits nontrivial topological properties. While  it is common for physical systems  to support defect modes concentrated at points  or interfaces, these modes are in general not stable against significant perturbations
 of the structure. However, it has been recognized that topological characteristics in the bulk (Floquet-Bloch) band structure can give rise to modes which are robust against large (but localized) perturbations of the system. The role of band structure topology in wave physics was first recognized in the context of condensed matter physics, e.g. the integer quantum Hall effect \cite{klitz80} and topological insulators \cite{has10}.
 The hallmark of topological materials is the presence of  
topologically protected edge states. These states are localized at interfaces (line defects, facets), propagate unidirectionally and are robust against localized - even large - imperfections in the system.
 Many of the topological wave phenomena observed in these contexts were subsequently realized in engineered metamaterial systems in  photonics \cite{hal08,photon}, acoustics \cite{acoustic}, electronics \cite{Alu17} and elasticity  \cite{elastic}
 which, in the regime of linear phenomena, are characterized by a linear band structure. There is very wide interest in technologies based on topologically protected states due to the potential for extraordinarily robust energy and information transfer in communications and computing.

 Such systems can naturally be probed in the nonlinear regime via strong excitation, and so it is of great interest to study whether topological properties 
  persist in the regime where nonlinear effects are present and whether perhaps different topological phenomena emerge \cite{Alu17, theo19, Theo21, delplace, kiv20}.
  
  Among the simplest models exhibiting {\it topological phases} is the Su-Schrieffer-Heeger (SSH) model \cite{ssh79}, a discrete (tight binding) model on one-dimensional lattice in which two "atoms" per cell (dimers) are linearly and nearest-neighbor coupled. 
\begin{figure}[H]
  \centering
     \includegraphics[scale=0.6]{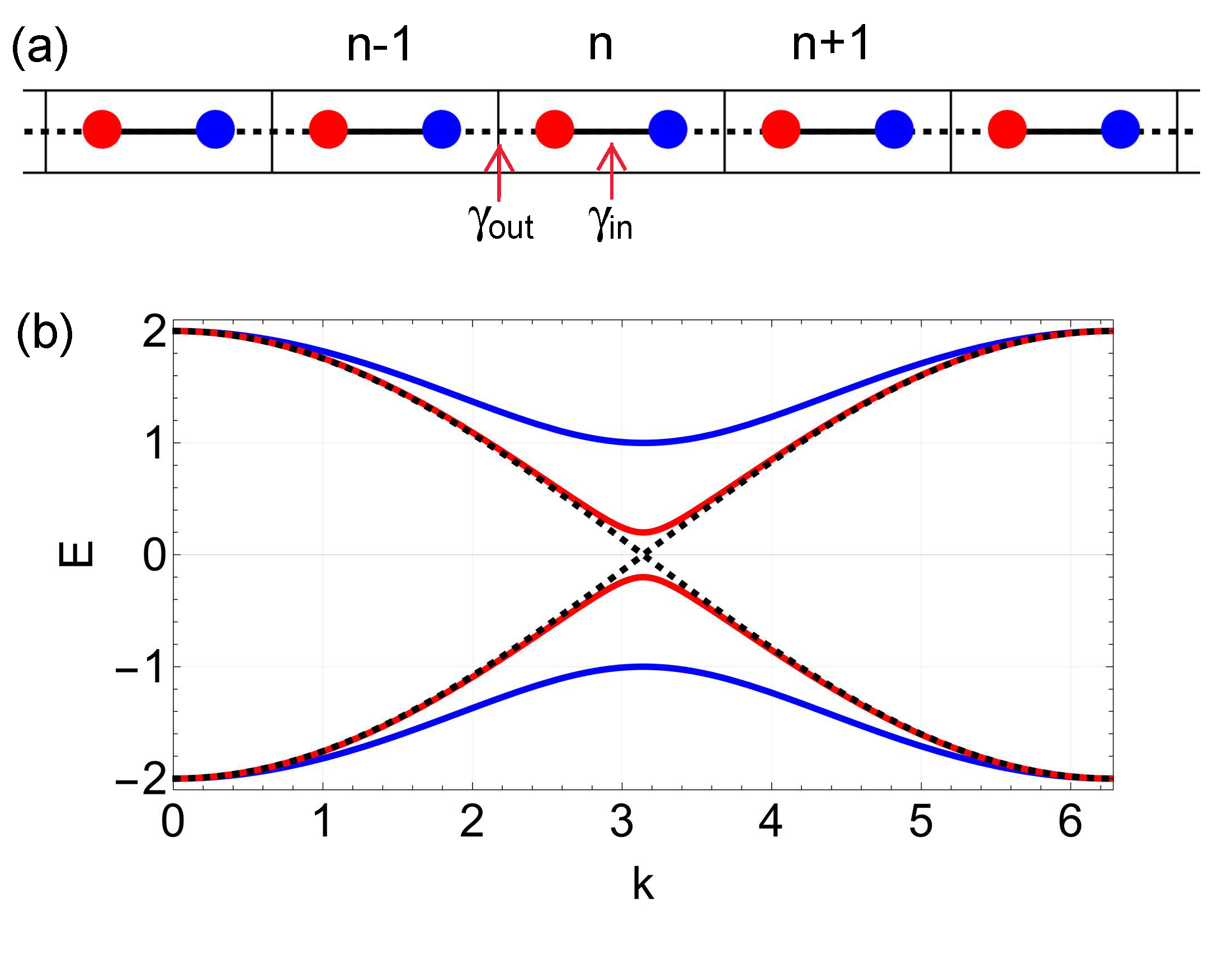}
     \vspace{-0.4cm}
  \caption{(a) schematic of the infinite dimer lattice considered with A (B) sites in red (blue) and intra (inter) cell coupling $\gamma_{\text{in}}$ ($\gamma_{\text{out}}$); (b) band structure of the lattice for various ratios $|\gamma_{\text{in}}/\gamma_{\text{out}}|$. A gap in the energy spectrum opens if and only if  $|\gamma_{\text{in}}/\gamma_{\text{out}}|\ne1$. }
  \label{fig:dimer}
\end{figure}
Figure \ref{fig:dimer} displays a schematic of SSH array of dimers. The red sites are called $A-$ sites and the blue sites are called $B-$ sites. Each $A-$site has two nearest neighbor $B-$sites and each $B-$site has two nearest neighbor $A-$sites. In-cell (intra-cell) and out-of-cell (inter-cell) nearest neighbors are coupled via  {\it hopping coefficients}
 $\gamma_{\rm in}\in\mathbb{R}$ and $\gamma_{\rm out}\in\mathbb{R}$, respectively: for $n\in\mathbb{Z}$,
 \begin{align}\label{ssh}
 E\psi^A_n &= \gamma_{\rm in}\psi_n^B+\gamma_{\rm out}\psi_{n-1}^B\\
  E\psi^B_n &= \gamma_{\rm in}\psi_n^A+\gamma_{\rm out}\psi_{n+1}^A.
 \end{align}
The spectrum of the SSH-Hamiltonian, $H_{\rm SSH}$, acting in the space of discrete wave functions, $l^2(\mathbb{Z})$, consists of two real intervals (bands), swept out by the two eigenvalues $E_-(k)\le 0\le E_+(k)$,  as the quasimomentum $k$ varies over the interval $[0,2\pi]$,
of the family of Bloch Hamiltonian $H_{SSH}(k)=\sigma_1 h_1(k)+\sigma_2 h_2(k)$ (obtained by discrete Fourier transform, $\sigma_j$ denote Pauli matrices); see Figure \ref{fig:dimer}. 
A gap in the spectrum occurs for $|\gamma_{\text{in}}/\gamma_{\text{out}}|\ne1$ and the bands touch in a linear crossing for $|\gamma_{\rm in}/\gamma_{\rm out}|=1$.

The two distinct topological phases correspond to $|\gamma_{\rm in}/\gamma_{\rm out}|>1$ (trivial) and $|\gamma_{\rm in}/\gamma_{\rm out}|<1$ (non-trivial), and are identified with the cases where the {\it Zak phase}, a winding number (about the origin) associated with the variation of the vector field $h(k)=(h_1(k),h_2(k))$ as $k$ varies over $S^1=\mathbb{R}/2\pi\mathbb{Z}$, is equal to zero or one.  The topological character is also manifested in the spectrum of states for a terminated (semi-infinite) structure; there exists a zero energy edge state which decays exponentially into the bulk if and only if one is in the topologically non-trivial phase ($|\gamma_{\rm in}/\gamma_{\rm out}|<1$).
See, for example, \cite{shen17,shapiro-weinstein}. 

In this paper we study a non-linear variant of the SSH-model, introduced in \cite{Theo21}. In this model, each ``atom'' of the array corresponds to a nonlinear mass-spring oscillator described by a Newtonian law: $\ddot{x}= -V'(x)$,
 with an even anharmonic potential, $V(x)$. For illustrative purposes, it will be convenient at times to work with the specific potential:
 \begin{equation}
V(x)= \dfrac{3}{2}x^2 + \dfrac{\Gamma}{4}x^4.
\label{Vdef} \end{equation}
The case when $\Gamma>0$ is referred to as the case of a \textit{hardening} nonlinearity and the case when $\Gamma<0$ as a \textit{softening} nonlinearity.

Within a fixed dimer / cell, the two mass spring systems with amplitudes $x^A$ and $x^B$ are linearly coupled via the in-cell coupling coefficient $\gamma_{\text{in}}$ :
\begin{align}
&\ddot{x}^A=-V'(x^A)+\gamma_{\text{in}}x^B \label{cell-eq}\\ 
&\ddot{x}^B=-V'(x^B)+\gamma_{\text{in}}x^A
\nonumber\end{align}
The coupled system \eqref{cell-eq} is the fundamental unit with which we build  up a nonlinear SSH-network. The system \eqref{cell-eq}
 will be assumed to have {\it non-resonant} and {\it non-degenerate} time-periodic orbits in a sense which we shall make precise in Theorem 
 \ref{thm:breather} below.

We build an SSH-
%We focus on the situation where \eqref{cell-eq} has a %family of periodic orbits.
 network of nonlinear dimers by  coupling each oscillator to its out-of-cell nearest neighbors via a second coupling coefficient, $\gamma_{\rm out}$. This gives the system:
\begin{align}\label{eq1A}
&\ddot{x}_{n}^A=-V'(x_{n}^A)+ \lambda\gamma_{\text{out}}x_{n-1}^B+\gamma_{\text{in}}x_{n}^B\\ \nonumber
&\ddot{x}_{n}^B=-V'(x_{n}^B)+\gamma_{\text{in}}x_{n}^A+
\lambda\gamma_{\text{out}}x_{n+1}^A,\quad n\in\mathbb{Z}.
\end{align}
 In \eqref{eq1A} we make explicit the $B-$ site terms which interact with the $x_n^A$ and the $A-$ site terms which interact with $x_n^B$. $V'$ consists of both linear and non-linear onsite contributions. The non-negative parameter, $\lambda$ has been inserted in order to interpolate
 between the {\it anti-continuous limit} ($\lambda=0$) and globally coupled models. 
 
 More generally, we study
\begin{equation}
\begin{pmatrix}
\ddot{x}_{n}^A+V'(x_{n}^A)-\gamma_{\text{in}}x_{n}^B\\
\ddot{x}_{n}^B+V'(x_{n}^B)-\gamma_{\text{in}}x_{n}^A 
\end{pmatrix} 
+ \lambda
\left[{\bf R}\begin{pmatrix}x^A\\ x^B\end{pmatrix}\right]_n = \begin{pmatrix}0\\ 0\end{pmatrix},\quad {n\in\mathbb{Z}},
\label{nloc-case}\end{equation}
where ${\bf R}$ is a bounded linear operator on $l^2(\mathbb{Z};\mathbb{R}^2)$. The coupling operator, ${\bf R}$, can couple sites beyond nearest-neighbor, but its defining matrix elements are assumed to be exponentially decaying away from the diagonal; see \eqref{Rml-decay}). The case of nearest neighbor interactions (see \eqref{eq1A}) corresponds to:
 \[\left[{\bf R}\begin{pmatrix}x^A\\ x^B\end{pmatrix}\right]_n = -\gamma_{\text{out}}\begin{pmatrix}
     x^B_{n-1} \\  x^A_{n+1} \end{pmatrix}.\]

Consider the band structure of the linearized dynamics for \eqref{eq1A} about the zero state ($x^A_n=x^B_n=0$, $n\in\mathbb{Z}$), determined by the set of non-trivial plane wave states: $e^{i(kn-\omega t)}\xi,\ 0\ne\xi\in\mathbb{C}^2$.
In terms of  $E=\omega^2-\omega_0^2$ vs. $k$, the band spectrum is 
a re-centering about $\omega_0^2=V''(0)$ of the SSH band spectrum:
\begin{equation}\label{phonon}
   E_\pm(k) =  \left(\omega^2\right)_{\pm}(k)-\omega_0^2=\pm\sqrt{(\gamma_{\rm in}-\lambda\gamma_{\rm out})^2+4\gamma_{\rm in}\lambda\gamma_{\rm out}\cos^2\left({k\over2}\right)},
    \quad \omega_0^2\equiv V''(0).
\end{equation}
Figure \ref{fig:dimer} displays the graphs of these band functions. The two spectral bands are separated by a gap for $|\lambda\gamma_{\rm out}/\gamma_{\rm in}|\ne1$
  and touch at a linear crossing for $|\lambda\gamma_{\rm out}/\gamma_{\rm in}|=1$. It is therefore natural to 
contrast the properties of the  $\lambda-$ parametrized family of equations \eqref{eq1A} for
   the (linearly) topologically distinct regimes
      $ |\lambda\gamma_{\rm out}/\gamma_{\rm in}|<1$ and 
      $  |\lambda\gamma_{\rm out}/\gamma_{\rm in}|>1$.
 \begin{remark}[Phonon gaps]\label{rmk:pgap}
 In terms of the frequency parameter $\omega$, there are two pairs of dispersion curves, symmetric about $\omega=0$, each pair having  {\it phonon gap} for $|\lambda\gamma_{\rm out}/\gamma_{\rm in}|\ne1$:
 one about $\omega_0$ and one about $-\omega_0$, each of width $\approx \omega_0^{-1}|\epsilon|$, where
 $\epsilon=\gamma_{\rm in}-\lambda\gamma_{\rm out}$
 \end{remark}
 
\subsection{Summary of the article and results} 
We study the existence and properties of {\it discrete breathers},  solutions of the infinite lattice nonlinear system \eqref{nloc-case}, which are periodic in time and localized on the discrete lattice $\mathbb{Z}$. We outline the key points of this paper:
  
  \begin{enumerate}
      \item {\it Existence of discrete breathers; Theorem \ref{thm:breather}.} Assume that the anharmonic potential in \eqref{nloc-case} satisfies $V(-x)=V(x)$. Let $t\mapsto X_*(t)$ 
      denote a non-resonant and non-degenerate $T_b=\frac{2\pi}{\omega_b}-$ periodic solution of the limiting ($\lambda=0$)
      infinite dimer array, associated with \eqref{cell-eq}. Then,
       for all $\lambda$ sufficiently small and non-zero,
       there is a  unique $T_b-$ periodic solution
      \[ t\mapsto X^\lambda(t) = \left[ 
      \begin{pmatrix}x_n^A(t)\\ x_n^B(t)\end{pmatrix}
      \right]_{n\in\mathbb{Z}}\]
      with $X^0=X_*$
       of the globally coupled lattice equations  \eqref{nloc-case}.
       This solution lies in the space 
        $\mathcal{H}^2_{T_b}$, consisting of sequences $X(t)$,
        which satisfy $X(t)=X(-t)$, and together with derivatives up to order $2$, are $T_b-$ periodic and square integrable over $[0,T_b]$, and square summable (spatially) over $\mathbb{Z}$:
         \[ \|X^\lambda\|^2_{\mathcal{H}^2_{T_b}} \equiv \sum_{n\in\mathbb{Z}} 
         \int_0^{T_b} \sum_{j=0}^2\Big|\frac{d^j}{dt^j}\begin{pmatrix}x_n^A(t;\lambda)\\ x_n^B(t;\lambda)\end{pmatrix}\Big|^2 dt<\infty. \]
        Furthermore, the mapping $\lambda\mapsto \|X^\lambda\|_{\mathcal{H}^2_{T_b}}$ is smooth.
        For a discussion of the behavior when the non-resonance hypothesis is violated,
        see in particular Remark \ref{non-res} and Figure \ref{resonant}.
        
      \quad Our proof is based on a Poincar\'e continuation strategy, used in the pioneering article \cite{ma94} on discrete breathers. The richer structure of the building-block isolated dimer dynamical system \eqref{cell-eq} (anti-continuous limit) allows for richer behaviors in the global array. Finally recall that for general nonlinear autonomous dynamical systems the period of the solution varies along the continuation. Here, the symmetry condition on $V(x)$ enables us to restrict our study to time-reversible solutions with fixed period. The analysis can be adapted to more general potentials, $V(x)$, by incorporating the determination of the discrete breather  period as a function of $\lambda$. 
      \item {\it Applications of Theorem \ref{thm:breather}.} In Section \ref{sec:application} we apply Theorem \ref{thm:breather} to obtain discrete breather solutions which are continuations of two classes of solutions to the isolated dimer dynamical system \eqref{cell-eq}: in-phase (Type I)  and out-of-phase (Type II) solutions. We verify the non-resonance and non-degeneracy assumptions of Theorem \ref{thm:breather} by a combination of rigorous analysis and numerical computation. 
\item {\it Exponential spatial decay of discrete breathers, Theorem \ref{thm:exp_decay}}
The breather solutions obtained via Theorem \ref{thm:breather} have the square-summable decay behavior of functions in $\mathcal{H}^2_{T_b}$. Hence they are only guaranteed to decay at infinity in a mild sense. In Section \ref{sec:localization} we prove Theorem \ref{thm:exp_decay}, a general result on exponential spatial decay of the discrete breathers, which applies to those constructed in Theorem \ref{thm:breather}. Our proof uses ideas underlying Combes and Thomas discrete operator estimates (see, for example, \cite{aizen15}), and offers a different perspective on the earlier decay results in \cite{ma94}. 

\item {\it Numerical simulations of discrete breathers: ranging from the highly discrete (anti-continuous) regime to the nearly continuum regime.} For \[
0<\lambda<\lambda_\star\equiv |\gamma_{\rm in}/\gamma_{\rm out}|,\] the phonon spectrum (linearized spectrum about the zero state) has an open spectral gap centered about the linearized ``atomic'' frequency,  $\omega_0=\sqrt{V''(0)}$; see Remark \ref{rmk:pgap}. The gap width is of order one for $\lambda$ near zero and shrinks down to the point $\omega_0$ as $\lambda$ approaches $\lambda_*$.  
Using a numerical method, outlined in Appendix A, we construct solutions
 corresponding to  fine grid of $\lambda-$ values starting at $\lambda=0$ and continued, when possible, till 
 very close to $\lambda_*$. The initializing
  discrete breather, which is supported on the  $n=0$ dimer, is taken to be a periodic orbit with frequency, $\omega_b$ corresponding to one of  the following cases:
  \begin{enumerate}
      \item[(A)] $\omega_b\approx\omega_0$ in the phonon gap, 
      \item[(B)] $\omega_b$ just above the optical branch of the phonon spectrum and
      \item[(C)] $\omega_b$ just below the acoustic branch of the phonon spectrum.
      \end{enumerate}
      
      Figure \ref{fig:continuation} presents a summary of our continuation results for {\it in-phase} (Type I) periodic orbits  $\omega_b$ in Case (A); this terminology is introduced in Section \ref{sec:application}.
      
      For $\lambda>0$ sufficiently  small, the breather is strongly localized on a few lattice sites; this behavior is captured by
Theorem  \ref{thm:breather}. For $\lambda$ less than but near $\lambda_\star$, where the parameter
 \[\epsilon\equiv\gamma_{\rm in}-\lambda\gamma_{\rm out}\]
 is small, the spectral (phonon) gap is small, and we expect the discrete breather spatial profile to decay very slowly on the lattice length-scale.
 This behavior is clearly indicated in panels (d)-(f) of Figure \ref{fig:continuation}.
  
 \begin{figure}
  \centering
  \includegraphics[scale=0.8]{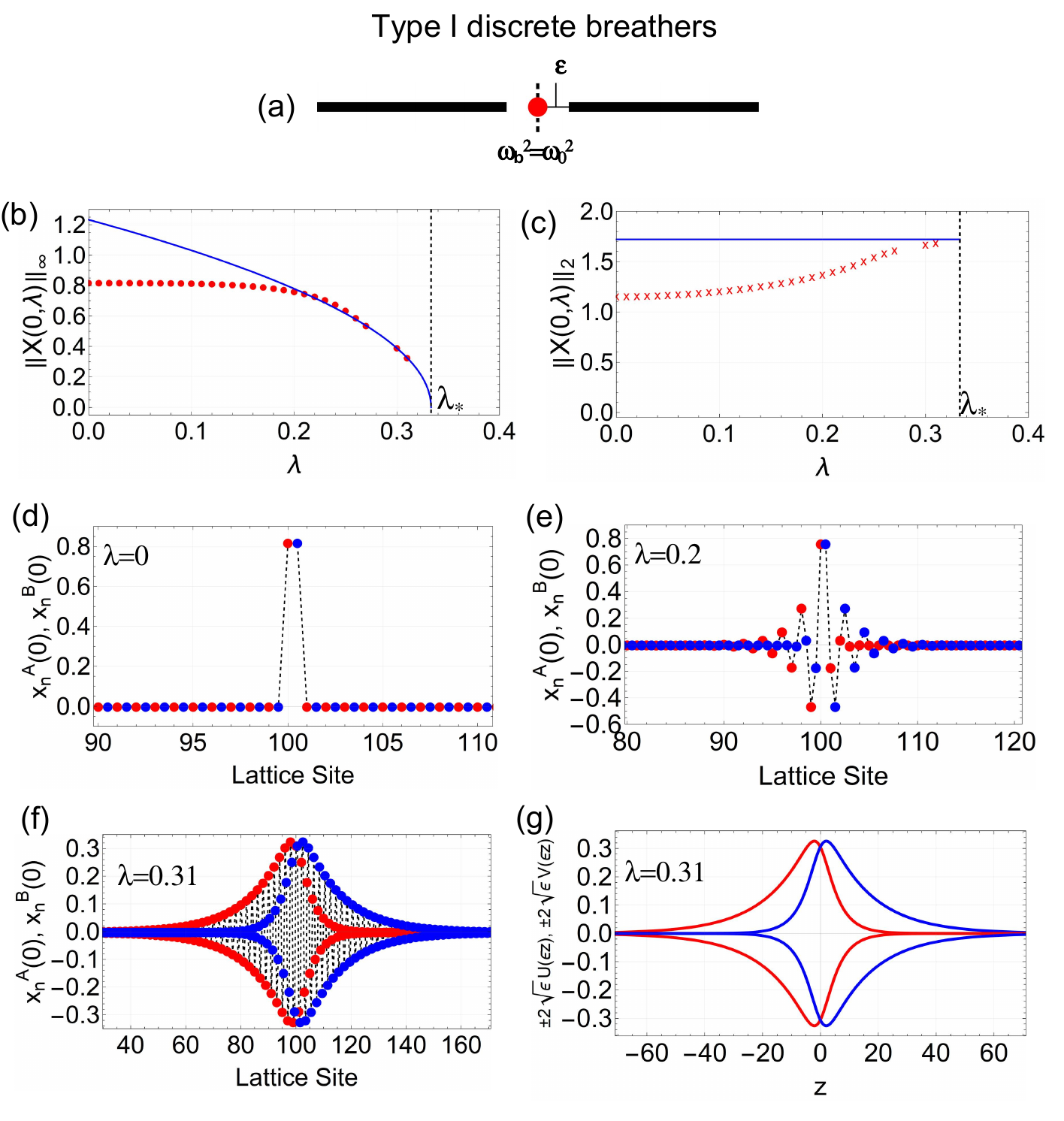}
  \vspace{-0.4cm}
  \caption{(a) phonon spectrum schematic (black) with breather frequency, $\omega_b$, (red); (b) $l^\infty$ norm of computed (dotted) discrete breather and its approximation from the weakly nonlinear long wave theory \eqref{envelope}; (c) $l^2$ norm of computed (x's) discrete breather and its approximation from the weakly nonlinear long wave theory \eqref{envelope}; (d), (e) and (f) show numerically computed discrete breather profiles for $\lambda=0,.2,.31$, respectively; and (g) shows envelope obtained from analytical approximation $z\mapsto (U(z),V(z))$, a homoclinic orbit of the system \eqref{eps7}. The continuation is initialized, for $\lambda=0$,
   with a anti-continuum in-phase periodic orbit of the nonlinear dimer \eqref{eq4} of frequency $\omega_b=\omega_0=\sqrt{V''(0)}=\sqrt{3}$, corresponding to the initial value parameter $a_*\sim 0.82$; see Section \ref{2classes}. Parameter values: $\Gamma=1$, $\gamma_{\text{in}}=0.5$, $\gamma_{\text{out}}=1.5$ ($\lambda_*=1/3$), $N=201$.}
  \label{fig:continuation}
\end{figure}

\item {\it Continuum envelope theory in the small phonon gap regime, $\lambda\to\lambda_\star$} 
As $\lambda$ approaches
 $\lambda_\star\equiv|\gamma_{\rm in}/\gamma_{\rm out}|$ ($\epsilon\to0$) the discrete breather has the structure of a weakly nonlinear wave-packet, whose amplitude decreases
  and width increases. 
  In Section \ref{sec:continuum} we consider separately the limits: $\lambda\uparrow\lambda_\star\ (\epsilon\downarrow0)$ and 
 $\lambda\downarrow\lambda_\star\ (\epsilon\uparrow0)$, which correspond, respectively, to the vanishing gap limit 
  in the topologically trivial 
  and topologically non-trivial linear phases.
  
In each of these scenarios we construct, by multiple scale asymptotic analysis, weakly nonlinear wave packets comprised of bulk spectral components corresponding to energies near the band crossing (see Figure \ref{fig:dimer}b). 
  This multiple scale expansion describes discrete breathers as a bifurcation from
  the phonon spectrum into the gap. Central to the construction are asymptotic gap soliton envelope equations corresponding to cases $\lambda\uparrow\lambda_\star$ and $\lambda\downarrow\lambda_\star$. These are related by 
  an emergent symmetry:
\begin{align} 
&\textrm{\it if $(U,V)^\top$ is a gap soliton which gives the discrete breather envelope}\label{emgnt}\\
&\textrm{\it for the regime $\lambda\uparrow\lambda_\star$, then the $(-V,U)^\top=-i\sigma_2(U,V)^\top$ defines }\nonumber\\
&\textrm{\it the envelope of a  discrete breather in the regime $\lambda\downarrow\lambda_\star$;}
\nonumber\end{align}
  see the further discussion below.
   
   We emphasize that in \cite{JW16,JW17} the phonon band edge is at a fixed frequency and the frequency of the bifurcating discrete breather (solitary standing wave) moves into the (semi-infinite) gap below the phonon spectrum. In contrast, in the present work, we prescribe the discrete breather frequency, $\omega_b$, which is fixed and outside the phonon spectrum. Our discrete breathers have this frequency all along the bifurcation curve, and it is the phonon spectrum that approaches $\omega_b$.

  \item {\it Agreement of continuum  envelope theory with numerical simulations for $\lambda\to\lambda_\star$, and $l^2-$ excitation threshold for in-gap discrete breathers. } The numerically computed discrete breather is very well approximated by the leading term of our asymptotic expansion for $0<\epsilon\ll1$. More precisely, consider the case where $\epsilon$, and hence the gap about $\omega_0$, is small and fixed and that the breather frequency, $\omega_b$, is
  in this gap.  For $\epsilon$ small, the phonon gap about frequency $\omega_0$ is approximately of width $\epsilon |\omega_0|^{-1}$ (Remark \ref{rmk:pgap}), and hence we may write:
\begin{equation} \omega_b= \omega_0 - \epsilon\frac{\nu}{2\omega_0},\label{nu-def}\end{equation}
where $-1<\nu<1$ measures the offset of $\omega_b$ from the center of the gap. We demonstrate that discrete breathers with frequencies in this small spectral gap are very well approximated by the wave form:
  \begin{equation}
\begin{pmatrix}
    x^A_n(t,\epsilon)\\  x^B_n(t,\epsilon) \end{pmatrix} \approx 2\sqrt{\epsilon} (-1)^n 
    \begin{pmatrix} U(\epsilon n;\nu)\\ V(\epsilon n;\nu) 
    \end{pmatrix} \cos\left[\ \left(\omega_0-\epsilon\frac{ \nu}{2\omega_0}\right)t\right],\quad 0<\epsilon\ll1.
        \label{envelope} \end{equation} 
The pair $(U(z;\nu), V(z;\nu))$ is a {\it gap soliton} solution of a coupled system of  nonlinear dispersive equations governing the slowly varying envelope; see \eqref{env-eqns}. We expect that such expansions can be made rigorous via bifurcation theory methods; see, for example, the derivation of discrete breathers of the discrete nonlinear Schr\"odinger equation in
   \cite{JW16,JW17}; see also \cite{IW10} and articles cited therein, as well as results concerning discrete breathers on diatomic Fermi-Pasta-Ulam-Tsingou (FPUT) lattices\cite{james04}.  

Representative phase portraits of the planar dynamical system governing $(U(\cdot;\nu),V(\cdot;\nu))$
 are plotted for representative values of $\nu$ in Figure \ref{phase}. Gap solitons are homoclinic orbits which connect $(0,0)$ to itself.  We obtain an excellent numerical fit to \eqref{envelope} with computed discrete breathers for
$\lambda$ less than and near  $\lambda_\star$; see Figure \ref{fig:continuation}.\\ 
Figure \ref{fig:continuation} also shows that while the amplitude ($l^\infty$ norm) of ``in-gap'' discrete breathers tends to zero as $\lambda\to\lambda_\star$ (for the cubic nonlinear lattice model)
there is a strictly positive {\it excitation threshold}
 with respect to the $l^2$ norm; there is a minimum $l^2$ norm below which there are no discrete breathers near the continuum limit; see \cite{We:99}. In the present setting, this is a consequence of the asymptotic nonlinear Dirac equation with cubic nonlinearity, which has a \underline{critical} dilation scaling;
 see Section \ref{sec:nonlin} for a discussion in the context of  general nonlinearities. In contrast, for the cubic nonlinear model, states whose frequencies bifurcate ``out-of-gap" (above the optical or below the acoustic bands) have $l^2$ which tend to zero as the frequency approaches the edge; their envelopes are governed by a nonlinear Schroedinger equation for which the cubic nonlinearity has \underline{subcritical} scaling properties \cite{SulemSulem,Fibich}.

\item{\it Bifurcation of discrete breathers into the (linear) topologically non-trivial regime.} Note, in view of the symmetry \eqref{emgnt} of the continuum envelope equations, we have that from \eqref{envelope} we obtain a bifurcation of discrete breathers,
for small and \underline{ negative} $\epsilon=\gamma_{\rm in}-\lambda\gamma_{\rm out}$, corresponding to the topologically non-trivial phase of the linear SSH:
\begin{equation} -i\sigma_2\ \begin{pmatrix}
    x^A_n(t,-\epsilon)\\  x^B_n(t,-\epsilon) \end{pmatrix},\quad\quad  0<-\epsilon\ll1. 
\label{DB-top}\end{equation}
While the continuation from the anticontinuous limit (linearly topologically trivial band structure) breaks down as $\lambda\uparrow\lambda_\star$ (the phonon gap closes), the emergent continuum symmetry \eqref{emgnt} provides a means for continuation into the regime where there is a linear topologically non-trivial band structure.

\item{\it Chirality of mid-gap discrete breathers.} 
 Finally, we note a {\it chiral} feature of ``mid-gap'' discrete breathers displayed in Figure \ref{fig:continuation}. The decay of discrete breathers  as $|n|$ tends to infinity is determined by the available decaying solutions of the asymptotic linear problem. Since the discrete breather frequency in this case is at the center of the gap, the corresponding linear states are those of the linear SSH model of zero energy; \eqref{ssh} with $E=0$. Zero energy solutions which decay as $n\to+\infty$ and zero energy solutions which decay as $n\to-\infty$ are concentrated on distinct sublattices and this is precisely reflected in the large $|n|$ behavior of Figure \ref{fig:continuation}f. This dichotomy at mid-gap is also reflected in the continuum theory via the tangency of invariant manifolds at the origin; Section \ref{sec:numerics} provides a very detailed discussion. This is consistent with modeling and experiments in \cite{solny,bloch}.
  \end{enumerate}
 
\subsection{Relation to previous work}\label{prev-work}
There is an extensive theoretical and applied literature devoted to the study of discrete breathers on a wide class of lattice structures; see for instance the review \cite{flach-gorbach-08}.  Discrete breathers on diatomic FPUT-like lattices are studied in \cite{livi97}, \cite{man-03}, and \cite{james04}. The nonlinearities we consider act ``on-site" but the underlying linear band structure of such FPUT systems and those we consider are similar.  We also mention the experimental works \cite{boech-10} and \cite{bloch} which study discrete breathers and gap solitons in dimerized granular crystals and photonic lattices modeled by FPUT lattice systems, respectively. The current study is motivated by recent works in the physics literature investigating the interplay between topological band-structures and nonlinear effects, primarily in the context of discrete and continuum SSH and other type photonic and mechanical systems; see, for example, \cite{Alu17}, \cite{theo19}, \cite{Theo21}, \cite{Ruzz18}, \cite{delplace}. Note that the topological character of the discrete SSH is due to a chiral symmetry which emerges in the tight binding limit of a class of continuum dimer models; see \cite{shapiro-weinstein}. Finally, related to the weakly nonlinear continuum theory and gap solitons we derive in the small phonon gap regime, we note earlier work on gap solitons in nonlinear periodic optical media, governed by nonlinear Dirac type models; see, for example, \cite{aceves}, \cite{christ}, \cite{goodman}, and the recent studies \cite{smirnova}, \cite{chaunsali} and references cited therein.

\section*{Acknowledgments}
A.H. was supported in part by the {\it Simons Collaboration of Extreme Wave Phenomena Based on Symmetries} and the Air Force Office of Scientific Research with Grant No. FA9550-23-1-0144.  M.I.W. and H.L. were supported in part by National Science Foundation grants DMS-1620418, DMS-1908657 and DMS-1937254 as well as Simons Foundation Math + X Investigator Award \#376319.
The authors would like to thank Andrea Al\`{u}, Yakir Hadad and Panayotis Kevrekides for informative and stimulating discussions.

\section{Discrete breathers in the anti-continuous regime}
\label{sec:existence}

We seek time periodic and spatially localized solutions of coupled nonlinear lattice system:
\begin{align}\label{eq5-gen}
\begin{pmatrix}
\ddot{x}_{n}^A+V'(x_{n}^A)-\gamma_{\text{in}}x_{n}^B\\
\ddot{x}_{n}^B+V'(x_{n}^B)-\gamma_{\text{in}}x_{n}^A 
\end{pmatrix} 
+ \lambda
\left[{\bf R}\begin{pmatrix}x^A\\ x^B\end{pmatrix}\right]_n = \begin{pmatrix}0\\ 0\end{pmatrix},\quad n\in\mathbb{Z},
\end{align}
for $\lambda$ real non-zero and sufficiently small.
The mapping $x\mapsto {\bf R}[x]$ is assumed to be a bounded linear map on $l^2(\mathbb{Z};\mathbb{C}^2)$ with exponentially decaying matrix elements:
\begin{equation}\label{Rml-decay}
\|R_{ml}x\|_{\mathbb{C}^2}\le C |\nu_R|^{|m-l|}\|x\|_{\mathbb{C}^2},\quad m,l\in\mathbb{Z},
\end{equation}
where $0<|\nu_R|<1$ and $0<C<\infty$.
The case of nearest neighbor interactions \eqref{eq1A} is an example.
 We assume that the potential, $V$,  in \eqref{eq5-gen} satisfies:
\begin{equation}\label{pot}
  V\in C^2(\mathbb{R}),\quad V(-x)=V(x).  
\end{equation} 
\begin{remark}
For Theorem \ref{thm:breather} on the existence of discrete breathers we only require that $x\mapsto {\bf R}[x]$ is bounded linear map on $l^2(\mathbb{Z};\mathbb{C}^2)$. The exponential decay hypothesis on the coupling matrix $R_{ml}$
 is used in Section \ref{sec:localization} to prove spatial exponential decay of discrete breathers.
\end{remark}

For $\lambda=0$  there is no coupling among the
individual dimers in the array; this is the {\it anti-continuous limit}:
 \begin{align}\label{eq3a}
&\ddot{x}_{n}^A=-V'(x_{n}^A)+\gamma_{\text{in}}x_{n}^B\\ \nonumber
&\ddot{x}_{n}^B=-V'(x_{n}^B)+\gamma_{\text{in}}x_{n}^A, \quad n\in\mathbb{Z}.
\end{align}

We consider the simplest type of solution \eqref{eq3a} in which only the  $n=0$ oscillators are excited and all other dimer amplitudes, $n\neq 0$, are set to zero.  Hence, we  seek solutions to the system
\begin{equation}
\begin{aligned}
&\ddot{x}_{0}^A=-V'(x_{0}^A)+\gamma_{\text{in}}x_{0}^B\\ 
&\ddot{x}_{0}^B=-V'(x_{0}^B)+\gamma_{\text{in}}x_{0}^A.
\end{aligned}
\label{eq4}
\end{equation}
A general analysis of \eqref{eq4} requires a study of a four-dimensional phase space. In Section \ref{2classes} we consider two classes of periodic orbits, in-phase and out-of-phase; each leads to a reduction of \eqref{eq4} to a two-dimensional phase space.
For now, we assume that $(x^A_*(t),x^B_*(t))$ denotes a periodic solution of \eqref{eq4} of period $T_b$ (frequency $\omega_b=2\pi/T_b$).
Hence, for $\lambda=0$, the infinite  dimer array \eqref{eq5-gen} has  a breather solution 
\begin{equation}\label{Xstar}
    X_*(t)=\Big\{\cdots,0,0,\begin{pmatrix} x_{*}^A(t)\\x_{*}^B(t)\end{pmatrix},0,0,\cdots\Big\}.
\end{equation}
Equivalently, in terms of the mapping
\begin{align}\label{eq5}
\begin{pmatrix}
\ddot{x}_{n}^A+V'(x_{n}^A)-\gamma_{\text{in}}x_{n}^B\\
\ddot{x}_{n}^B+V'(x_{n}^B)-\gamma_{\text{in}}x_{n}^A 
\end{pmatrix} 
+ \lambda
\left[{\bf R}\begin{pmatrix}x^A\\ x^B\end{pmatrix}\right]_{n\in\mathbb{Z}},
\end{align}
we have 
\[ F(X_\star,0) = 0.\]
Our first goal is to construct a mapping $\lambda\mapsto X^\lambda(t)$, defined 
 for all  real $\lambda\ne0$ and sufficiently small in a Banach space of  
  $T_b-$ periodic in time, spatially decaying sequences, such that 
 \begin{equation}
 \label{F lambda implicit equation}
      F(X^\lambda,\lambda) = 0.
 \end{equation}

We now introduce a function space framework appropriate for an application of the implicit function theorem.
Let $ \mathcal{H}^2_{_{T_b}}$ denote the  Banach space of  infinite sequences of 
 time-periodic $H^2=H^2([0,T_b];\mathbb{R}^2)-$ functions or {\it loop space } given by
\begin{align}
    \mathcal{H}^2_{_{T_b}}&=\Big\{ X(t)= \{x_n(t)\}_{n\in\mathbb{Z}} :  x_n\in H^2(\mathbb{R}/\mathbb{Z}T_b),\ X(t)=X(-t) \Big\}\nonumber \\
  &  \label{loop}
\end{align}
endowed with the norm given by
 \[ \|X(t)\|_{_{\mathcal{H}^2}}^2 =  \sum_{n\in\mathbb{Z}} \|x_n\|^2_{H^2} = \sum_{n\in\mathbb{Z}} 
 \Big\|\begin{pmatrix} x^A_n\\ x^B_n\end{pmatrix}\Big\|_{H^2}^2 ,\]
 and $\mathcal{H}^0_{_{T_b}}$ given by
 \begin{align}
   \mathcal{H}^0_{_{T_b}}&=\Big\{ X(t)= \{x_n(t)\}_{n\in\mathbb{Z}} :  x_n\in L^2(\mathbb{R}/T_b\mathbb{Z}), X(t)=X(-t)\ \textrm{a.e.}\  \Big\}\nonumber \\
  &  \label{M-def}
\end{align}
with norm
 \[ \|X(t)\|_{_{\mathcal{H}^0}}^2 =  \sum_{n\in\mathbb{Z}} \|x_n\|^2_{L^2} = \sum_{n\in\mathbb{Z}} 
 \Big\|\begin{pmatrix} x^A_n\\ x^B_n\end{pmatrix}\Big\|_{L^2}^2 ,\]
%  \footnote{\textcolor{red}{The condition that $\dot{X}(0)=0$ in \eqref{loop} effectively %fixes the phase of the periodic functions living in $\mathcal{L}_{T_b}$}}
 %By the hypotheses on $V(z)$ 
 %
 %\footnote{ \textcolor{red}{$V\in C^2$ and EVEN?? -- Have we spelled out hypotheses on %$V$?}}
 %
 we have that 
 \[ F:  \mathcal{H}^2_{_{T_b}}\times\mathbb{R}\to  \mathcal{H}^0_{_{T_b}}\quad \textrm{is a $C^1$ mapping.}\]

\begin{figure}[H]\label{resonant}
  \centering
  \includegraphics[scale=0.6]{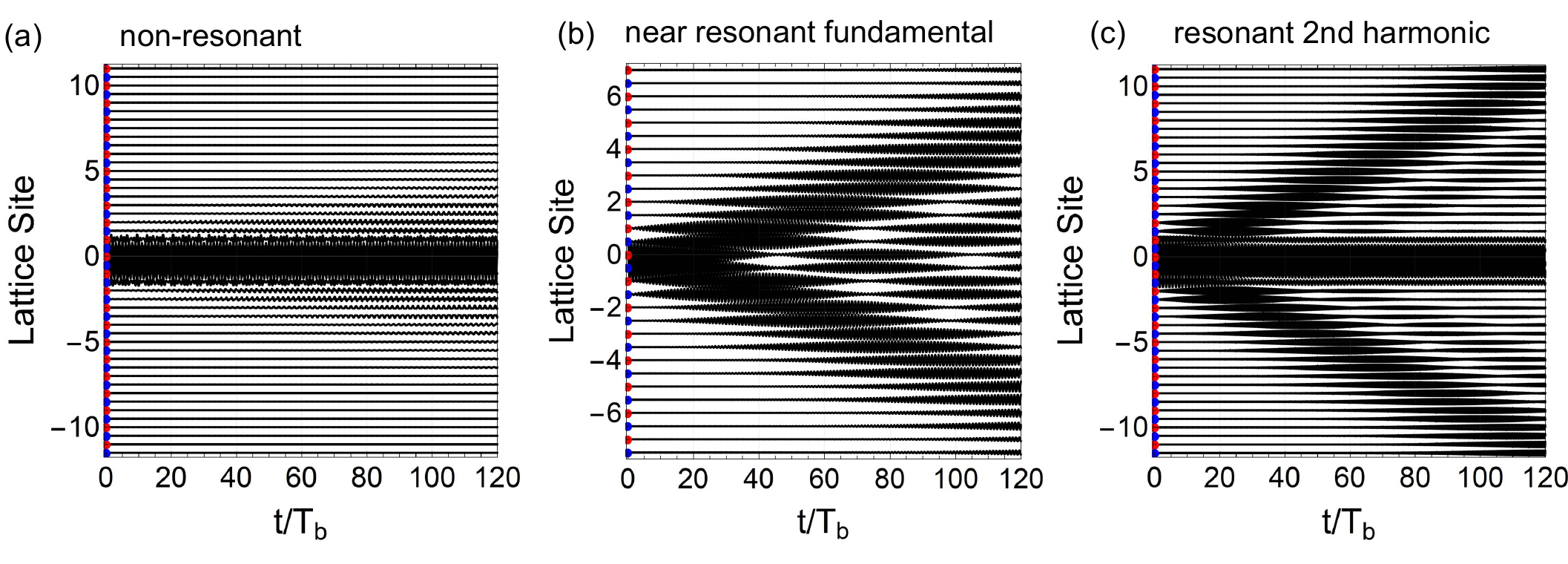}
   \caption{Panel (a) shows the spatio-temporal evolution of non-resonant Type I seeded data (see Section \ref{sec:application} for definition) with weak coupling ($\lambda=0.1$) over 120 anti-continuum breather periods. In contrast, panels (b) and (c) break the non-resonance condition, \eqref{nonres1}, and therefore cannot be continued. Here, the phonon coupling is evident.  Panel (b) is very close to the linear regime and nearly resonantes with the fundamental ($n=1$ in the theorem).  Panel (c) resonates with the second harmonic ($n=2$ in the theorem).}
\end{figure}

We now state a theorem on the existence and uniqueness of discrete breather solutions to \eqref{nloc-case}.
\begin{theorem}[Breathers near the anticontinuum limit]\label{thm:breather}
 Consider the nonlinear dimer array \eqref{eq5-gen} with potential, $V$, satisfying \eqref{pot}.   Fix a periodic solution, $X_*=(x_*^A,x_*^B)^\top$,  of period $T_b$ associated with the isolated ($\lambda=0$) dimer \eqref{eq4}. 
  We make the following two additional hypotheses:
  \begin{enumerate}
\item[(a)] Non-resonance: 
\begin{equation}
  (n\omega_b)^2 \ne V''(0)\pm\gamma_{\rm in},\quad \textrm{for all $n\in\mathbb{Z}$.}
   \label{nonres1}   
   \end{equation}
%\begin{equation}
%  \left |V''(0)-\left(n\omega_b\right)^2\right|\neq|\gamma_{\text{in}}%|,\quad \omega_b\equiv \dfrac{2\pi n}{T_b}\quad\textrm{for all}\ %n\in\mathbb{Z}
%   \label{nonres}   
%   \end{equation}
   %
  %
   \item[(b)] Non-degeneracy:  The nullspace of the operator
 \begin{equation}  
 L_*=
 \begin{pmatrix}
    \dfrac{d^2}{dt^2}+V''(x_*^A(t)) & 0\\ 
   0 & \dfrac{d^2}{dt^2}+V''(x_*^B(t)) \end{pmatrix}
   -\gamma_{\text{in}}
   \begin{pmatrix} 0&1\\ 1&0\end{pmatrix}
 \label{L*def}\end{equation}
acting in the space $ \mathcal{H}^2_{_{T_b}}$ is \underline{empty}.
(The operator  $L_*$ is the linearized operator of 
    the isolated dimer dynamical system \eqref{eq4} about the periodic orbit, $X_*$.)
    \end{enumerate}
Then, under hypotheses (a) and (b) there exists $\lambda_b>0$ and $C^1$ curve
\[\textrm{
$\lambda\in[0,\lambda_b)\mapsto X^\lambda\in
 \mathcal{H}^2_{_{T_b}}$ such that $X^0=X_*$ and  $F(X^{\lambda};\lambda)=0$ for all $0\leq\lambda<\lambda_b$.} \]
\end{theorem}
\begin{remark}[On the non-resonance condition \eqref{nonres1}]\label{non-res}
Let $\lambda=0$. The linearization of the dimer system is given by the block-diagonal system:
\begin{align}
\ddot\eta_0^A &=-V''(x_*^A(t))\eta_0^B+\gamma_{\text{in}}\eta_0^B\\ 
\ddot\eta_{0}^B &=-V''(x_*^B(t))\eta_{0}^A+\gamma_{\text{in}}\eta_{0}^A,\
\label{decup-dimer}
\end{align}
in the $n=0$ dimer and, for $n\ne0$
\begin{equation}
\begin{aligned}
&\ddot{\eta}_{n}^A=-V''(0)\eta_{n}^B+\gamma_{\text{in}}\eta_{n}^B\\ 
&\ddot{\eta}_{n}^B=-V''(0)\eta_{n}^A+\gamma_{\text{in}}\eta_{n}^A.
\end{aligned}
\label{decup-dimer1}
\end{equation}
Each $n\ne0$ block has the four distinct frequencies, $\tilde\omega$, given by
 the solutions of: $\tilde\omega^2= V''(0)+\gamma_{\rm in}$ and 
$\tilde\omega^2= V''(0)-\gamma_{\rm in}$. Hence, the infinite system 
 has the identical four distinct frequencies, each now having infinite multiplicity,
with corresponding modes supported on distinct dimer cells. For $\lambda\ne0$ and small, these infinite multiplicity frequencies 
perturb to {\it phonon bands}, intervals of spectra, corresponding to the time harmonic solutions, $\xi e^{-i\omega t},\ \xi\in l^2(\mathbb{Z})$, of 
\begin{align}\label{lin-lam}
\begin{pmatrix}
\ddot{x}_{n}^A+V''(0)x_{n}^A-\gamma_{\text{in}}x_{n}^B\\
\ddot{x}_{n}^B+V''(0)x_{n}^B-\gamma_{\text{in}}x_{n}^A 
\end{pmatrix} 
+ \lambda 
\left[{\bf R}\begin{pmatrix}x^A\\ x^B\end{pmatrix}\right]_n
= \begin{pmatrix}0\\ 0\end{pmatrix}.
\end{align}
The non-resonance condition \eqref{nonres1} ensures that nonlinearity-induced harmonics of the breather frequency $\omega_b$ do not resonate 
with the phonon spectrum. Such resonances are known to lead to the slow
resonant radiation damping of coherent structures; see Figure \ref{resonant}, and also, for example, 
\cite{SW:99,SW:04,SW:05,W:15}.
\end{remark}
\begin{remark}[On the non-degeneracy condition]
Below, in Section \ref{sec:application}, we show in a concrete family of examples, that the non-degeneracy hypothesis holds generically. Its verification in any individual case can easily be addressed numerically; see Figure \ref{fig:phaseports}.
    \end{remark}
\begin{proof}[Proof of Theorem \ref{thm:breather}] 
The proof is based on the implicit function theorem; see, for example, \cite{nirenberg}.  Clearly we have $F(X_*,0)=0$.  
To apply the implicit function theorem we must check that differential of the mapping $F$
 at $(X,\lambda)=(X_*,0)$, 
 \[ F_X(X_*,0):  \mathcal{H}^2_{_{T_b}}\to  \mathcal{H}^0_{_{T_b}}\] 
 is one to one and onto, and that the inverse is bounded. 
The differential with respect to $\lambda$ at any $(X,\lambda)$ is 
\begin{equation}
\label{eq: F diff wrt lambda}
   \left[ \frac{\partial F(X;\lambda) }{\partial \lambda} \right]_n
    %\left\{ \begin{pmatrix}
  %  -\gamma_\text{out} x^B_{n-1} \\ -\gamma_\text{out} x^A_{n+1} %\end{pmatrix} \right\}_{n\in\mathbm Z}
   = \left[{\bf R}\begin{pmatrix}x^A\\ x^B\end{pmatrix}\right]_n .
\end{equation}
 
The differential with respect to $X$ is given by the block diagonal operator 
\begin{equation}
F_X(X_*;0)Y=
 \begin{pmatrix}
   \ddots &  &  & & \\ 
   & L_0 &  & &\\ 
   &  &  L_* & &\\ 
   &  &   & L_0 & \\
   &  &   &  & \ddots
 \end{pmatrix}
\begin{pmatrix}
\vdots\\
y_{-1}\\
y_{0}\\
y_{1}\\
\vdots
\end{pmatrix},\quad y_j\in H^2(\mathbb{R} / \mathbb{Z}T_b;\mathbb{R}^2),\ j\in\mathbb{Z},
\label{F_X}
\end{equation}
where the $2\times2$ block operators are $L_*$, displayed in \eqref{L*def}, and 
\begin{align}\label{eq7}
    &L_0=\left[\dfrac{d^2}{dt^2}+V''(0)\right] I_{2\times2}-\gamma_{\text{in}}
     \begin{pmatrix} 0&1\\ 1&0\end{pmatrix}\ .
\end{align}
Here, $I_{2\times2}=\sigma_0$ denotes the $2\times2$ identity matrix.

We first claim that the operators $L_0$ and $L_*$ both map 
\begin{equation} 
H^2_{_{T_b}}\equiv \{x\in H^2(\mathbb{R}/\mathbb{Z} T_b;\mathbb{R}^2):\ x(-t)=x(t)\}
\label{H2Tb}\end{equation}
to
\[
H^0_{_{T_b}}\equiv\{x\in L^2(\mathbb{R}/\mathbb{Z} T_b;\mathbb{R}^2):\ x(-t)=x(t)\ {\rm a.e.}\ \}.
\]
We further claim that these maps are one to one, onto and have a bounded inverse:
\begin{align} 
\|L_0^{-1} f \|_{H^2_{_{T_b}}} &\le C_0  \|f \|_{H^0_{_{T_b}}} \label{L0-1bd}\\
\|L_*^{-1} f \|_{H^2_{_{T_b}}} &\le C_*  \|f \|_{H^0_{_{T_b}}},\label{L*-1bd}\
\end{align}
where $C_0$ and $C_*$ are constants.

The bound \eqref{L0-1bd} on $L_0^{-1}$ follows by an explicit Fourier series calculation using 
the Non-resonance Hypothesis \eqref{nonres1}. Indeed, consider the equation
 \[ L_0y=f,\quad {\rm where}\quad f\in L^2\]
 is given by a Fourier series:
 \begin{equation}
\begin{pmatrix}
f^A(t)\\
f^B(t)
\end{pmatrix}
=\sum_{n\in\mathbb{Z}}
\begin{pmatrix}
\hat{f}^{n,A}\\
\hat{f}^{n,B}
\end{pmatrix}
e^{2n\pi i t/T_b}.
\end{equation}
We seek a solution 
\begin{equation}
y(t)=\begin{pmatrix}
y^A(t)\\
y^B(t)
\end{pmatrix}
=\sum_{n\in\mathbb{Z}}
\begin{pmatrix}
\hat{y}^{n,A}\\
\hat{y}^{n,B}
\end{pmatrix}
e^{2n\pi i t/T_b}
\end{equation}
 and find that a solution  $y(t)\in H_{_{T_b}}^2$ can be constructed for arbitrary $f\in L^2_{_{T_b}}$
  if and only if 
\begin{equation}\label{eq8}
\text{det}
    \begin{pmatrix}
    -\left(n\omega_b\right)^2+V''(0) & -\gamma_{\text{in}}\\
    -\gamma_{\text{in}} & -\left(n\omega_b\right)^2+V''(0)
    \end{pmatrix}
    \neq 0\quad \textrm{for all $n\in\mathbb{Z}$}.
\end{equation}
Equivalently, $\left( (n\omega)^2-V''(0) \right)^2 - \gamma_{\rm in}^2\ne0$ or 
\[ \left( (n\omega)^2-V''(0)-\gamma_{\rm in}\right)
\ \left( (n\omega)^2-V''(0)+\gamma_{\rm in}\right)\ne0,\quad 
\textrm{for all $n\in\mathbb{Z}$}. \]
This is precisely the Nonresonance Hypothesis \eqref{nonres1} so the proof of  the bound \eqref{L0-1bd} is complete.

The bound \eqref{L*-1bd} on $L_*^{-1}$ follows from 
 the Non-degeneracy Hypothesis; see \eqref{L*def}. Indeed, the spectrum of $L_*$ acting in $H^0_{_{T_b}}$ is discrete, and non-degeneracy implies that $0\not\in{\rm spec}(L_*)$. Hence, 
  $\|L_*^{-1}f\|_{L^2_{T_b}}\le [{\rm dist}\left(0,{\rm spec}(L_*)\right)]^{-1} \|f\|_{L^2_{T_b}}$. Standard elliptic theory
   implies the bound \eqref{L*-1bd}.

The inverse of $F_X(X_*;0)$ acting on sequences \[ Y=(\dots,y_{-1},y_0,y_1,\dots)\in \mathcal{H}^0_{_{T_b}}\]
is then given by
\begin{equation}
F_X(X_*;0)^{-1}Y=
 \begin{pmatrix}
   \ddots &  &  & & \\ 
   & L_0^{-1} &  & &\\ 
   &  &  L_*^{-1} & &\\ 
   &  &   & L_0^{-1} & \\
   &  &   &  & \ddots
 \end{pmatrix}
 \begin{pmatrix}
\vdots\\
y_{-1}\\
y_{0}\\
y_{1}\\
\vdots
\end{pmatrix},\quad y_j\in H^0(\mathbb{Z}T_b;\mathbb{R}^2),\ j\in\mathbb{Z},
\label{F_X-inv}
\end{equation}
which satisfies the bound
\[ \|F_X(X_*;0)^{-1}Y\|_{_{\mathcal{H}^2_{_{T_b}}}}^2 \le 
C_1 \sum_{n\in\mathbb{Z}} \|y_j\|_{L^2}^2 = C_1 \|Y\|_{_{\mathcal{H}^0_{_{T_b}}}}^2,
\]
where $C_1=\max\{C_0,C_*\}$. We may now apply the implicit function theorem
 to obtain the existence of a curve $\lambda\in[0,\lambda_b)\mapsto X^\lambda\in  \mathcal{H}^2_{_{T_b}} $ such that $X^\lambda=X_*$ for $\lambda=0$ and $F(X^\lambda,\lambda)=0$ for all $\lambda\in[0,\lambda_b)$. This completes the proof of
 Theorem \ref{thm:breather}.

\section{Application of Theorem \ref{thm:breather} }\label{sec:application}
In this section we introduce two classes of periodic solutions of isolated dimer dynamical system \eqref{eq3a}. We then verify the resonance and non-degeneracy hypotheses of Theorem \ref{thm:breather} to obtain curves
 of discrete breathers in the weak coupling (anti-continuous) regime.
 \subsection{Two classes of periodic orbits}\label{2classes}

\subsection*{\bf Type I States (in-phase):}
\begin{equation} a_*\equiv x_{*}^A(0)=x_{*}^B(0)\neq 0 \text{ and } \dot{x}_{*}^A(0)=\dot{x}_{*}^B(0)=0.
\label{ic1}\end{equation}
Let $a_*$ be such that 
\begin{equation}
\ddot{z}=-V'(z) + \gamma_{\text{in}}z
\label{z+eqn}
\end{equation}
with initial data $z(0)=a_*$ and $\dot{z}(0)=0$
has a periodic solution $z_*^{(I)}(t)$
Then,
 \[ \begin{pmatrix} x_*^A(t) \\ x_*^B(t) \end{pmatrix} = 
z_*^{(I)}(t)\ \begin{pmatrix} 1 \\ 1\end{pmatrix} \] 
 is a periodic solution of 
 \eqref{eq3a} with initial conditions \eqref{ic1}.
\subsection*{\bf Type II States (out-of-phase):}
\begin{equation}
 a_*=x_{*}^A(0)=-x_{*}^B(0)\neq 0 \text{ and } \dot{x}_{*}^A(0)=\dot{x}_{*}^B(0)=0.
\label{ic2}
\end{equation}
Let $a_*$ be such that 
\begin{equation}
\ddot{z}=-V'(z) - \gamma_{\text{in}}z
\label{z-eqn}
\end{equation}
with initial data $z(0)=a_*$ and $\dot{z}(0)=0$
has a periodic solution $z_*^{(II)}(t)$
Then, 
\[ \begin{pmatrix} x_*^A(t) \\ x_*^B(t) \end{pmatrix} = 
z_*^{(II)}(t)\ \begin{pmatrix} 1 \\ -1\end{pmatrix} \] 
is a periodic solution of 
 \eqref{eq3a} with initial conditions \eqref{ic2}.
Note, the requirement that $V$ is even in \eqref{pot}, implies that there are Type II dimer periodic solutions. 

 \begin{figure}[H]
  \centering
  \includegraphics[scale=0.71]{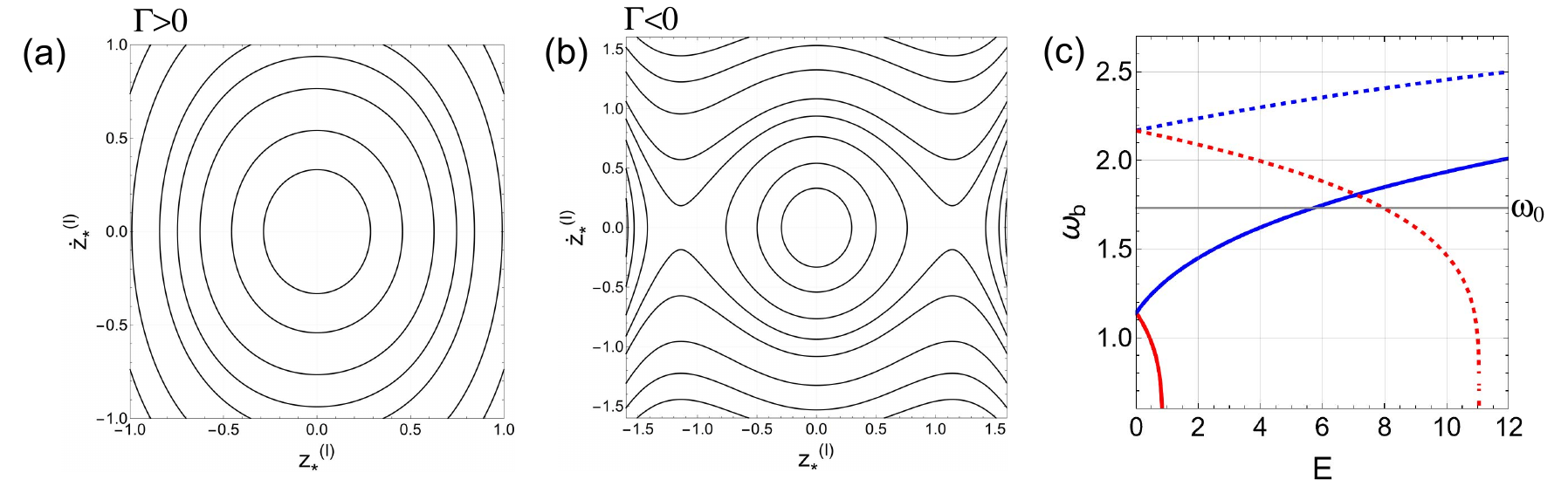}
  \vspace{-.5cm}
  \caption{Phase plots of typical Type I or II states for the potential $V(x)=\frac32 x^2+\frac{\Gamma}4 x^4$. Panel a: hardening case, $\Gamma>0$; Panel b: softening case, $\Gamma<0$. Panel c: Representative plots of the isolated dimer frequency as a function of energy for Type I states (solid-curves) and Type II states (dashed curves). Curves for hardening nonlinearity are in blue and for softening nonlinearity in red.  The horizontal line depicts the midgap phonon frequency, $\omega_0=\sqrt{V''(0)}=\sqrt{3}$, for reference.}
\label{fig:phaseports}\end{figure}

The phase portraits in Figure \ref{fig:phaseports} (a)-(b) display periodic solutions of Type I to the anti-continuum problem, for a hardening and softening quartic potential, respectively. The period $T_b=T_b(a_*)$ is given by the expression:
%\begin{equation}\label{theo_anti}
%    T_b^{(I,II)}=4\int_0^{a_*}\dfrac{dz}{\sqrt{2\left(V(a_*)-V(z)\right)\mp %\left\gamma_{\text{in}}(a_*^2-z^2\right)}}.
%\end{equation}
\begin{equation}\label{theo_anti}
    T_b^{(I,II)}(E)=4\int_0^{a_*(E)}
    \dfrac{dz}{\sqrt{2\left(E-V(z)\right)\pm \gamma_{\text{in}}z^2}},\quad 
     E-V(a_*)\pm\gamma_{\rm in}\frac{a_*^2}{2}=0.
\end{equation}
\noindent
The upper choice of sign in \eqref{theo_anti} corresponds to in-phase (Type I) periodic solutions and the expression with the lower choice of sign corresponds to out-of-phase (Type II) periodic solutions. In panel (c) of Figure \ref{fig:phaseports}, the isolated dimer angular frequency, $\omega_b(E)=2\pi/T_b(E)$, is plotted as a function of its energy, $E$, using \eqref{theo_anti}, for Type I (solid-curves) and Type II (dashed-curves) data, for both a softening (red) and hardening (blue) potential, at a fixed of value of $\gamma_{\text{in}}$.

 We now discuss the application of Theorem \ref{thm:breather}  to prove that isolated dimer solutions 
 of Type I and Type II  continue to discrete breather solutions (exciting all lattice sites) for all nonzero $\lambda$, which are sufficiently small. To apply Theorem \ref{thm:breather},
  we assume the Non-resonance Condition \eqref{nonres1} (see Figure \ref{resonant} for an illustration of when this condition is and is not met and also the discussion in Remark \ref{non-res}) and then need only verify the Non-degeneracy Condition (b) that
 \[\textrm{ the nullspace of $L_*$, acting in the space ${H}^2_{_{T_b}}$, is given by $\{0\}$.}
 \]
 \subsection{Verification of the non-degeneracy condition (b) for Type I and Type II anticontinuum periodic orbits}
The linearized operator about the solutions, $z_*^{(I)}$ and $z_*^{(II)}$, of Type I and Type II
 are given by the expression
 \begin{equation}
 L_*=
\left( \dfrac{d^2}{dt^2}+V''(z_*^{(J)}(t)) \right) \sigma_0 
   -\gamma_{\text{in}} \sigma_1
 \label{L*def-J}\end{equation}
 where $J=I$ corresponds to states of Type I and $J=II$ corresponds to states of Type II.
 
 To study the nullspace of $L_*$, acting in the space ${H}^2_{_{T_b}}$, we first diagonalize it using the eigenvectors of $\sigma_1$; let 
 $M=\begin{pmatrix} 1&1\\1&-1\end{pmatrix}$. Then, $\sigma_1M=M{\rm diag}(1,-1)$ and 
  hence
 \begin{equation}\label{L*sim}
M^{-1}L_*M =
\begin{pmatrix} 
\dfrac{d^2}{dt^2}+V''(z_*^{(J)}(t)) -\gamma_{\rm in} & 0\\
0& \dfrac{d^2}{dt^2}+V''(z_*^{(J)}(t)) +\gamma_{\rm in} 
\end{pmatrix}
 \end{equation} 
Therefore, determining the nullspace of $L_*$, acting in the space ${H}^2_{_{T_b}}$, reduces to separately determining the ${H}^2_{_{T_b}}$- nullspaces of the scalar operators
\begin{align}\label{decomp}
L_{*-} &= \dfrac{d^2}{dt^2}+V''(z_*^{(J)}(t)) -\gamma_{\rm in},\quad {\rm and} \nonumber\\
L_{*+} &= \dfrac{d^2}{dt^2}+V''(z_*^{(J)}(t)) +\gamma_{\rm in}
    \end{align}
    
    \subsubsection{The nullspaces of $L_{*+}$ and $L_{*-}$ for Type I states}\label{sec:typeI}
    
    {\ } \medskip
    
 \noindent   {\bf ${H}^2_{_{T_b}}$- nullspace of $L_{*-}$ : }\
     The ${H}^2_{_{T_b}}$- nullspace of $L_{*-}$ has dimension no larger than two.
     We first produce two linearly independent solutions which span the 
      set of all solutions to the second order ODE $L_{*-}Z=0$ and then investigate whether any element in this two-dimensional space qualifies as an element of the 
        ${H}^2_{_{T_b}}$- nullspace of $L_{*-}$, in particular whether any of these solutions is even and $T_b$ periodic. 
        
     Type I states correspond to periodic solutions, $z_*^{(I)}(t)$, of the equation
$\ddot{z}=-V'(z) + \gamma_{\text{in}}z$. Differentiation with respect to $t$ yields that 
$Z=\dot{z}_*^{(I)}(t)$ satisfies $L_{*-}Z=0$ with initial conditions $Z(0)=0$ and $\dot{Z}(0)\ne0$. However, since $z_*^{(I)}(t)$ is an even function of $t$, $Z(t)$ is odd and hence $Z$ does not belong to the ${H}^2_{_{T_b}}$- nullspace of $L_{*-}$. 

To obtain a second, linearly independent, solution we proceed as follows. Denote
 by $z(t,a)$ the solution of the initial value problem
 \begin{equation} \ddot{\zeta}=-V'(\zeta) + \gamma_{\text{in}}\zeta,\quad \zeta(0)=a,\quad \dot{\zeta}(0)=0.\label{zeta-a}\end{equation}
 Then, of course we have $\zeta(t,a_*) = z_*^{(I)}(t)$. We have from \eqref{zeta-a}
  that 
  \[ \frac{(\dot{\zeta}(t,a))^2}{2} + V(\zeta(t,a) ) - 
  \frac{\gamma_{\rm in}}{2}(\zeta(t,a) )^2 = V(a)-\frac{\gamma_{\rm in}}{2} a^2 \equiv E(a).\]
  Let $W(t)=\partial_a\zeta(t,a)\Big|_{a=a_*}$. Differentiation of \eqref{zeta-a} with respect to $a$ and setting $a=a_*$ yields
   \[ L_{*-}W=0,\quad W(0)=1,\quad \dot{W}(0)=0.\]
   Since $E^\prime(a_*)= V^\prime(a_*)-\gamma_{\rm in}a_*\ne0$ ($z_*^{(I)}(t)$ is not an equilibrium point), the map $a\mapsto E(a)$ is invertible near $a=a_*$ and we may equivalently write
    $\zeta(t,E(a))$ for $a$ near $a_*$ with $\zeta(t,E_*)=z_*^{(I)}(t)$ for $E=E(a_*)$.
    
    We claim now that $\partial_a\zeta(t,E(a))\Big|_{a=a_*} = 
    (\partial_E\zeta)(t,E(a_*))E^\prime(a_*) = W(t)$ is not $T_b-$ periodic and is therefore not in the ${H}^2_{_{T_b}}$- nullspace of $L_{*-}$. It suffices to check
      that $\partial_E\zeta(t,E_*)$ is not $T_b-$ periodic. 
       For all $E$ in an open interval, $\zeta(t,E)$ is periodic of some period $T(E)$ which is a smooth function of $E$ and such that $T(E_*)=T_b$. Differentiation of the relation $\zeta(t+T(E),E)=\zeta(t,E)$ with respect to $E$ and setting $E=E_*$ yields
       \[
       \dot{\zeta}(t,E_*)\frac{\partial T}{\partial E}(E_*) + \frac{\partial \zeta}{\partial E}(t+T_b,E_*) = \frac{\partial \zeta}{\partial E}(t,E_*).
       \]
       Since $\zeta(t,E_*)$ is a non-constant periodic solution, it follows 
        that $\partial_E\zeta(t,E_*)$ is not $T_b-$ periodic if  $E\mapsto T(E)$
        satisfies:
        \begin{equation} \frac{\partial T}{\partial E}(E_*) \ne0.\label{Tprime}\end{equation}
        
        For the case of hardening nonlinearity, the relation \eqref{Tprime} is
        proved in \cite{levi91}. Figure 2 (c) displays representative dimer frequency plots $E\mapsto\omega_b(E)=2\pi T_b(E)^{-1}$ for  periodic orbits of Type I  (solid-curves) and Type II (dashed-curves), for both hardening and softening nonlinearities. In all our simulations the condition \eqref{Tprime} is seen to hold.  This concludes our verification that $H^2_{_{T_b}}$- nullspace of $L_{*-}$ is equal to $\{0\}$.

\begin{remark}\label{cl-action}
Condition \eqref{Tprime} is equivalent to the condition of  \cite{ma94} on the non-degeneracy of the classical action.
\end{remark}

\noindent{\bf${H}^2_{_{T_b}}$- nullspace of $L_{*+}$ :} 
 Since $L_*$ is a second-order ordinary differential operator its nullspace is at most two-dimensional. Further, $V''(z_*(t))$ is an even function, and so the nullspace is the direct sum of orthogonal eigenspaces of  even and odd  functions. Since $L_*$ is considered on the space ${H}^2_{_{T_b}}$, consisting of even functions (see \eqref{H2Tb}), it follows that either $0$ is not an eigenvalue or $0$ is a simple eigenvalue. As we move continuously among the phase curves of periodic orbits by varying the ``energy'', $E$ (see Figure \ref{fig:phaseports}), the periodic orbits and hence 
the coefficients of $L_*$ vary analytically. Hence the simple eigenvalues of the family of self-adjoint operators $E\mapsto L_*(E)$ vary  analytically. Therefore, zero can be an eigenvalue for only a discrete set of energies.

Summarizing, we have that for Theorem \ref{thm:breather} on the existence of discrete breathers 
applies to all Type I states (in-phase periodic orbits),  except possibly for an exceptional discrete set of energies, $E$.

  %   \footnote{   \textcolor{blue}{A remark (not sure if it's useful): We have $L_{*-}Z=0$ and $Z(0)=Z(T_b)=0$. Moreover, from the phase portrait we see that $Z(t)$ has one interior node (zero) on the interval $(0,T_b)$. Therefore $0$ is the 2nd eigenvalue of $L_{*-}$ on $[0,T_b]$ with Dirichlet BCs. Therefore $L_{*-}$ on $[0,T_b]$ with Dirichlet BCs has exactly one strictly negative eigenvalue.}}
     
       \subsubsection{The nullspaces of $L_{*\pm}$ for Type II states}
       Type II states correspond to periodic solutions, $z_*^{(II)}(t)$, of the equation
$\ddot{z}=-V'(z) - \gamma_{\text{in}}z$;
 $+\gamma_{\text{in}}$ in the ODE for Type I states is replaced by 
 $-\gamma_{\text{in}}$. 
 The corresponding replacement of $+\gamma_{\text{in}}$ by $-\gamma_{\text{in}}$ in the linearized analysis shows that verifying the non-degeneracy hypothesis of Theorem \ref{thm:breather} on the nullspace of
  $L_*$ reduces to studying the nullspace of the diagonal operator
   ${\rm diag}(L_{*+}, L_{*-})$; compare with \eqref{L*sim}.
   Hence for Type II states, the nullspace of $L_*$ reduces to the nullspace of the same two scalar linear operators as in the case of Type I states. 
   and hence the arguments presented in Section \ref{sec:typeI} can be applied to Type II states as well.  We conclude that 
   Theorem \ref{thm:breather} 
applies to all Type II states (out-of-phase periodic orbits),  except at a possibly discrete set of energies, $E$. Of course, for any fixed $E$, it is easy to numerically verify the non-degeneracy hypothesis of Theorem \ref{thm:breather}.

\end{proof}

\section{Exponential spatial localization of discrete breathers}
\label{sec:localization}

The breather solutions constructed in Section \ref{sec:existence} 
 lie in a Banach with weak spatial decay. 
 Let $\lambda\mapsto X^\lambda\in\mathcal{H}$ be $C^1$ with respect to $\lambda$ and solve $F(X^\lambda,\lambda)=0,\quad \lambda\in[0,\lambda_\star)$.
Here, 
  \begin{equation} X^\lambda =  \{ x_n(\cdot;\lambda)\},\quad x_n(t;\lambda)\in H^2_{T_b},\quad  
   \sum_{n\in\mathbb{Z}} \|x_n(\cdot,\lambda)\|_{H^2_{T_b}}^2<\infty,\label{db}\end{equation}
   and $X^0=X_*$ is the anticontinuum limit solution.
 In this section we prove the following result on exponential spatial decay. 
 \begin{theorem}\label{thm:exp_decay}
Consider the setting of Theorem \ref{thm:breather} and let $\lambda\mapsto X^\lambda$ denote a curve of discrete breathers, defined for $\lambda\in[0,\lambda_b)$. Then, there exist constants $C_1>0$, $C_2>0$ and $|\mu|<1$ such that for all $\lambda\in[0,\lambda_b)$
for all $n\in\mathbb{Z}$ 
\[ \|x_n(\cdot,\lambda)\|_{H^0_{T_b}} \le C_1 \exp(C_2\lambda)\ \mu^{|n|}.\]
 \end{theorem}\bigskip
 
 Our arguments are related to those 
    presented in \cite{ma94}. In our proof, we reduce spatial decay to a Combes-Thomas type bound given in Proposition \ref{CT-bd} below.

  Differentiation of equation   $F(X^\lambda,\lambda)=0$ (see \eqref{eq5}) yields 
  \begin{equation}
\label{eq:var-eqn}
    \frac{d X^\lambda }{d\lambda} = -F_X\left(X^\lambda;\lambda\right)^{-1}\frac{\partial F(X^\lambda;\lambda) }{\partial \lambda},\qquad X^\lambda\Big|_{\lambda=0}=X_*.
\end{equation}
Written out componentwise, we have for all $n\in\mathbb{Z}$
\begin{align}
\label{eq:var-eqn1a}
    \frac{d x_n(\cdot, \lambda) }{d\lambda} &= -\sum_{m\in\mathbb{Z}} \left[F_X\left(X^\lambda;\lambda\right)^{-1}\right]_{nm} \left(\frac{\partial F(X^\lambda;\lambda) }{\partial \lambda} \right)_m\\
   &= -\sum_{m\in\mathbb{Z}} \left[F_X\left(X^\lambda;\lambda\right)^{-1}\right]_{nm} \left({\bf R}x\right)_m \nonumber\\
   &=-\sum_{m\in\mathbb{Z}} \left[F_X\left(X^\lambda;\lambda\right)^{-1}\right]_{nm}
   \sum_{l} R_{ml}x_l\nonumber\\
   &=-\sum_l \left[\sum_m\left[F_X\left(X^\lambda;\lambda\right)^{-1}\right]_{nm} R_{ml}\right]\ x_l\nonumber\\
   &\equiv -\sum_l Q_{nl}(X^\lambda,\lambda)\ x_l,
\nonumber\end{align}
where we have used \eqref{eq: F diff wrt lambda}.
Our goal is to prove exponential decay of the sequence of norms \[ \Big\{\|x_n(\cdot,\lambda)\|_{H^2_{T_b}}\Big\}_{n\in\mathbb{Z}}.\]
Since the mapping $\lambda\mapsto \{ x_n(\cdot,\lambda)\}$ is $C^1([0,\lambda_\star); \mathcal{H}^2)$ we have 
\begin{equation}
\label{eq:var-est}
    \frac{d \|x_n(\cdot, \lambda)\|_{H^0_{T_b}} }{d\lambda} \le \sum_{l\in\mathbb{Z}} \Big\| Q_{nl}(X^\lambda,\lambda)\Big\|_{\mathcal B\left(H^0_{T_b}\right)}\cdot  \big\| x_l(\cdot,\lambda)\big\|_{H^0_{T_b}}
\end{equation}
The key to estimating $\|x_n(\cdot, \lambda)\|_{H^0_{T_b}}$ from \eqref{eq:var-est} is:
\begin{proposition}\label{CT-bd}
There are constants $M>0$ and $\mu_b\in(0,1)$ such that for all $\lambda\in[0,\lambda_\star)$, and all $n,m\in\mathbb{Z}$ 
\begin{equation}\label{dF-inv-bd}
    \left\| \left(F_X \left(X^\lambda;\lambda\right)^{-1}\right)_{n,m} \right\|_{\mathcal B\left(H^0_{T_b}\right)} < M \mu_F^{|n-m|}.
\end{equation}
\end{proposition}
The proof of Proposition \ref{CT-bd} is presented below in Section \ref{sec:CT-bd}.
Using \eqref{dF-inv-bd} together with the assumed bound on ${\bf R}$ in \eqref{Rml-decay} we have:
\begin{align*}
 \|Q_{nl}(X^\lambda,\lambda)\| &=
\Big\|\left(F_X \left(X^\lambda;\lambda\right)^{-1}\right)_{n,m} R_{ml}\Big\|\\
&\le M \mu_F^{|n-m|}\times C \nu_R^{|m-l|}\le MC \mu_b^{|n-l|},
\end{align*}
where
\[  0<\mu_b\equiv \max\{\mu_b,\nu_R\}<1.\]

 Applying
 \eqref{dF-inv-bd}  in \eqref{eq:var-est} we have 
 \begin{equation}
\label{eq:var-est1}
 \frac{d \|x_n(\cdot, \lambda)\|_{H^0_{T_b}} }{d\lambda} \le \Big\| \frac{d x_n(\cdot, \lambda)}{d\lambda}\Big\|_{H^0_{T_b}}  \le   M^\prime\  \sum_{m\in\mathbb{Z}}  J_{|n-m|} \| x_{m}(\cdot,\lambda) \|_{H^0_{T_b}},
\end{equation}
where 
\begin{equation}\label{eq:Jdef}
 J_{|n-m|} = \mu_b^{|m-n+1|} + \mu_b^{|m-n-1|}.
 \end{equation}
Recall that $0<\mu_b<1$.
We shall use \eqref{eq:var-est1} to show that for any $\mu$ such that $\mu_b<\mu<1$, the sequence 
\[
\{ w_n(\lambda) \} \equiv \Big\{ \mu^{-|n|}\|x_n(\cdot,\lambda)\|_{H^0_{T_b}}
\Big\}\]
 is uniformly bounded for $n\in\mathbb{Z}$.
 
 Since, {\it a priori}, we only know that $\{\|x_n(\cdot,\lambda)\|_{H^0_{T_b}}\}$ is square summable we work with a cut-off sequence. For each $N\ge1$, define
 \[
\{ w^{(N)}_n(\lambda) \} \equiv \Big\{ [\rho^{(N)}(n)]^{-1} \|x_n(\cdot,\lambda)\|_{H^0_{T_b}}
\Big\}\]
where 
\begin{equation}
    \rho^{(N)}(n)=\begin{cases} \mu^{|n|} & |n|<N\\ \mu^{N} & |n|\ge N \end{cases}
\end{equation}
 
 Multiplying \eqref{eq:var-est1} by 
  $[\rho^{(N)}(n)]^{-1}$ we obtain
   \begin{equation}
\label{eq:var-est2}
    \frac{d w^{(N)}_n(\lambda) }{d\lambda} \le \ M^\prime\  \sum_{m\in\mathbb{Z}} [\rho^{(N)}(n)]^{-1} J_{|n-m|} \rho^{(N)}(m)w^{(N)}_m(\lambda),
\end{equation}
where $ w^{(N)}_n(0)=[\rho^{(N)}(n)]^{-1}\|x_{*,n}\|_{H^0_{T_b}}$.
One checks easily, using the form of \eqref{eq:Jdef} and $0<\mu_b<\mu<1$, that there is a constant, $C_J>0$, which is independent of $N$, such that:
\[
\sup_{n\in\mathbb{Z}}\quad \sum_{m\in\mathbb{Z}} [\rho^{(N)}(n)]^{-1} J_{|n-m|} \rho^{(N)}(m) \le\ C_J.
\]
Integrate  \eqref{eq:var-est2} with respect to $\lambda$ over the interval $[0,\lambda)$, $\lambda<\lambda_\star$, to obtain
\[
w^{(N)}_n(\lambda) \le [\rho^{(N)}(n)]^{-1}\|x_{*,n}\|_{H^0_{T_b}} + 
 M^\prime\ C_J\ \int_0^\lambda \sup_{m\in\mathbb{Z}} w^{(N)}_m(\lambda') d\lambda^\prime
\]
Finally, let $[w^{(N)}](\lambda)\equiv  \sup_{m\in\mathbb{Z}} w^{(N)}_m(\lambda)$
 and note that for all $N\ge1$
 \[ [w^{(N)}](0)= \sup_{n\in\mathbb{Z}}[\rho^{(N)}(n)]^{-1}\ \|x_{*,n}\|_{H^0_{T_b}}\le \sup_{n\in\mathbb{Z}} \mu^{-|n|} \|x_{*,n}\|_{H^0_{T_b}} \equiv c_*<\infty.\]
 Then, 
 \[ 
[w^{(N)}](\lambda) \le c_* +  M^\prime\ C_J\ \int_0^\lambda  [w^{(N)}](\lambda^\prime) d\lambda^\prime,\quad \lambda\in[0,\lambda_\star).
 \]
 It follows from Gronwall's inequality that 
\begin{equation} [w^{(N)}](\lambda) \le c_* \exp(
M^\prime C_J\lambda),\quad \lambda\in[0,\lambda_\star).\label{brw}
\end{equation}
Since $N\ge1$ is arbitrary and the right hand side of \eqref{brw} is independent of $N$ we have that
\[ \|x_n(\cdot,\lambda)\|_{H^0_{T_b}} \le c_* \exp(M^\prime C_J\lambda)\ \mu^{|n|},\ n\in \mathbb{Z},\quad \lambda\in[0,\lambda_\star).\] This completes the proof of exponential decay, modulo Proposition \ref{CT-bd}, which we prove 
in the following subsection.
\subsection{Proof of Proposition \ref{CT-bd}}\label{sec:CT-bd}

We prove the bound \eqref{dF-inv-bd} using a strategy of proof 
for the Combes-Thomas estimate in \cite{aizen15}.

From expression for $F(X,\lambda)$ given in \eqref{eq5} we have
\begin{align}
\label{eq: F diff wrt X}
    F_X(X(\lambda), \lambda) &= {\bf \Delta}(\lambda) - \lambda {\bf R}
    \end{align}
  where   ${\bf \Delta}(\lambda)$ is the block diagonal operator
  \[
 {\bf \Delta}(\lambda) =  \begin{bmatrix}
    \ddots &&&&&& \\
    &0& L^{(-1)}(\lambda) & 0 && & \\
    &&0 & L^{(0)}(\lambda) & 0 && \\
    &&& 0 & L^{(1)}(\lambda)& 0& \\
    &&&&&& \ddots
    \end{bmatrix}\ ,
  \]
with 
\begin{equation}  
 L^{(j)}(\lambda)= 
 \begin{pmatrix}
    \dfrac{d^2}{dt^2}+V''(x_j^A(t,\lambda)) & 0\\ 
   0 & \dfrac{d^2}{dt^2}+V''(x_j^B(t,\lambda)) \end{pmatrix}
   -\gamma_{\text{in}}
   \begin{pmatrix} 0&1\\ 1&0\end{pmatrix},
 \label{def: L j lambda}\end{equation}
 and 
 ${\bf R}$ satisfies the bound \eqref{Rml-decay}. For the special case of nearest neighbor interactions, corresponding to the model \eqref{eq1A}, 
\begin{align}
 {\bf R} \equiv 
    \begin{bmatrix}
    \ddots &&&&&& \\
    &R^\top& 0 &R && & \\
    &&R^\top & 0 & R && \\
    &&& R^\top & 0& R& \\
    &&&&&& \ddots
    \end{bmatrix}\ ,
\end{align}
with  \[ R =  \gamma_\text{out} \begin{pmatrix}
 0&0\\1 &0
\end{pmatrix}.
\]

Note that $ {\bf \Delta}(0)= F_X(X_\star;0)$ is invertible by non-resonance and non-degeneracy hypotheses of Theorem \ref{thm:breather}:
\[ \|{\bf \Delta}(0)^{-1}\|_{\mathcal{H}^0_{T_b}\to\mathcal{H}^0_{T_b}}\le \frac{1}{c_{gap}}.\]
By smoothness of $F$ and $\lambda\mapsto X^\lambda$ we have, for some $0<\lambda_1\le \lambda_\star$, that $ {\bf \Delta}(\lambda)$ has a bounded inverse
satisfying
\[ \|{\bf \Delta}(\lambda)^{-1}\|_{\mathcal{H}^0_{T_b}\to\mathcal{H}^0_{T_b}}\le \frac{2}{c_{gap}},\quad \lambda\in[0,\lambda_1).\]

We shall study decay of the $(m,n)$ matrix element of $ F_X(X(\lambda), \lambda)^{-1}$ by studying
the solution of the equation 
\[ \left( {\bf \Delta}(\lambda) - \lambda{\bf R} \right) u = v\]
in an exponentially weighted space. Fix $n\in\mathbb{Z}$ and define   $f(m)=\eta|m-n|$, for all $m\in\mathbb{Z}$, where $\eta>0$ is to be chosen.
We denote by $e^f$ the multiplication operator on $\mathcal{H}^0_{T_b}$ given 
by $\{\alpha_m\}\mapsto \{e^{f(m)}\alpha_m\}$.
\[ e^f\left( {\bf \Delta}(\lambda) - \lambda{\bf R} \right)e^{-f} e^f u = e^fv
\quad\textrm{or}\quad \left( {\bf \Delta}(\lambda) - \lambda\ e^f{\bf R}e^{-f}\right)\ e^fu\ =\ e^f v\]
Therefore,
\[  e^f \left( {\bf \Delta}(\lambda) - \lambda {\bf R} \right)^{-1} v\ =\ \left( {\bf \Delta}(\lambda) - \lambda\ e^f{\bf R}e^{-f}\right)^{-1}\ e^f v.\]
Coordinatewise, we have 
\begin{equation}
 e^{f(m)} [\left( {\bf \Delta}(\lambda) - \lambda {\bf R} \right)^{-1}]_{_{mj}} v_j\ =\ [\left( {\bf \Delta}(\lambda) - \lambda\ e^f{\bf R}e^{-f}\right)^{-1}]_{_{mj}}\ e^{f(j)} v_j,
 \label{mj-wtd}
\end{equation}
where we sum over repeated indices.

Now take $v=e^{(n)}$, whose only non-zero entry is a one in the $n^{th}$ slot. Then,
 \eqref{mj-wtd} becomes
 \begin{equation} e^{ f(m)} [\left( {\bf \Delta}(\lambda) - \lambda {\bf R} \right)^{-1}]_{_{mn}} = [\left( {\bf \Delta}(\lambda) - \lambda\ e^f{\bf R}e^{-f}\right)^{-1}]_{_{mn}}
 \label{mj-wtd1}\end{equation}
Since $\Delta(\lambda)^{-1}$ is invertible, we have from \eqref{mj-wtd}
 and that $e^{f(m)}=e^{\eta|m-n|}$:
\begin{equation} e^{\eta|m-n|} 
[\left( {\bf \Delta}(\lambda) - \lambda {\bf R} \right)^{-1}]_{_{mn}} = [\left( I - \lambda\ {\bf \Delta}(\lambda)^{-1}\ e^f{\bf R}e^{-f}\right)^{-1}{\bf \Delta}(\lambda)^{-1}]_{_{mn}}
 \label{mj-wtd2}\end{equation}
We have that
\begin{align}\label{pert-bd}
   \Big\| \lambda\ {\bf \Delta}(\lambda)^{-1}\ e^f{\bf R} e^{-f} \Big\|\ 
   &\le |\lambda| \times \| {\bf \Delta}(\lambda)^{-1}\|\times \| e^f{\bf R} e^{-f}\|\nonumber\\
   &\quad \le  C
   \ |\lambda|\ \frac{2}{c_{gap}}\ \| e^f{\bf R} e^{-f}\|,
\end{align}
 %\footnote{OLD:\\
%\begin{align}\label{pert-bd}
 %  \Big\| \lambda\ {\bf \Delta}(\lambda)^{-1}\ %e^f{\bf R} e^{-f} \Big\|\ 
 %  \le |\lambda| \times \| {\bf %\Delta}(\lambda)^{-1}\|\times \| e^f{\bf R} %e^{-f}\| \le  C
 %  \ |\lambda|\ \frac{2}{c_{gap}}\ e^{\eta_0},
%\end{align}
%with norm in the space %$\mathcal{B}(\mathcal{H}^0_{T_b})$ and 
%where $\eta_0>0$ is arbitrary and fixed, and %$\eta\in(0,\eta_0)$. 
%}
with norm in the space $\mathcal{B}(\mathcal{H}^0_{T_b})$. 
Let's now bound $\| e^f{\bf R} e^{-f}\|$. 
With summation over repeated indices implied, we have:
\begin{align*}
    \left(e^f{\bf R} e^{-f}X\right)_m &= 
    e^{f(m)}R_{mj}e^{-f(j)}X_j = R_{mj}e^{\eta(|m-n|-|j-n|)}X_j\\
    &\le R_{mj}e^{\eta|m-j|}|X_j|\le \mu_R^{|m-j|}e^{\eta|m-j|}|X_j|\le (\mu_R e^\eta)^{|m-j|}\ |X_j|.
\end{align*}
Since $|\mu_R|<1$, by taking $0<\eta<\eta_0$ sufficiently small we have by Young's inequality that 
\[\|e^f{\bf R} e^{-f}X\|_{\mathcal{H}^0_{T_b}}\le C \|X\|_{\mathcal{H}^0_{T_b}}.\]

Restricting $\lambda>0$ possibly further, by taking   $0<\lambda<\lambda_2$
sufficiently small $(\lambda_2\le\lambda_1$), depending on $\eta_0$, we have
that $[\left( I - \lambda\ {\bf \Delta}(\lambda)^{-1}\ e^f{\bf R}e^{-f}\right)^{-1}$
is invertible on $\mathcal{H}^0_{T_b}$ and hence, 
\begin{equation}
[\left( {\bf \Delta}(\lambda) - \lambda {\bf R} \right)^{-1}]_{_{mn}} 
\le  M\ e^{-\eta|m-n|} =  M\ \mu^{|m-n|},
 \label{mj-wtd3}\end{equation}
where $\mu\equiv e^{-\eta}<1$ with $0<\eta<\eta_0$. This completes 
 the proof of Proposition \ref{CT-bd}.

\section{Analysis of the weakly nonlinear long wave regime - continuum theory}
\label{sec:continuum}
From Figure \ref{fig:continuation} (b) and (c) we have that,  as $\lambda$ tends to $\lambda_\star$: the maximum amplitude of the discrete breather decreases toward zero
 and its spatial width becomes large on the scale of the lattice-spacing.
 To describe this behavior precisely, we  use a multiple scale analysis 
designed to capture the  weakly nonlinear long wave regime. To carry this out it is useful making a suitable rescaling and recentering of \eqref{eq1A}.
%In this section we derive a system of PDEs which approximates the discrete %system \eqref{eq1} in the long-wave limit in the regime of a small bandgap.  %The derived model describes the evolution of wavepackets with slowly-varying %envelopes which are spectrally localized around the quasi-momentum at the %center of the Brillouin zone of the discrete system, where the spectral gap is %the narrowest.  To achieve this we will employ a multiple-scale analysis in %the small parameter $\epsilon$, which we define below. 
%
\subsection*{Introducing the natural small parameter} Since we are interested in the regime where the band gap width tends toward zero, we introduce the  parameter:
\begin{equation}\epsilon\equiv\gamma_{\text{in}}-\lambda
\gamma_{\text{out}}
\label{epsilon-def}\end{equation}
which tends to zero as the phonon gap width tends to zero at $\lambda_\star=|\gamma_{\rm in}/\gamma_{\rm out}|$,
and we rewrite \eqref{eq1A} as
\begin{align}\label{eps1}
&\ddot{x}_{n}^A=-\omega_0^2x_{n}^A+\gamma_{\text{in}}(x_{n}^B+x_{n-1}^B)+(V''(0)x_n^A-V'(x_n^A))-\epsilon x_{n-1}^B\\ \nonumber
&\ddot{x}_{n}^B=-\omega_0^2x_{n}^B+\gamma_{\text{in}}(x_{n}^A+x_{n+1}^A)+(V''(0)x_n^B-V'(x_n^B))-\epsilon x_{n+1}^A.
\end{align}
Here, we set $\omega_0^2\equiv V''(0)$. The frequency $\omega_0$ is at the center of the phonon gap. We study breather solutions whose frequency $\omega_b$ lies within the order $\epsilon$ width gap and we express this as:
\begin{equation} \omega_b = \omega_0 - \frac{\epsilon \nu}{2\omega_0}.\label{nu-def1}
\end{equation}
The order one parameter, $\nu$, in \eqref{nu-def1} determines the offset from $\omega_0$ 
within the $\epsilon-$ width gap .
\subsection*{Balancing weak nonlinearity with linear (phonon) dispersion} Since $V(z)$ is assumed to be smooth with $V'(-z)=-V'(z)$, and $V'(0)=0$, the
 leading order nonlinearity is cubic;
$ V'(z) = V''(0) z + \frac16 V''''(0) z^3 +\dots$.
To fix an example  we choose the leading order behavior: $V(z) = 3z + \Gamma z^3 +\dots $; see \eqref{Vdef}.

Let us assume that  $\epsilon=\gamma_{\rm in}-\lambda\gamma_{\rm out}$ is strictly positive and small; we shall comment on the case where $\epsilon$ is negative and small below.  In order to balance the linear phonon dispersion with nonlinearity we rescale the amplitude:
\begin{equation} \begin{pmatrix} x_n^A\\ x_n^B\end{pmatrix} =
\sqrt{\epsilon}\begin{pmatrix} y_n^A\\ y_n^B\end{pmatrix},
\label{x_to_y}
\end{equation}
and we obtain:
\begin{align}\label{y-eqn}
&\ddot{y}_{n}^A=-\omega_0^2y_{n}^A+\gamma_{\text{in}}(y_{n}^B+y_{n-1}^B)-\epsilon \Gamma(y_n^A)^3-\epsilon y_{n-1}^B\\ \nonumber
&\ddot{y}_{n}^B=-\omega_0^2y_{n}^B+\gamma_{\text{in}}(y_{n}^A+y_{n+1}^A)-\epsilon \Gamma(y_n^B)^3-\epsilon y_{n+1}^A,
\end{align}
where we have dropped terms of order $\epsilon^2$ and higher.

In terms of our parameter $\epsilon$, defined in \eqref{epsilon-def}, the two  band functions \eqref{phonon} may be re-expressed as:
\begin{equation}\label{phonon1}
    (\omega^2)_{\pm}(k)-\omega_0^2=
    \pm\sqrt{
    \epsilon^2+
    4\gamma_{\rm in}(\gamma_{\rm in}-\epsilon)
    \cos^2\left({k\over2}\right)
    },
    \quad \omega_0^2\equiv V''(0).
\end{equation}
As noted in the Introduction, for $\epsilon\ne0$, there are two disconnected intervals of spectrum,.
The interval of spectrum associated with the $+$branch is called the {\it optical band} and that associated with the $-$branch is called the {\it acoustic band}. The open interval of energies lying between these bands is called the phonon gap; see Figure \ref{fig:dimer}b
 with $E=\omega^2-\omega_0^2$.
The maximum and minimum of $(\omega^2)_\pm(k)-\omega_0^2$ (top of the optical band) occur for $k=0$, and phonon gap width is at its smallest for $k=\pi$. We next carry out, for $\epsilon$ small, an asymptotic study which yields
discrete breathers in three regimes
 \begin{enumerate}
     \item[(A)] $\omega_b^2$ in the $\mathcal{O}(\epsilon)$ spectral gap,
     \item[(B)]  $\omega_b^2$ below and near the minimum of the acoustic band, and 
     \item[(C)] $\omega_b^2$ above and near the maximum of the optical band.
 \end{enumerate} 
 In all three regimes, the solution will be shown to have the structure of slow modulation of rapidly oscillatory plane wave states. Since the dispersion relation for regimes (B) and (C) is approximately quadratic, the governing envelope equations will be of (nonlinear) Schroedinger type. And since the dispersion relation for regime (A) is that of a gapped linear crossing (Dirac point), 
 the governing envelope equations will be of (nonlinear) massive Dirac type.
 
\subsection{Asymptotic study of discrete breathers in Regime (A); 
$\omega^2$ in the $\mathcal{O}(\epsilon)$ spectral gap}\label{a_dbs}
\subsection*{Centering the analysis near the asymptotic linear band crossing} 

Recall that as $\epsilon\sim 0$ ($\lambda\sim\lambda_\star$) the spectral gap of the linear band structure is narrowest in a neighborhood of $k=\pi$ and closes at quasimomentum $k=\pi$ as $\epsilon\to0$.
The asymptotic solution we seek is of the form of a wave-packet, spectrally localized at $k=\pi$ and hence we set
\begin{equation}
\begin{pmatrix} y_n^A\\ y_n^B\end{pmatrix} \equiv
 e^{i\pi n}\begin{pmatrix} Y_n^A\\ Y_n^B\end{pmatrix}
 =(-1)^n\begin{pmatrix} Y_n^A\\ Y_n^B\end{pmatrix},
 \label{y_to_Y}
 \end{equation}
 and we obtain
\begin{align}\label{Y-eqn}
&\ddot{Y}_{n}^A=-\omega_0^2Y_{n}^A+\gamma_{\text{in}}(Y_{n}^B-Y_{n-1}^B)-\epsilon\Gamma(Y_n^A)^3+\epsilon Y_{n-1}^B\\ \nonumber
&\ddot{Y}_{n}^B=-\omega_0^2Y_{n}^B+\gamma_{\text{in}}(Y_{n}^A-Y_{n+1}^A)-\epsilon\Gamma(Y_n^B)^3+\epsilon Y_{n+1}^A.
\end{align}
In the next section, we embark on an asymptotic analysis construction of nonlinear standing wave states of the rescaled and recentered system \eqref{Y-eqn}.

The solutions we seek are to be spectrally concentrated on the set of momentum
 $q\equiv k-\pi\approx0$ over a band width of order $\epsilon$. We now deduce a continuum approximation flowing from this requirement. Using the discrete Fourier inversion formula we write, for $J=A,B$:
  \[ Y^J_{n\pm1}-Y^J_n =\frac{1}{\epsilon}\int_{-\pi}^{\pi} e^{inq} (e^{\pm iq}-1) \chi^J\left({q\over\epsilon}\right) dq, \]
  where $\chi^J(z)$ is rapidly decaying  away from $z=0$ and smooth. The overall factor of $\epsilon^{-1}$ ensures that $Y_n$ is of order one for small $\epsilon$
   (consistent with the multiple scale expansion below). Changing variables ($q=\epsilon Q$) and using the approximation $e^{\pm i\epsilon Q}-1\approx \pm i\epsilon Q$,  we have
  \begin{align*} Y^J_{n\pm1}-Y^J_n &\approx \epsilon \int_{-\pi/\epsilon}^{\pi/\epsilon} e^{iQ(\epsilon n)} \left(\pm iQ\right) \chi^J(Q) dQ \approx 
  \epsilon \int_{-\infty}^{\infty} e^{iQ(\epsilon n)} \left(\pm iQ\right) \chi^J(Q) dQ\\
  &\equiv \pm \epsilon \partial_Z u_J(Z)\vert_{Z=\varepsilon n} = \pm\partial_z u_J(n), \quad Z=\epsilon z.\end{align*}
Here, $z$ and $Z=\epsilon z$ are fast and slow continuum spatial scales, which  we systematically introduce  below as independent variables in a multiple scale analysis.

This continuum approximation gives
\begin{align}\label{eps3}
&\partial_t^2u_A=-\omega_0^2u_A+\gamma_{\text{in}}\partial_zu_B-\epsilon\Gamma u_A^3+\epsilon(u_B-\partial_zu_B)\\ \nonumber
&\partial_t^2u_B=-\omega_0^2u_B-\gamma_{\text{in}}\partial_zu_A-\epsilon\Gamma u_B^3+\epsilon(u_A+\partial_zu_A).
\end{align}

We now seek a solution to
 \eqref{eps3} having a multi-scale structure, depending on  both fast space and time variables $z$ and $t$, as well as newly introduced  ``slow" space and time variables $Z$ and $T$, all treated as independent variables: 
 \begin{equation}
u_{A,B}=u_{A,B}(t,z,T,Z),\quad {\rm where}\quad  Z=\epsilon z,\quad T=\epsilon t\label{m-scale}\end{equation} 
We rewrite \eqref{eps3}
 in terms of the extended set of variables, by making the replacements 
 %\begin{equation}
 $\partial_t\to \partial_t + \epsilon\partial_T,\quad \partial_z\to \partial_z + \epsilon\partial_Z.$
 %\label{new-partial}
 %\end{equation}
 This gives
 \begin{align}\label{eps3-A}
&\left(\partial_t+\epsilon\partial_T\right)^2u_A=-\omega_0^2u_A+\gamma_{\text{in}}\left(\partial_z+\epsilon\partial_Z\right) u_B-\epsilon\Gamma u_A^3+\epsilon(u_B-\left(\partial_z+\epsilon\partial_Z\right)u_B)\\ \nonumber
&\left(\partial_t+\epsilon\partial_T\right)^2u_B=-\omega_0^2u_B-\gamma_{\text{in}}\left(\partial_z+\epsilon\partial_Z\right)u_A-\epsilon\Gamma u_B^3+\epsilon(u_A+\left(\partial_z+\epsilon\partial_Z\right)u_A).
\end{align}
 
Further, we expand $u_{A,B}=u_{A,B}(t,z,T,Z)$ in powers of $\epsilon$: 
\begin{equation}
    u_{A,B}=u_{A,B}^{(0)}(z,t,Z,T)+\epsilon u_{A,B}^{(1)}(z,t,Z,T)+\epsilon^2 u_{A,B}^{(2)}(z,t,Z,T)+\cdots
\label{taylor}\end{equation}
Recall that we are in the regime of an order $\mathcal{O}(\epsilon)-$ frequency gap around $\omega_0$. Since we seek breather-like (spatially localized) states
 we impose the boundary condition at infinity:
 \begin{equation}
u_{A,B}^{(j)}(z,t,Z,T)\to 0 \quad {\rm as}\quad |Z|\to\infty,\quad j=0,1,\dots
\label{BC-pert}\end{equation}
 
We then substitute  \eqref{taylor} into \eqref{eps3-A} and obtain a hierarchy of equations of order  $\epsilon^j,\ j=1,2,\dots$. Each equation is of the form
\begin{equation}\label{hierarchy}
    \left(\partial_t^2-\mathcal{L}_0\right)U^{(j)}
    = F^{(j)},
    \end{equation}
  where  $U^{(j)}=(u_{A}^{(j)},u_{B}^{(j)})^\top$ and 
$F^{(j)}=(F_A^{(j)},F_{B}^{(j)})^\top$. Here,
\begin{equation}
\mathcal{L}_0\equiv
 \begin{pmatrix}
    -\omega_0^2 & \gamma_{\text{in}}\partial_z\\
    -\gamma_{\text{in}}\partial_z & -\omega_0^2
    \end{pmatrix}.
\end{equation}
We view each equation in the hierarchy \eqref{hierarchy} as a PDE with respect to the fast variables $z$ and $t$ and seek bounded solutions
at each order. This imposes solvability conditions on the source terms $F^{(j)}$ which, along with the decay condition \eqref{BC-pert}, prescribes the behavior with respect to the slow variables, $Z,T$.

We now implement this expansion procedure. Here, we only require only the first two equations in this hierarchy:
 the equations arising at order $\epsilon^0$ and $\epsilon^1$. 

At order $\epsilon^0$,  we have the system
\begin{equation}\label{zero}
    \left(\partial_t^2-\mathcal{L}_0\right)
    \begin{pmatrix}
    u_A^{(0)}\\
    u_B^{(0)}
    \end{pmatrix}
    =0.
\end{equation}

We solve \eqref{zero} by taking a time-harmonic solution
\begin{equation}\label{zero-solve}
   \begin{pmatrix}
    u_A^{(0)}(z,t,Z,T)\\
    u_B^{(0)}(z,t,Z,T)
    \end{pmatrix}=
    \begin{pmatrix}
    u^{(0)}(Z,T)\\
    v^{(0)}(Z,T)
    \end{pmatrix}
    e^{i\omega_0 t}+\text{c.c.},
\end{equation} 

 % \footnote{\textcolor{red}{I cleaned up the notation here, as we had a few errors.  Reviewer 1 also comments about $U$ and $V$ having one argument in line 785, but later we include the parameter $\nu$ as another argument, see line 796.  Maybe we should just include $U(z,\nu)$ and $V(z,\nu)$ into line 785?  Reviewer 1 also asks if it is possible to construct a proof for the existence of discrete breathers in the long-wave regime based on our numerical continuations from the anti-continuum. Do you think this is possible?  Could one just apply the Implicit Function Theorem again to these solutions? \textcolor{blue}{It is possible to set the problem up via IFT but the analysis is rather different.
 % One is now bifurcating from the zero solution at the end of the continuous spectrum. This what I meant by the remark midway through item 6 in the introduction which refers to 17,18 and 15. The referee implicitly suggests that one of our references (for a discrete FPU) problem should be added as well. Let's look and discuss.}}}
  
where $\text{c.c.}$ denotes the complex conjugate of the first term. The amplitudes 
$u^{(0)}(Z,T)$ and $v^{(0)}(Z,T)$, which are constant on the fast scales, will be determined at the next order.

At order $\epsilon^1$ we have the system
\begin{equation}\label{first}
    \left(\partial_t^2-\mathcal{L}_0\right)
    \begin{pmatrix}
    u_A^{(1)}\\
    u_B^{(1)}
    \end{pmatrix}
    =
    \begin{pmatrix}
    F_1(z,t,Z,T)\\
    F_2(z,t,Z,T)
    \end{pmatrix}
\end{equation}
where the right hand side forcing term depends on $u^{(0)}_{A,B}$ and is given by:
\begin{align}\label{eps5}
&F_1(z,t,Z,T)=-2\partial_t\partial_Tu_A^{(0)}+u_B^{(0)}+\gamma_{\text{in}}\partial_Zu_B^{(0)}-\partial_zu_B^{(0)}-\Gamma\left(u_A^{(0)}\right)^3\\ \nonumber
&F_2(z,t,Z,T)=-2\partial_t\partial_Tu_B^{(0)}+u_A^{(0)}-\gamma_{\text{in}}\partial_Zu_A^{(0)}+\partial_zu_A^{(0)}-\Gamma\left(u_B^{(0)}\right)^3
\end{align}

 Substituting \eqref{zero-solve} into \eqref{eps5} gives
\begin{align}
\hspace{-.5cm}
&F_1=-2\partial_T[i\omega_0u^{(0)}e^{i\omega_0 t}+\text{c.c.}]+(1+\gamma_{\text{in}}\partial_Z)[v^{(0)}e^{i\omega_0 t} +\text{c.c}]
-\Gamma\left[u^{(0)}e^{i\omega_0 t}+\text{c.c}\right]^3\\ \nonumber
&F_2=-2\partial_T[i\omega_0v^{(0)}e^{i\omega_0 t}+\text{c.c.}]+(1-\gamma_{\text{in}}\partial_Z)[u^{(0)}e^{i\omega_0 t} +\text{c.c}]
-\Gamma\left[v^{(0)}e^{i\omega_0 t}+\text{c.c}\right]^3
\label{F12}\end{align}
Equation \eqref{first} may be expressed as:
\begin{equation}\label{new}
    \left(\partial_t^2-L_0\right)
    \begin{pmatrix}
    u_A^{(1)}\\
    u_B^{(1)}
    \end{pmatrix}
    =
    \begin{pmatrix}
    f_1(Z,T)\\
    f_2(Z,T)
    \end{pmatrix}e^{i\omega_0 t}
    +
      \begin{pmatrix}
    g_1(Z,T)\\
    g_2(Z,T)
    \end{pmatrix}e^{2i\omega_0 t}
    +
      \begin{pmatrix}
    h_1(Z,T)\\
    h_2(Z,T)
    \end{pmatrix}e^{3i\omega_0 t}
    +
    \text{c.c.}
\end{equation}
The source terms proportional to $e^{i\omega_0 t}$ and $e^{-i\omega_0 t}$ are resonant and the others are not.  Hence, a necessary and sufficient condition for the solution to be bounded in $t$ is that $f_1(Z,T)=f_2(Z,T)=0$.  Thus, 
\begin{align}\label{env-eqns}
&2i\omega_0\partial_T u^{(0)}=(1+\gamma_{\text{in}}\partial_Z)v^{(0)}-3\Gamma|u^{(0)}|^2u^{(0)}\\ \nonumber
&2i\omega_0\partial_T v^{(0)}=(1-\gamma_{\text{in}}\partial_Z)u^{(0)}-3\Gamma|v^{(0)}|^2v^{(0)}
\end{align}
\subsection*{Gap solitons}
In order that \[\begin{pmatrix} u_A(z,t,Z,T)\\ u_B(z,t,Z,T)\end{pmatrix} \approx \begin{pmatrix} u_A^{(0)}(Z,T) \\ u_B^{(0)}(Z,T)\end{pmatrix} e^{i\omega_0t}\ +\ c.c. \]
approximate a solution with frequency $\omega_b$ given by \eqref{nu-def1}, we seek
 time-harmonic solutions of the following form:
\begin{equation}
  \begin{pmatrix}
    u^{(0)}(Z,T)\\
    v^{(0)}(Z,T)
    \end{pmatrix}
    =e^{-i\nu T/2\omega_0}
    \begin{pmatrix}
    U(Z;\nu)\\
    V(Z;\nu)
    \end{pmatrix}.
\end{equation}
Then, $(U,V)$ coupled system of ODEs:
\begin{align}\label{eps6}
&\nu U=(1+\gamma_{\text{in}}\partial_Z)V-3\Gamma|U|^2U\\ \nonumber
&\nu V=(1-\gamma_{\text{in}}\partial_Z)U-3\Gamma|V|^2V.
\end{align}
We consider the case where $(U,V)$ are real-valued:
\begin{align}\label{eps7}
&\gamma_{\text{in}}U'=U-\nu V-3\Gamma V^3\\ \nonumber
&\gamma_{\text{in}}V'=-V+\nu U+3\Gamma U^3.
\end{align}

The above expansion leads to formal asymptotic solutions of \eqref{eq1A} (equivalently \eqref{eps1}), for $\epsilon=\gamma_{\rm in}-\lambda\gamma_{\rm out}$ positive and small ($\lambda\approx\lambda_\star$):
\begin{equation}\label{eps8}
   \begin{pmatrix}
   x_n^A(t)\\
   x_n^B(t)
    \end{pmatrix}
   \sim 2\sqrt{\epsilon}\ (-1)^{n}\ \begin{pmatrix}
    U(\epsilon n; \nu)\\
    V(\epsilon n; \nu)
    \end{pmatrix}\ \cos\left( \left[\omega_0 -\frac{\epsilon \nu}{2\omega_0}\right] t\right),\quad\quad 0<\epsilon\ll1.
\end{equation}

Since we shall use \eqref{eps8} as an analytical approximation for  discrete breathers, we focus on the orbits of \eqref{eps7} which are homoclinic to $(0,0)$, {\it i.e.} solutions $(U,V)=(U(Z;\nu),V(Z;\nu))$ of \eqref{eps8} for which $(U(Z;\nu),V(Z;\nu))$ tends to $(0,0)$ as $Z\to\pm\infty$. In the following subsection we discuss the phase portrait of \eqref{eps7}, giving special attention given to these homoclinic orbits. Comparison of the wave form \eqref{eps8} with 
the numerical continuation of breathers is presented in Section \ref{sec:numerics}.

We conclude this subsection with a remark on the case where $\epsilon$ is negative and small. In this case, we replace \eqref{x_to_y} by 
\begin{equation} \begin{pmatrix} x_n^A\\ x_n^B\end{pmatrix} =
\sqrt{-\epsilon}\begin{pmatrix} y_n^A\\ y_n^B\end{pmatrix},
\label{x_to_y1}
\end{equation}
which leads to the following gap soliton envelope system analogous to \eqref{eps7}:
\begin{align}\label{eps7a}
&\gamma_{\text{in}}U'=-U-\nu V-3\Gamma V^3\\ \nonumber
&\gamma_{\text{in}}V'=V+\nu U+3\Gamma U^3.
\end{align}

The system \eqref{eps7a} has an emergent symmetry-- a symmetry not present in the original discrete model.
If $(U,V)^\top$ is a solution of \eqref{eps7} which is homoclinic to $(0,0)$, 
then $i\sigma_2(U,V)^\top$ 
is a solution of \eqref{eps7a} which is homoclinic to $(0,0)$. It follows that for $\epsilon$ small and negative we have discrete breathers approximated by the expression:

\begin{equation}\label{eps8a}
   \begin{pmatrix}
   x_n^A(t)\\
   x_n^B(t)
    \end{pmatrix}
   \sim 2\sqrt{-\epsilon}\ (-1)^{n}\ \begin{pmatrix}
    -V(-\epsilon n; \nu)\\
    U(-\epsilon n; \nu)
    \end{pmatrix}\ \cos\left( \left[\omega_0 -\frac{(-\epsilon) \nu}{2\omega_0}\right] t\right),\quad\quad 0<-\epsilon\ll1.
\end{equation}

\subsubsection{Phase portraits and symmetries of midgap asymptotic description}
\begin{figure}[H]\label{phase}
  \centering
  \includegraphics[scale=0.7]{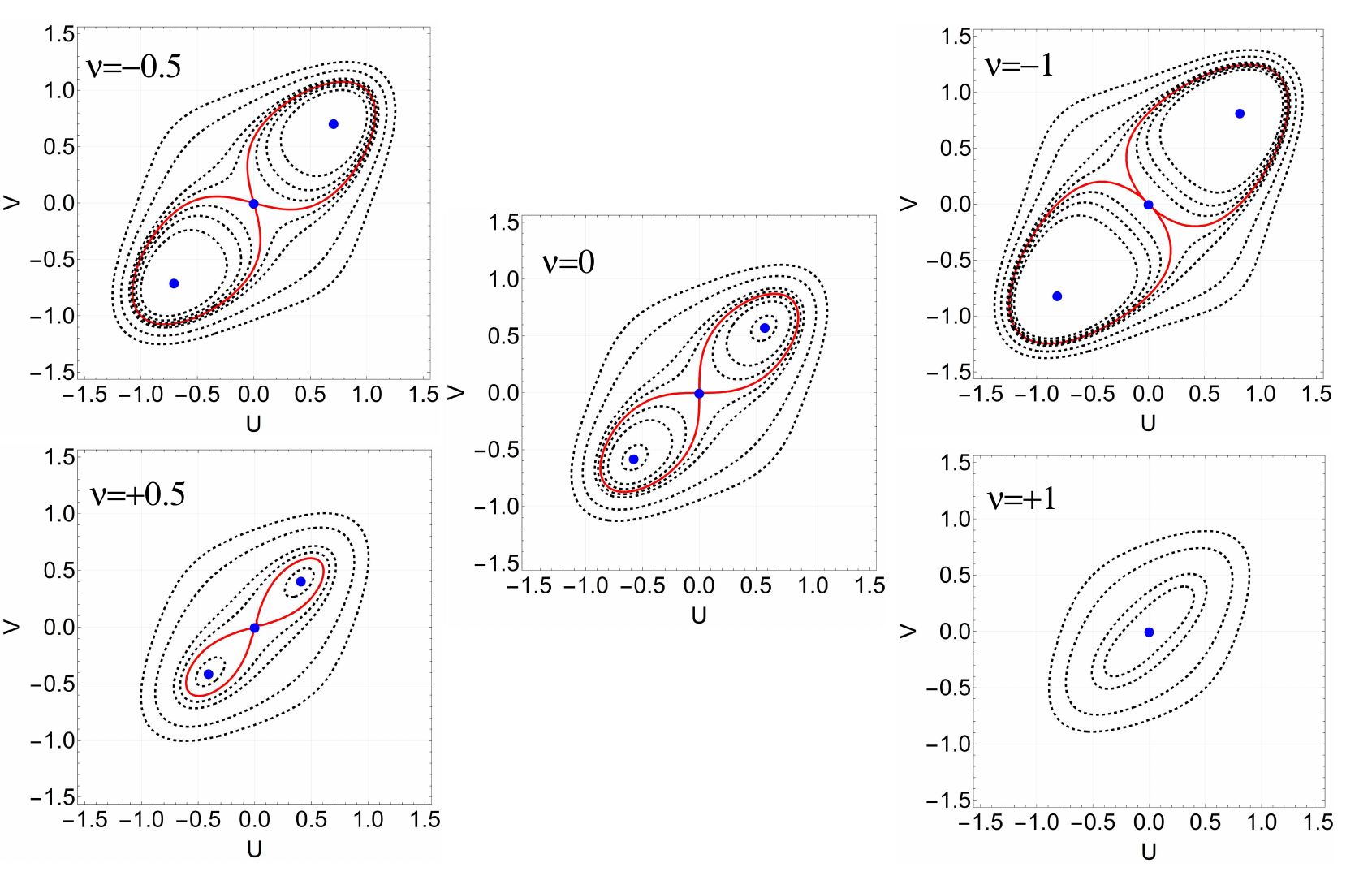}
  \caption{Phase portraits of  \eqref{eps7} for $\Gamma>0$ and different values of $\nu$ are given by the level sets of the Hamiltonian $H(U,V)$. Equilibrium points are marked (blue). Homoclinic orbits to the equilibrium at $(U,V)=(0,0)$, which correspond to the level set $H(U,V;\nu)=0$ (red curves) are indicated in subpanels.  For frequencies $|\nu|>1$, there are no homoclinic orbits.}
\end{figure}

The two-dimensional phase portrait of the system \eqref{eps7} is
given by the family of level curves of the 
Hamiltonian:
\begin{equation}\label{ham}
    \gamma_{\text{in}}H(U,V;\nu)\equiv UV-\dfrac{1}{2}\nu\left(U^2+V^2\right)-\dfrac{3}{4}\Gamma\left(U^4+V^4\right).
\end{equation}
 In  Figure \ref{phase}, we display phase portraits for representative values of $\nu$, and $\Gamma>0$. Darkened (blue) points are equilibria. As can be seen from the linearization of \eqref{eps7} about the zero solution, homoclinic (exponentially decaying) solutions to \eqref{eps7} can exist only if $|\nu|<1$,
  and in fact do exist for all $|\nu|<1$.
  Homoclinic orbits  (level sets $H(U,V;\nu)=0$, displayed as red contours), corresponding to different choices of $\nu$, 
  are displayed in Figure \ref{phase}. 
 
As $\nu$ is continuously increased from $\nu=0$ toward $\nu=+1$ the homoclinic figure-eight contracts to a point, and as $\nu$ is continuously decreased from $\nu=0$ toward $\nu=-1$ the homoclinic figure-eight expands till the two lobes of the figure become tangent at $(U,V)=(0,0)$.
\subsubsection{Preparation for comparison of asymptotic wave form \eqref{eps8}
 with numerical discrete breather in the continuum regime}
 \label{prep4comp}
With a view toward approximating the continuation of discrete breathers, whose breather frequency lie within $\mathcal{O}(\epsilon)$ width of the phonon gap ($\lambda$ near $\lambda_\star$), we take a breather frequency, $\omega_b$, of the form
\begin{equation} \omega_b \equiv \omega_0 -\frac{\epsilon \nu}{2\omega_0}.
\label{omega-fit}\end{equation}
Here $\nu$ specifies the frequency offset from the center of the phonon gap at $\omega_0$.

From \eqref{phonon} and the relation $\epsilon=\gamma_{\rm in}-\lambda\gamma_{\rm out}$, the width of the phonon gap can be computed at $k=\pi$; it is given to first-order by 
\begin{equation}
    \omega_+(\pi)-\omega_-(\pi)-=\sqrt{\omega_0^2+\epsilon}-\sqrt{\omega_0^2-\epsilon}\approx\dfrac{\epsilon}{\omega_0}
\end{equation}
Therefore, $\omega_b$ is in the spectral gap
 if and only if \[ |\omega_b-\omega_0|<\frac{\epsilon}{2\omega_0}\]
 or equivalently, using \eqref{omega-fit}, $|\nu|<1$.
 
Thus for $\epsilon$ small,  a choice of breather frequency \eqref{omega-fit} selects a distinguished homoclinic orbit (specified by the parameter $\nu$) which defines the slowly varying envelope in the multi-scale approximation \eqref{eps8} to the discrete breather.
In  Section \ref{sec:numerics}, we assess the accuracy 
of this analytic approximation through a comparison with discrete breathers
which are numerically continued from the highly discrete (anti-continuous) regime. As we shall see, the approximation is excellent in representative examples.

\subsection{
Asymptotics of discrete breathers in Regimes (B) and (C); $\omega^2$ just below / just above the acoustic / optical bands}
\label{nls}

We now search for asymptotic solutions of \eqref{y-eqn} spectrally localized just below/above the minimum/maximum of the phonon bands at $k=0$, again as $\epsilon\sim 0$ ($\lambda\sim\lambda_*$). We again introduce the continuum variables $u_{A,B}(z,t)$ and here expand to second-order in the spatial derivative.

In analogy with the discussion in Section \ref{a_dbs}, we consider $(y_n)$ spectrally concentrated near $k=0$ over a band width of order $\sqrt{\epsilon}$. We write
  \[ y^J_{n\pm1}-y^J_n =\frac{1}{\sqrt{\epsilon}}\int_{-\pi}^{\pi} e^{ink} (e^{\pm ik}-1) \chi^J\left({k\over\sqrt{\epsilon}}\right) dk, \]
  We set $k=\sqrt{\epsilon}K$ and make the approximation, $\left(e^{i\sqrt{\epsilon}K}-1\right)\approx\pm i\sqrt{\epsilon}K-\epsilon K^2/2$, due to the parabolic behavior of the bands near $k=0$.  Now we have
    \begin{align*} y^J_{n\pm1}-y^J_n \approx\sqrt{\epsilon}\int_{-\infty}^{\infty} e^{iK(\sqrt{\epsilon}n)} \left(\pm iK-\sqrt{\epsilon}\dfrac{K^2}{2}\right)\chi^J\left(K\right) dK\qquad\qquad\\
  \equiv \left(\pm \sqrt{\epsilon} \partial_Z u_J(Z)+\dfrac{\epsilon}{2}\partial_Z^2 u_J(Z) \right)\vert_{Z=\sqrt{\epsilon} n}=\pm\partial_z u_J(n)+\dfrac{1}{2}\partial_z^2 u_J(n), \quad Z=\sqrt{\epsilon} z.\end{align*}
  This motivates the multiple-scale scaling below.  Plugging the expansion
\[
y^{J}_{n\pm 1}=u_{J}\pm \partial_zu_{J}+\dfrac{1}{2}\partial^2_zu_{J}\]
into \eqref{y-eqn}, gives
\begin{align}\label{nls1}
\partial_t^2u_A=&-\omega_0^2u_A-\epsilon\Gamma u_A^3-\epsilon(u_B-\partial_zu_B+\dfrac{1}{2}\partial^2_zu_B)\\ \nonumber
&+\gamma_{\text{in}}\left(2u_B-\partial_zu_B+\dfrac{1}{2}\partial^2_zu_B\right)\\ \nonumber
\partial_t^2u_B=&-\omega_0^2u_B-\epsilon\Gamma u_B^3-\epsilon(u_A+\partial_zu_A+\dfrac{1}{2}\partial^2_zu_A)\\ \nonumber
&+\gamma_{\text{in}}\left(2u_A+\partial_zu_A+\dfrac{1}{2}\partial^2_zu_A\right).
\end{align}

We introduce slow spatial and temporal scales (contrast with 
\eqref{m-scale})
\begin{equation}
Z=\sqrt{\epsilon}z,\quad T=\epsilon t
 \label{m-scale1}   \end{equation} 
 and expand $u_{A,B}=u_{A,B}(t,z,T,Z)$ in powers of $\sqrt{\epsilon}$: 
\begin{equation}
    u_{A,B}=u_{A,B}^{(0)}(z,t,Z,T)+\sqrt{\epsilon} u_{A,B}^{(1)}(z,t,Z,T)+\sqrt{\epsilon}^2 u_{A,B}^{(2)}(z,t,Z,T)+\cdots.
\label{nls2}\end{equation}
We now implement the same expansion procedure as before to obtain a hierarchy of PDEs with respect to the fast variables, $z$ and $t$, at orders $\sqrt{\epsilon}^{0}$, $\sqrt{\epsilon}^{1}$, $\sqrt{\epsilon}^{2}$,\dots. We solve (the first three equations of) the hierarchy recursively, subject to the condition that $(z,t)\mapsto u(z,t,\cdot,\cdot)$ is bounded. This imposes non-resonance conditions on the source terms in this hierarchy which then constrain the dependence on the slow variables $Z$ and $T$.

At order $\sqrt{\epsilon}^{0}$, we have the system
\begin{equation}\label{nls3}
    \mathcal{L}
    \begin{pmatrix}
    u_A^{(0)}\\
    u_B^{(0)}
    \end{pmatrix}
    =0
\end{equation}
where 
\begin{equation}
\mathcal{L}\equiv\left(\partial_t^2+\omega_0^2\right)\sigma_0-\left(2\gamma_{\text{in}}+\dfrac{\gamma_{\text{in}}}{2}\partial^2_z\right)\sigma_1+\left(i\gamma_{\text{in}}\partial_z\right)\sigma_2.
\end{equation}
Here $\sigma_j$ are standard Pauli $2\times2$ matrices. Our solution, $u(z,t,\epsilon z,\epsilon t)$, will be a constructed as a slow modulation of a plane wave state of \eqref{nls3}. Since we shall seek a solution which is spectrally supported near maximum or minimum of the phonon spectrum,  we solve \eqref{nls3} by seeking a time-harmonic solution with $k=0$ of the form: $R(Z,T)\xi e^{i\omega t}$, $\xi\in\mathbb{C}^2$. Here, $R(Z,T)$ is constant with respect to the fast variables $z,t$. Substitution into \eqref{nls3} yields
$\omega=\omega_\pm$, where
\begin{equation}\label{nls5}
\omega_{\pm}^2\equiv \omega_0^2\pm2\gamma_{\text{in}}\quad\textrm{corresponding to}
\quad  \xi^{(\pm)}=  \begin{pmatrix}
    1\\
    \mp 1
    \end{pmatrix}.
\end{equation}
\begin{remark}
For $\epsilon$ small and positive, and $\gamma_{\rm in}>0$,  we have from \eqref{phonon1} that the upper/lower limits of the phonon spectrum (in terms of $E=\omega^2(k)$, see Figure \ref{fig:dimer}) 
 are given by:
 \[ \omega_0^2 \pm 2\gamma_{\rm in}\mp \epsilon=\omega_\pm^2\mp\epsilon.\]
 Hence,  $\omega_+^2$ corresponds to a breather frequency $\mathcal{O}(\epsilon)$ above
  the optical band, and $\omega_-^2$ corresponds to a frequency $\mathcal{O}(\epsilon)$ below
  the acoustic band.
\end{remark}
  We have at order $\sqrt\epsilon^0=1$, solutions of the form
\begin{equation}
    u^{(0,\pm)}(z,t,Z,T)=R(Z,T)\xi^{(\pm)} e^{i\omega_{\pm}t}+\text{c.c.}
\end{equation}
At order $\sqrt{\epsilon}^{1}$, we have 
\begin{equation}\label{nls6}
    \mathcal{L}
    \begin{pmatrix}
    u_A^{(1)}\\
    u_B^{(1)}
    \end{pmatrix}
    =
    \begin{pmatrix}
    -\gamma_{\text{in}}\partial_Zu_B^{(0)}+\gamma_{\text{in}}\partial_z\partial_Z u_B^{(0)}\\
    \gamma_{\text{in}}\partial_Zu_A^{(0)}+\gamma_{\text{in}}\partial_z\partial_Z u_A^{(0)}
    \end{pmatrix}
    = \gamma_{\text{in}}\partial_ZR(Z,T) \left(\pm\xi^{(\mp)}\right) e^{i\omega_{\pm}t}.
   % =\gamma_{\text{in}}h\partial_ZR
   %  \begin{pmatrix}
  %  \pm 1\\
  %  \end{pmatrix}
  %  e^{i\omega_{\pm}t}+\text{c.c.}
\end{equation}
Note that $\left\langle\xi^{(\pm)},\xi^{(\mp)}\right\rangle_{\mathbb{C}^2}=0$, from which it follows that the right hand side of \eqref{nls6} is non-resonant. (Had it turned out to be resonant, this would have called for an additional time-scale, $\sqrt{\epsilon}t$.) A particular solution of \eqref{nls6} can be constructed in the form of a constant multiple of its forcing term:
\begin{equation}
u_p^{(1,\pm)}(z,t,Z,t) = -\frac{1}{4}\partial_ZR(Z,T)
\xi^{(\mp)} e^{i\omega_{\pm}t} + c.c.
\end{equation}

At order $\sqrt{\epsilon}^2$ we have the system
\begin{equation}\label{nls7}
    \mathcal{L}
    \begin{pmatrix}
    u_A^{(2,\pm)}\\
    u_B^{(2,\pm)}
    \end{pmatrix}
    =
    \begin{pmatrix}
    G_{1,\pm}(z,t,Z,T)\\
    G_{2,\pm}(z,t,Z,T)
    \end{pmatrix},
\end{equation}
where
\begin{align}\label{nls8}
G_1=&-2\partial_t\partial_Tu_A^{(0)}-\Gamma\left(u_A^{(0)}\right)^3-u_B^{(0)}+\partial_zu_B^{(0)}-\dfrac{1}{2}\partial_z^2u_B^{(0)}-\gamma_{\text{in}}\partial_Zu_B^{(1)}\\ \nonumber
&+\dfrac{\gamma_{\text{in}}}{2}\partial_Z^2u_B^{(0)}+\gamma_{\text{in}}\partial_z\partial_Zu_B^{(1)}\\ \nonumber
G_2=&-2\partial_t\partial_Tu_B^{(0)}-\Gamma\left(u_B^{(0)}\right)^3-u_A^{(0)}-\partial_zu_A^{(0)}-\dfrac{1}{2}\partial_z^2u_A^{(0)}+\gamma_{\text{in}}\partial_Zu_A^{(1)}\\ \nonumber
&+\dfrac{\gamma_{\text{in}}}{2}\partial_Z^2u_A^{(0)}+\gamma_{\text{in}}\partial_z\partial_Zu_A^{(1)}.
\end{align}
Substitution of the expressions of $u^{(0,\pm)}$
 and $u^{(1,\pm)}_p$ into \eqref{nls7}-\eqref{nls8}  gives
\begin{align}\label{nls9}
    \mathcal{L}u^{(2,\pm)}&=\left(-2i\omega_{\pm}\partial_{T}R-3\Gamma|R|^2R\pm R\mp\dfrac{\gamma_{\text{in}}}{4}\partial_Z^2R\right)\xi^{(\pm)}
    e^{i\omega_{\pm}t}+\text{c.c}\\
    &\quad + \textrm{non-resonant source terms.}
\end{align}
Boundedness of $u^{(2,\pm)}$ imposes the constraint that the envelope function, $R(Z,T)$, satisfies an equation of nonlinear Schroedinger (NLS) type:
\begin{equation}\label{nls10}
2i\omega_{\pm}\partial_{T}R = \mp\dfrac{\gamma_{\text{in}}}{4}\partial_Z^2R-3\Gamma|R|^2R\pm R
\end{equation}
We remark that the NLS equation \eqref{nls10} is of ``focusing"-type, and thus has spatially localized solitary standing wave solutions (solitons), provided
\begin{align*}
\Gamma>0&\quad \textrm{(hardening nonlinearity) for $\omega_+$, and }\\
 \Gamma<0&\quad \textrm{(softening nonlinearity) for $\omega_-$}.
\end{align*}

We choose a solution to \eqref{nls10} of the form $R(Z,T)=e^{i\nu T/2\omega_{\pm}}S(Z;\nu)$, respectively.  If $\Gamma>0$, we have a long-wave asymptotic description of the discrete dimer lattice at frequencies just above the optical band and if 
$\Gamma<0$, we have a long-wave asymptotic description of the discrete dimer lattice at frequencies just below the acoustic band, given respectively by
\begin{equation}\label{ab}
   \begin{pmatrix}
   x_n^A(t)\\
   x_n^B(t)
    \end{pmatrix}^{\pm}
   \sim 2\sqrt{\epsilon}S(\sqrt{\epsilon}n;\nu) \begin{pmatrix}
    1\\
    \mp 1
    \end{pmatrix}\ \cos\left(\left[\omega_{\pm}+\dfrac{\nu\epsilon}{2\omega_{\pm}}\right] t\right).
\end{equation}
Localized solutions of $S$ exist above the optical band as long as $\nu>-1$ and below the acoustic band when $\nu<1$. Of course, because of the homogeneity of the nonlinearity, one can rescale the positive decaying solution of $S''-S+S^3=0$ to obtain $S(Z;\nu,\gamma_{\rm in},\omega_\pm)$.

\section{
Global numerical continuation of discrete breathers and comparision with analytical results}
\label{sec:numerics}
In this section we bring together our analytical results with numerical simulations. In previous sections we analytically constructed discrete breather states in the regime of weak coupling, a regime in which very few lattice sites are significantly active (anticontinuum limit, $\lambda\sim0$).
In the weakly nonlinear, long wave regime (continuum limit, $\lambda\uparrow\lambda_*$) we provided an asymptotic (multiple scale) construction
of discrete breather states which bifurcate from the phonon band; see  \eqref{eps8}.
  
Our global numerical continuation, for $\lambda\in[0,\lambda_*)$, of discrete breathers for numerous initializing choices of isolated dimer frequencies, $\omega_b$,
   shows that weakly nonlinear limiting states are continuations of the highly discrete breathers of Type I or Type II or both Type I and Type II,  depending on the prescribed location of $\omega_b$, relative to the phonon spectrum.  We numerically construct discrete breather solutions to \eqref{eq1A} via the iterative Fourier method outlined in the Appendix. 
  In the following subsections we specify an isolated dimer periodic orbit of Type I (in-phase) or Type II (out-of-phase) - see Section \ref{sec:application} - and discuss 
  its continuation. In Section \ref{sec:stability} we discuss dynamical stability.
  
\subsection{Breather solutions of \eqref{eq1A} inside the phonon gap}
We numerically solve for discrete breathers with a specified frequency, $\omega_b=\omega_0-\epsilon\nu/(2\omega_0)$, where $|\nu|<1$,  in the phonon gap; see \eqref{omega-fit}.  For different choices of $\nu$ we find good agreement with the $\nu$-dependent ``gap solitons" and the constructed breathers in the following subsections.

\subsection*{Asymptotic localization of mid-gap discrete breathers on  sublattices:}\ It is interesting to note that discrete breathers with a frequency located at the center of the phonon gap ($\nu=0$) are, for $|n|$ sufficiently large, dominantly supported on either the $A-$ sublattice or on the $B-$ sub-lattice.
This asymmetric localization of the spatial \textit{tails} of midgap breathers is seen in Figure
\ref{fig:continuation} (e) and (f) of the introduction. 
 Recently, gap solitons showing related behavior were observed experimentally  in nonlinear SSH-like photonic lattices  \cite{solny,bloch}. 

This behavior can be explained as follows. We expect that the asymptotic (large $|n|$) spatial localization of discrete breathers with \underline{midgap} frequency is  determined by the zero energy, exponentially decaying solutions of SSH:
 \begin{align}\label{chiral}
     &0=\lambda\gamma_{\text{out}}x_{n-1}^B+\gamma_{\text{in}}x_n^B\\ \nonumber
     &0=\lambda\gamma_{\text{in}}x_{n}^A+\lambda\gamma_{\text{out}}x_{n+1}^A.
 \end{align}
 Here, we consider the regime: $\lambda\gamma_{\text{out}}<\gamma_{\text{in}}$ and look for exponential solutions of \eqref{chiral} of the form $\rho^n\left(\xi^A,\xi^B\right)^{\top}$. This leads to
 \begin{equation}
     \begin{pmatrix}
       0 \\
       0
     \end{pmatrix}
     =\begin{pmatrix}
       0 & \lambda\gamma_{\text{out}}\rho^{-1}+\gamma_{\text{in}}\\
       \gamma_{\text{in}}+\lambda\gamma_{\text{out}}\rho & 0
     \end{pmatrix}
     \begin{pmatrix}
       \xi^A\\
       \xi^B
     \end{pmatrix}.
 \end{equation}
 There exist nontrivial solutions if and only if $\lambda\gamma_{\text{out}}\rho^{-1}+\gamma_{\text{in}}=0$ or $\lambda\gamma_{\text{out}}\rho+\gamma_{\text{in}}=0$, which yields
 \begin{equation}
     \rho_1=-\dfrac{\lambda\gamma_{\text{out}}}{\gamma_{\text{in}}},\ \xi=\begin{pmatrix}0 \\ 1\end{pmatrix}
     \quad\text{and}\quad
     \rho_2=-\dfrac{\gamma_{\text{in}}}{\lambda\gamma_{\text{out}}}, \ \xi=\begin{pmatrix}1 \\ 0\end{pmatrix}.
 \end{equation}
 Exponential solutions are
 \begin{equation}\label{chiral_2}
     \rho_1^n\begin{pmatrix}0 \\ 1\end{pmatrix}
     \quad\text{and}\quad
     \rho_2^n\begin{pmatrix}1 \\ 0\end{pmatrix}.
 \end{equation}
 Since our parameter regime is $0<\lambda\gamma_{\text{out}}<\gamma_{\text{in}}$ we have that $|\rho_1|<1$ and $|\rho_2|>1$.  Hence, the first solution in \eqref{chiral_2} decays as $n\to+\infty$ and the second solution in \eqref{chiral_2} decays as $n\to-\infty$.  The above hypothesis therefore implies that midgap discrete breathers are concentrated on $B$ sites for $n\to+\infty$ and on $A$ sites for $n\to-\infty$. This  is corroborated by both the discrete profiles and our leading-order asymptotics shown in Figure \ref{fig:continuation}. Indeed, from the homoclinic orbit in Figure \ref{phase} with $\nu=0$ (see \eqref{ham}), we have that $U=\mathcal{O}(V^3)$ as $Z\to+\infty$ and $V=\mathcal{O}(U^3)$ as $Z\to-\infty$. 
 
 A numerical fit of the tails of the breather shown in Figure \ref{fig:continuation} (f) corroborates this prediction; we find $x_n^B\gg x_n^A\sim (x_n^B)^3$ as $n\to+\infty$ and $x_n^A\gg x_n^B\sim (x_n^A)^3$ as $n\to-\infty$.  Furthermore, when the breather's frequency is slightly tuned away from the center of the gap, $\nu\neq 0$ (see Figures \ref{gap_above} and \ref{gap_below}) we find that to leading-order $x_n^B\sim x_n^A$ as $n\to\pm\infty$, consistent with our asymptotics in \eqref{eps7} and \eqref{eps8} for $0<|\nu|<1$.  We remark that the discrete breathers in  regimes (B) and (C) from section \ref{sec:continuum} (where the breather frequency is not in the narrow phonon gap) do not exhibit the asymmetric sub-lattice concentration seen in Figure \ref{fig:continuation} at all; see Figures \ref{acoustic}-\ref{optical}.     

\subsubsection{Type I (in-phase) breathers with mid-gap frequency; $\omega_b=\omega_0$, $\nu=0$, $\Gamma>0$}{
Figure \ref{fig:continuation}, shown in the introduction, displays the numerical continuation of \eqref{eq1A}, seeded with Type I data at the central dimer cell, from $\lambda=0$ to just before the phonon gap closes at $\lambda_*=\gamma_{\rm in}/\gamma_{\rm out}=1/3$; see the caption of Figure \ref{fig:continuation} for all parameter values.   The breather frequency, $\omega_b$, is fixed at the center of the phonon gap, $\omega_0$, and we use a hardening on-site potential, $\Gamma>0$.  Our numerical scheme converges to a discrete breather up to $\lambda=0.31$.  The red dots in Figure \ref{fig:continuation} (b) chart the $\ell^{\infty}$-norm of the family of discrete breathers, $X^{\lambda}(t=0)$, as a function of $\lambda$.  As the phonon gap width is $\mathcal{O}(\epsilon)$, we expect that as $\lambda\to\lambda_*$ ($\epsilon\to 0$), the leading-order envelope approximation, \eqref{eps8}, becomes relevant. 
}

For comparison, the solid-blue curve in Figure \ref{fig:continuation} (b) displays the $L^{\infty}(\mathbb{R})$-norm of the asymptotic solution, \eqref{eps8}, as a function of $\lambda$.  Figure \ref{fig:continuation} (b) shows that the $\ell^{\infty}$-norm of $X^{\lambda}(0)$ are nearly constant for $\lambda$ small (around the anti-continuum limit, $\lambda=0$), but decreases quickly as $\lambda$ approaches $\lambda_*$, where there is excellent agreement with the norm derived from \eqref{eps8}.  
 
Similarly, the $\ell^2$-norm of $X^{\lambda}(0)$ is shown in Figure \ref{fig:continuation} (c) with red $x's$. As $\lambda$ approaches $\lambda_*$, we have from the continuum approximation \eqref{eps8} that
\begin{align}\label{2norm}
    \|X^{\lambda}(0)\|_2^2 &\equiv\sum_{n\in\mathbb{Z}}x_n^A(0)^2+x_n^B(0)^2\sim
    4\epsilon\sum_{n\in\mathbb{Z}}U(\epsilon n;\nu)^2+V(\epsilon n;\nu)^2\qquad\qquad \\\nonumber
    &\sim 4\int_{-\infty}^{\infty}\left[U(z;\nu)^2+V(z;\nu)^2\right]dz, \quad \text{for}\ \epsilon\ll 1,
\end{align}
where $U$ and $V$ are the homoclinic solutions of system $\eqref{eps7}$ when $|\nu|<1$.  The horizontal blue line in Figure \ref{fig:continuation} (c) is at the level $\sim 1.72$, predicted by \eqref{2norm} for $\nu=0$ (mid-gap discrete breather). The numerically continued discrete breather's $l^2$ norm, $\|X^{\lambda}(0)\|_2$, is consistent with this prediction, as $\lambda\to\lambda_*$. 

Panels (d)-(f) in Figure \ref{fig:continuation} show the corresponding spatial profiles of discrete breathers for various values of $\lambda$.  For $\lambda$ near zero, the discrete solution bears little resemblance to the envelope profile, but as $\lambda\to\lambda_*$, we find that the breather's spatial outline closely resembles the long-wave, midgap vector-soliton given in \eqref{eps8}.  For comparison, panel (g) in Figure \ref{fig:continuation} shows $U$ and $V$ in \eqref{eps7}, scaled with the same value of $\lambda$ (equivalently $\epsilon=\gamma_{in}-\lambda\gamma_{\rm out}$) as the plot to its left.  Panels (f) and (g) in Figure \ref{fig:continuation} shows that near the point where the linear phonon bands close, there is remarkable agreement between the two profiles, whose origins 
lie in very different scaling regimes. 

\subsubsection{Type II (out-of-phase) breathers with mid-gap frequency; $\omega_b=\omega_0$, $\nu=0$, $\Gamma<0$} 
\begin{figure}[H]\label{midgap_2}
  \centering
  \includegraphics[scale=0.8]{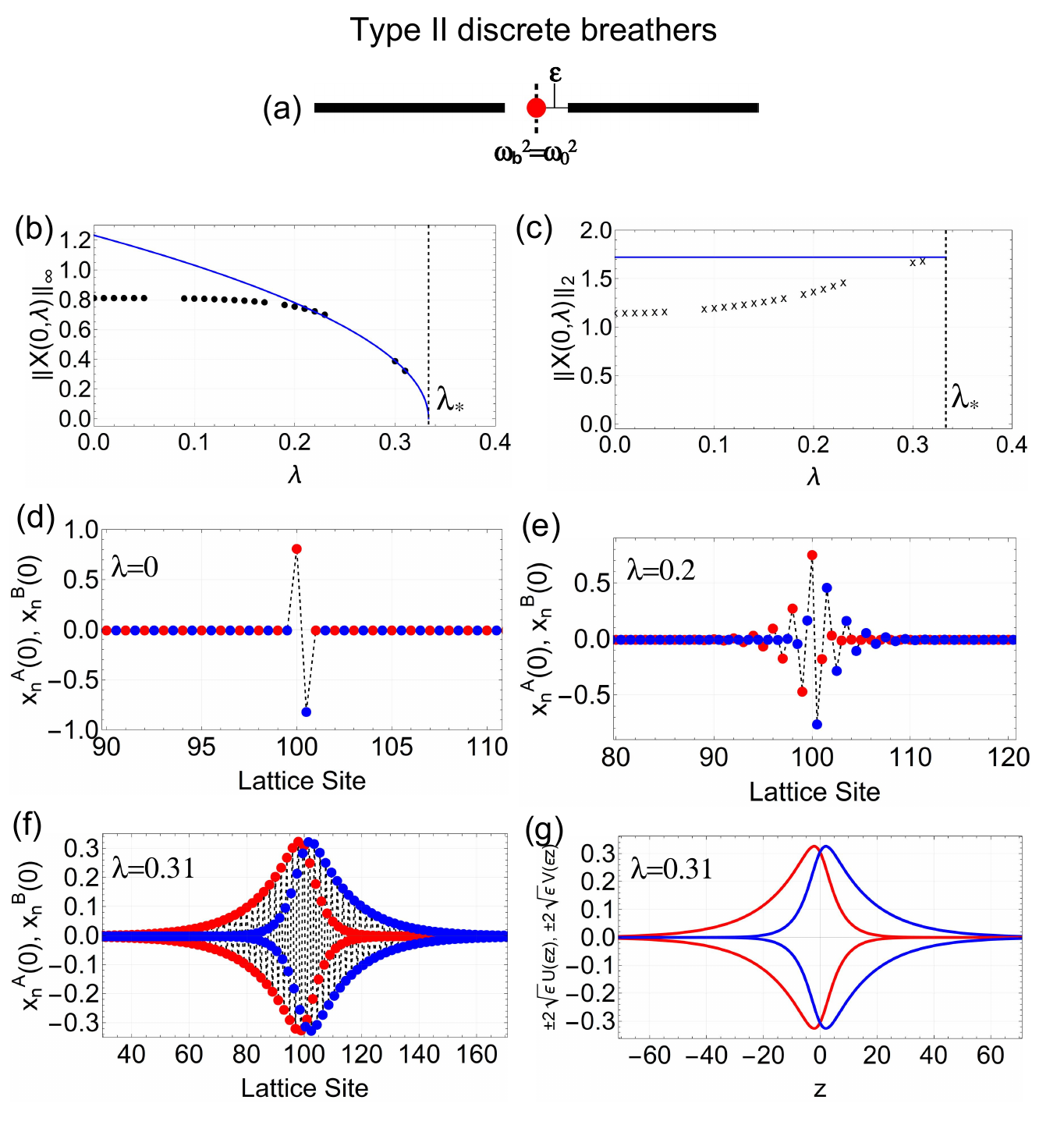}
  \caption{(a) phonon spectrum schematic (black) with breather frequency, $\omega_b$, (red); (b) $l^\infty$ norm of computed (dotted) discrete breather and its approximation from the weakly nonlinear long wave theory \eqref{envelope}; (c) $l^2$ norm of computed (x's) discrete breather and its approximation from the weakly nonlinear long wave theory \eqref{envelope}; (d), (e) and (f) show numerically computed discrete breather profiles for $\lambda=0,.2,.31$, respectively; and (g) shows envelope obtained from analytical approximation $z\mapsto (U(z),V(z))$, a homoclinic orbit of the system \eqref{eps7}. The continuation is initialized, for $\lambda=0$,
   with an anti-continuum out-of-phase periodic orbit of the nonlinear dimer \eqref{eq4} of frequency $\omega_b=\omega_0=\sqrt{V''(0)}=\sqrt{3}$, corresponding to the initial value parameter $a_*\sim 0.81$; see Section \ref{2classes}. Parameter values: $\Gamma=-1$, $\gamma_{\text{in}}=0.5$, $\gamma_{\text{out}}=1.5$, ($\lambda_*=1/3$), $N=201$.}
\end{figure}

Next, we continue discrete breathers into the long-wave regime, again with a breather frequency centered inside the phonon gap, but seeding the lattice with Type II (out-of-phase) data.  By \eqref{theo_anti} (and Figure \ref{fig:phaseports}), to initialize (at $\lambda=0$) an out-of-phase state  at the mid-gap frequency, $\omega_0$, we require  a softening nonlinearity ($\Gamma<0$).  Figure \ref{midgap_2} shows the results of this continuation, analogous to Figure \ref{fig:continuation}.    
\begin{remark}
  The absent sections in plots of  norms of $X^\lambda$ versus $\lambda$ in panels (b) and (c) of Figure \ref{midgap_2}, and likewise in other figures, are regions where the numerical scheme for the discrete breather failed to converge to the required tolerance. In these cases we find that the Jacobian
is nearly singular. We are presently investigating whether or not bifurcations occur. 
\end{remark} 

 For $\lambda$ near $\lambda_*$, the Type II discrete breathers displayed in Figure \ref{midgap_2} are very well-approximated by the same family of weakly-nonlinear, long-wave gap solitons
 as in Figure \ref{fig:continuation}. Note however, that this is the case even though for all $\lambda\in[0,\lambda_\star)$,  the lattice breathers maintain an odd-spatial-symmetry about their center (as seen in panels (d)-(f)), in contrast to the even spatial symmetry of the breathers shown in Figure \ref{fig:continuation}. On the microscale, the states are very different, but on the macroscale they agree.

 Note that, due to the odd-power nonlinearity in \eqref{eq1A}, both the original equations and system \eqref{eps7} have inversion symmetry--if $(U,V)$ is a solution then so is $(-U,-V)$.  System \eqref{eps7} has an additional symmetry--if $(U,V)$ is a solution then $(U,-V)$ is a solution to $\eqref{eps7}$ with $\Gamma\to-\Gamma$ and $\nu\to-\nu$. In particular, for a softening nonlinearity, the homoclinic orbits at $\nu=0$ shown in Figure \ref{phase} rotate into the second and fourth quadrants, explaining the nearly identical limiting profiles in Figures \ref{fig:continuation} and \ref{midgap_2} panel (f), having opposite spatial symmetry.  
\subsubsection{Breathers with frequency inside the phonon gap with frequency $\omega_b\ne\omega_0$ ($\nu\neq 0$), $\Gamma>0$} As in the previous subsections, we compare discrete breathers with the leading-order continuum approximation, \eqref{eps8}, but with fixed 
\begin{figure}[H]\label{gap_above}
  \centering
  \includegraphics[scale=0.8]{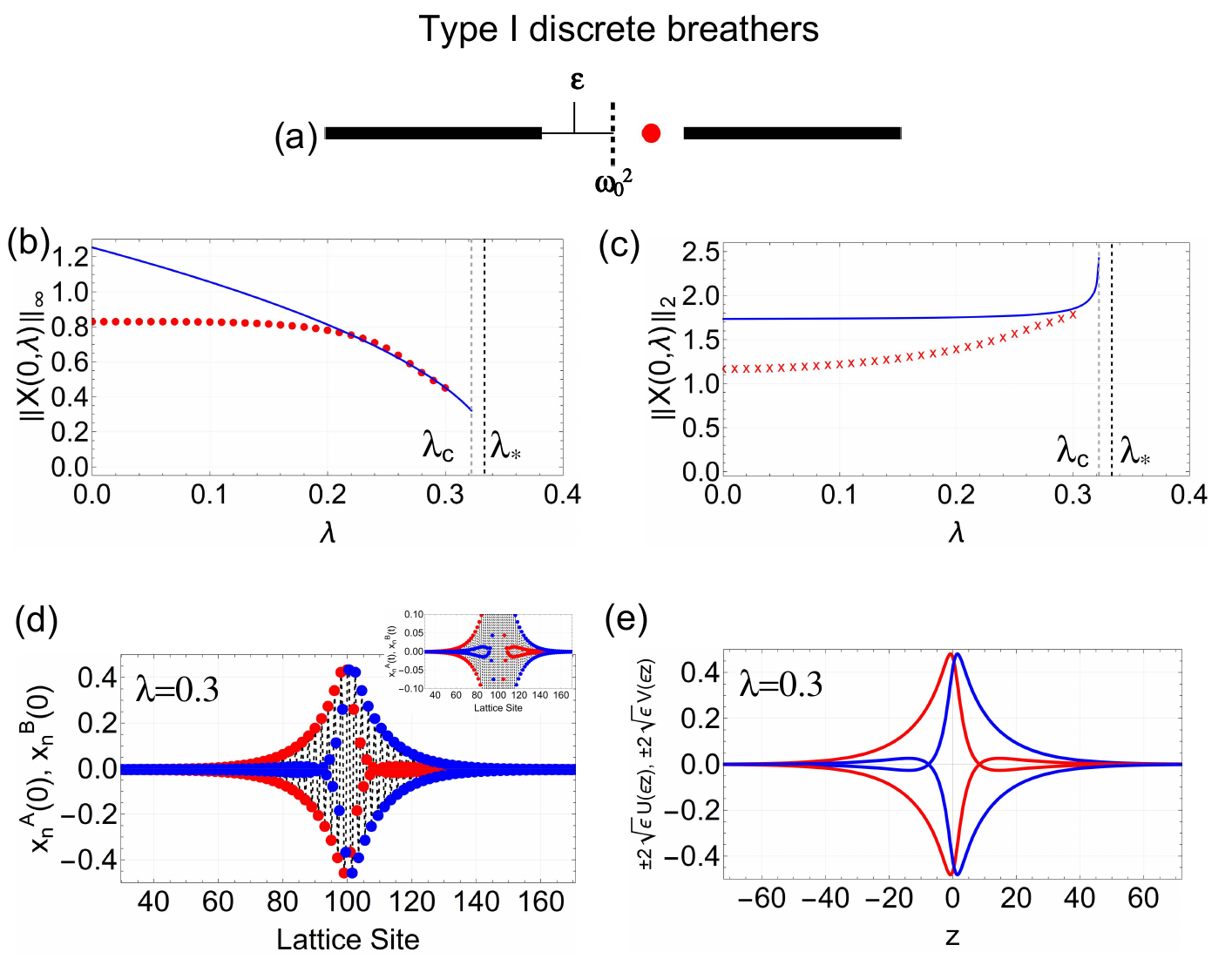}
  \caption{(a) phonon spectrum schematic (black) with breather frequency, $\omega_b$, (red); (b) $l^\infty$ norm of computed (dotted) discrete breather and its approximation from the weakly nonlinear long wave theory \eqref{envelope}; (c) $l^2$ norm of computed (x's) discrete breather and its approximation from the weakly nonlinear long wave theory \eqref{envelope}; (d) shows numerically computed discrete breather profile for $\lambda=0.3$ (the inset is simply a zoomed-in view); and (e) shows envelope obtained from analytical approximation $z\mapsto (U(z;\nu(\lambda)),V(z,\nu(\lambda)))$ at $\lambda=0.3$, a homoclinic orbit of the system \eqref{eps7}. The continuation is initialized, for $\lambda=0$,
   with an anti-continuum in-phase periodic orbit of the nonlinear dimer \eqref{eq4} with $\omega_b$ slightly above $\omega_0$, $|\omega_0-\omega_b|\sim 0.005$. Parameter values: $\Gamma=1$, $\nu(0.3)\sim-0.35$, $a_*=0.833$, $\gamma_{\text{in}}=0.5$, $\gamma_{\text{out}}=1.5$, ($\lambda_*=1/3$), $N=201$.}
\end{figure}
\noindent breather frequencies slightly above or below the center of the $\mathcal{O}(\epsilon)$ phonon gap ($\nu\neq 0$ in \eqref{eps7} and \eqref{eps8}).

Since the homoclinic trajectories in the phase portraits in Figure \ref{phase} deform asymmetrically with respect to the sign of $\nu$, we expect the continued discrete breather profiles to deform correspondingly, depending on whether the breather frequency is chosen inside the $\mathcal{O}(\epsilon)$ spectral gap with a value slightly above or below the gap's center.

To compare our envelope approximation \eqref{eps7}
 with numerically computed discrete breathers of frequency $\omega_b\ne\omega_0$, we use the homoclinic solution $(U(\cdot;\nu), V(\cdot;\nu))$ (see \eqref{eps7}),
where  $|\nu|<1$ is given by:
\begin{equation}\label{lam_c}
    \nu(\lambda)=\dfrac{2\omega_0(\omega_0-\omega_b)}{\gamma_{\text{in}}-\lambda\gamma_{\text{out}}},\text{ for }\lambda\in [0,\lambda_c),\text{ and }\lambda_c=\lambda_*-\dfrac{2\omega_0|\omega_0-\omega_b|}{\gamma_{\text{out}}}
\end{equation}
is the value of the coupling parameter, $\lambda$, for which the phonon band edge touches $\omega_b$.

Figure \ref{gap_above} shows the continuation of Type I seeded data with a breather frequency, $\omega_b$, just slightly above the center of the spectral gap located at $\omega_0^2$ (see schematic in Figure \ref{gap_above}a ).  Panels (b) and (c) again show the variations with $\lambda$ of the $\ell^{\infty}$ and $\ell^2$-norms of $X^{\lambda}(0)$, respectively, along with the corresponding $L^\infty$ and $L^2$ norms obtained from the continuum asymptotic theory.  

Figure \ref{gap_above}b shows that as the distance between $\omega_b$ and the upper band edge approaches zero ($\lambda\uparrow\lambda_c$), the $\ell^{\infty}$-norms closely follows the continuum-theory curve (computed using \eqref{2norm} and $\eqref{lam_c}$), which in this case limits to a nonzero value at $\lambda=\lambda_c$; the bifurcation from the phonon edge takes place a non-zero $l^\infty$ norm. 

Likewise, Figure \ref{gap_above}c shows that the distance of the upper band edge
to $\omega_b$ tends to zero, the $\ell^{2}$-norm approaches the continuum theory curve  from below. In contrast to the case of discrete breathers with $\omega_b$ in the center of the phonon gap, the computed $\ell^2$-norms and the approximating continuum theory in Figure \ref{gap_above} (c) do not level-off near $\lambda_c$. Panels (d) and (e) of Figure \ref{gap_above} display a comparison of the discrete breather spatial profile and the continuum theory envelope at a point near $\lambda_c$; they show remarkably good agreement (see the zoomed-in inset in panel (d)).

\begin{figure}[H]\label{gap_below}
  \centering
  \includegraphics[scale=0.8]{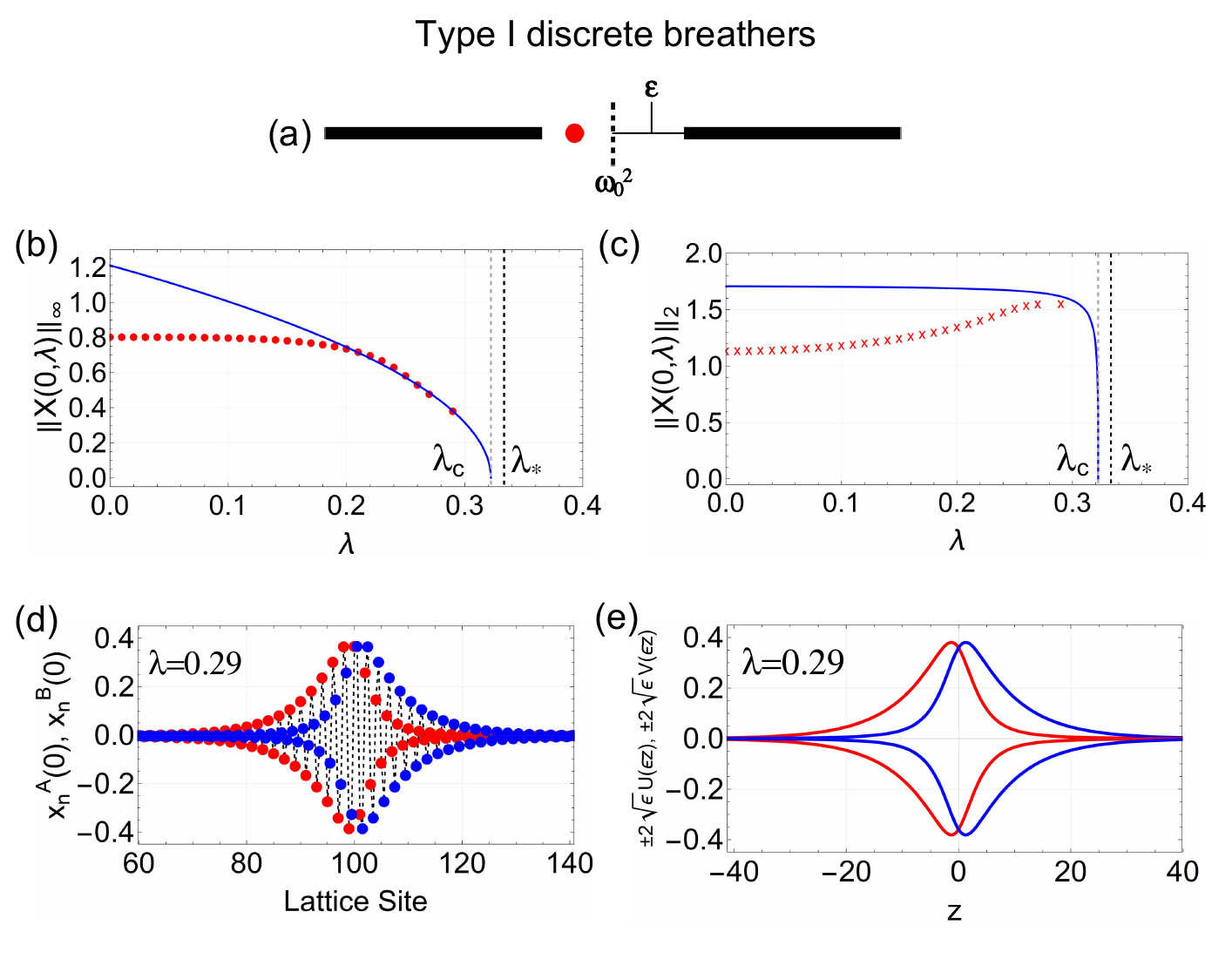}
  \caption{(a) phonon spectrum schematic (black) with breather frequency, $\omega_b$, (red); (b) $l^\infty$ norm of computed (dotted) discrete breather and its approximation from the weakly nonlinear long wave theory \eqref{envelope}; (c) $l^2$ norm of computed (x's) discrete breather and its approximation from the weakly nonlinear long wave theory \eqref{envelope}; (d) shows numerically computed discrete breather profile for $\lambda=0.29$; and (e) shows envelope obtained from analytical approximation $z\mapsto (U(z;\nu(\lambda)),V(z,\nu(\lambda)))$ at $\lambda=0.29$, a homoclinic orbit of the system \eqref{eps7}. The continuation is initialized, for $\lambda=0$,
   with an anti-continuum in-phase periodic orbit of the nonlinear dimer \eqref{eq4} with $\omega_b$ slightly below $\omega_0$, $|\omega_0-\omega_b|\sim 0.005$. Parameter values: $\Gamma=1$, $\nu(0.29)\sim 0.27$, $a_*=0.819$, $\gamma_{\text{in}}=0.5$, $\gamma_{\text{out}}=1.5$, ($\lambda_*=1/3$), $N=201$.}
\end{figure}

 Figure \ref{gap_below} shows the continuation of a family of discrete breathers for $\omega_b$ just below the center of the spectral gap. In this case, $\lambda_c<\lambda_*$, where $\lambda_c$ is the value of $\lambda$ for which the phonon band edge touches $\omega_b$. The continuum theory applies for $\lambda$ near and below $\lambda_c$. Note that the $\infty-$ norms approaches zero as $\lambda\to\lambda_c<\lambda_*$, as anticipated by the continuum theory.
 %from the phase portrait analysis of %system \eqref{eps7}. 
 In contrast to Figure \ref{gap_above}c, Figure \ref{gap_below}c shows that the  numerically computed $\ell^2$-norm and its continuum theory $L^2$ approximation of $X^{\lambda}(0)$ level-off below the constant mid-gap asymptotic solution ($\sim 1.72$); see Figures \ref{fig:continuation} and \ref{midgap_2}.   The breather profile  is compared with the asymptotic solution \eqref{eps8}, for $\lambda$ near $\lambda_c$, in Figure \ref{gap_below}. Panels  panels (d) and (e) show   agreement. 

\begin{remark} In both Figures \ref{gap_below} and \ref{gap_above} we have continued a Type I (in-phase) isolated dimer state periodic orbit with $\Gamma>0$.  In the case of a Type II (out-of-phase) initiating state with $\Gamma<0$, we can again continue breathers with frequencies slightly below or above the center of the spectral gap near $\lambda_c$. However, due to the symmetry mentioned in the concluding remarks of the previous subsection, the asymptotic homoclinic orbits deform with respect to $\nu$ in the opposite direction. \end{remark}  

\subsection{Discrete breathers with $\omega_b$ below/above the phonon spectrum- ``out-of-gap" discrete breathers}
We now turn to the continuation of discrete breathers of \eqref{eq1A} with $\omega_b$ fixed and $\mathcal{O}(\epsilon)$-distant (with $\epsilon\to 0$) below the acoustic (lower phonon) band or above the optical (upper phonon) band. This corresponds to regimes (B) and (C); see Section \ref{sec:continuum}.

As shown in Subsection \ref{nls}, the relevant  leading-order continuum approximation is given by the cubic nonlinear Schr\"odinger equation \eqref{nls10} and its family of explicit soliton solutions.  For example, for frequencies in regime (B) (whose frequenciesa are below the  acoustic band),  we have:
\begin{equation}\label{nls_sech}
    S(z;\nu)=\sqrt{\dfrac{2(\nu-1 )}{3\Gamma}}\text{sech}\left[2\sqrt{\dfrac{|\nu- 1|}{\gamma_{\text{in}}}}z\right],
\end{equation}
where $\nu\in(-\infty,1)$ and $\Gamma<0$. In analogy with \eqref{2norm}, we may use \eqref{nls_sech} to obtain  (in regime (B)) :
\begin{equation}\label{asymp_nls}
    \|X^{\lambda}(0)\|_2^2\sim 4\sqrt{\epsilon}\int_{-\infty}^{\infty}S(z;\nu)^2dz=\dfrac{8\sqrt{|\nu- 1|\gamma_{\text{in}}\epsilon}}{3|\Gamma |} \quad \text{for}\ 0<\epsilon\ll 1.
\end{equation}
Expressions \eqref{nls_sech} and \eqref{asymp_nls} also apply above the optical band in regime (C), with $\Gamma>0$ and  $\nu-1$ replaced by $\nu+1$ and  $\nu\in(1,\infty)$.

\begin{figure}[H]\label{acoustic}
  \centering
  \includegraphics[scale=0.8]{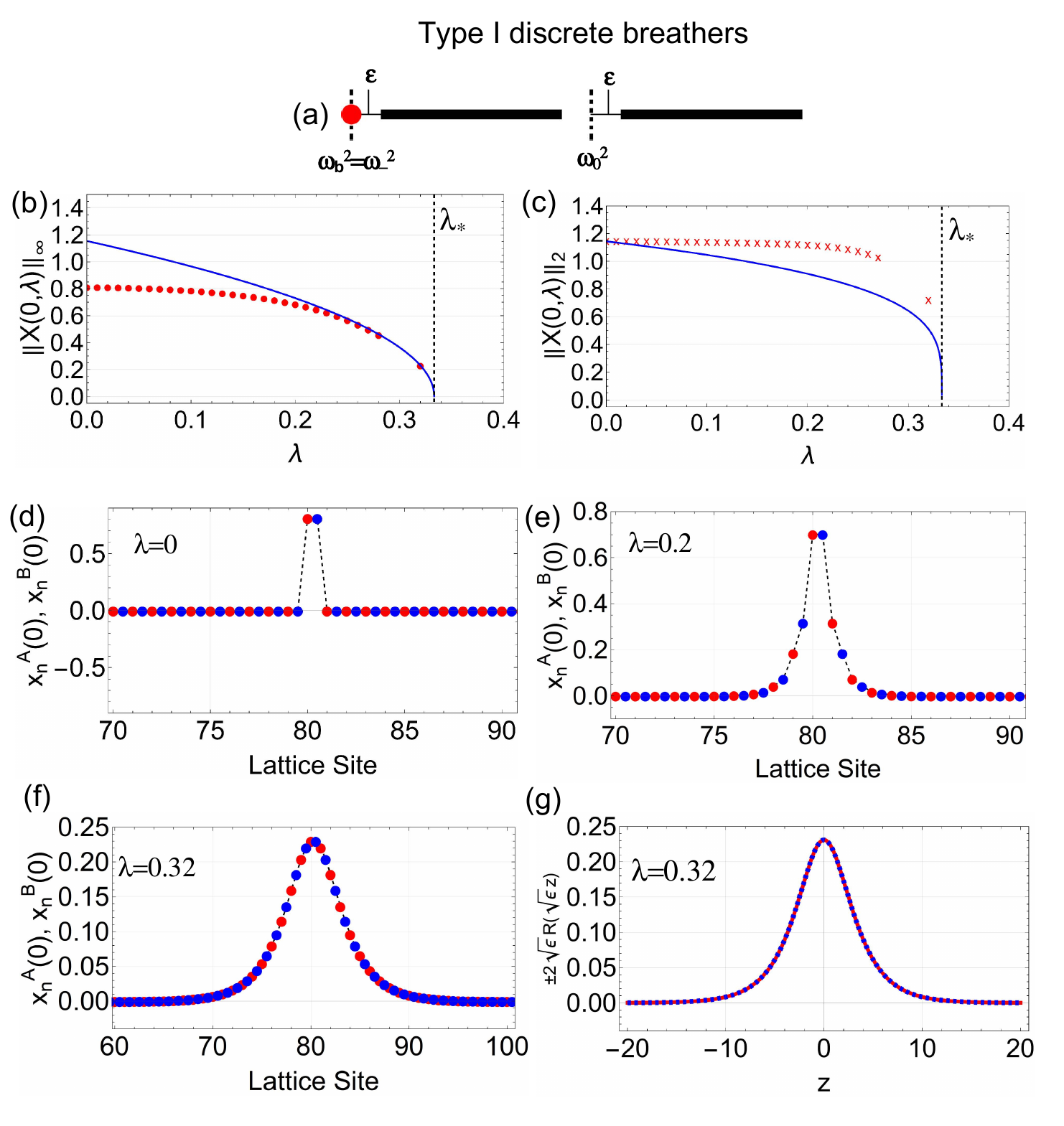}
  \caption{(a) phonon spectrum schematic (black) with breather frequency, $\omega_b$, (red); (b) $l^\infty$ norm of computed (dotted) discrete breather and its approximation from the weakly nonlinear long wave theory \eqref{ab}; (c) $l^2$ norm of computed (dotted) discrete breather and its approximation from the weakly nonlinear long wave theory \eqref{nls10}; (d), (e) and (f) show numerically computed discrete breather profiles for $\lambda=0,.2,.32$, respectively; and (g) shows envelope obtained from analytical approximation $z\mapsto R(z)$, a homoclinic orbit of the system \eqref{nls10}. The continuation is initialized, for $\lambda=0$,
   with a anti-continuum in-phase periodic orbit of the nonlinear dimer \eqref{eq4} of frequency $\omega_b=\omega_-=\sqrt{2}$, corresponding to the initial value parameter $a_*\sim 0.81$; see Section \ref{2classes}. Parameter values: $\Gamma=-1$, $\gamma_{\text{in}}=0.5$, $\gamma_{\text{out}}=1.5$ ($\lambda_*=1/3$), $N=161$.}
\end{figure}

\begin{figure}[H]\label{optical}
  \centering
  \includegraphics[scale=0.8]{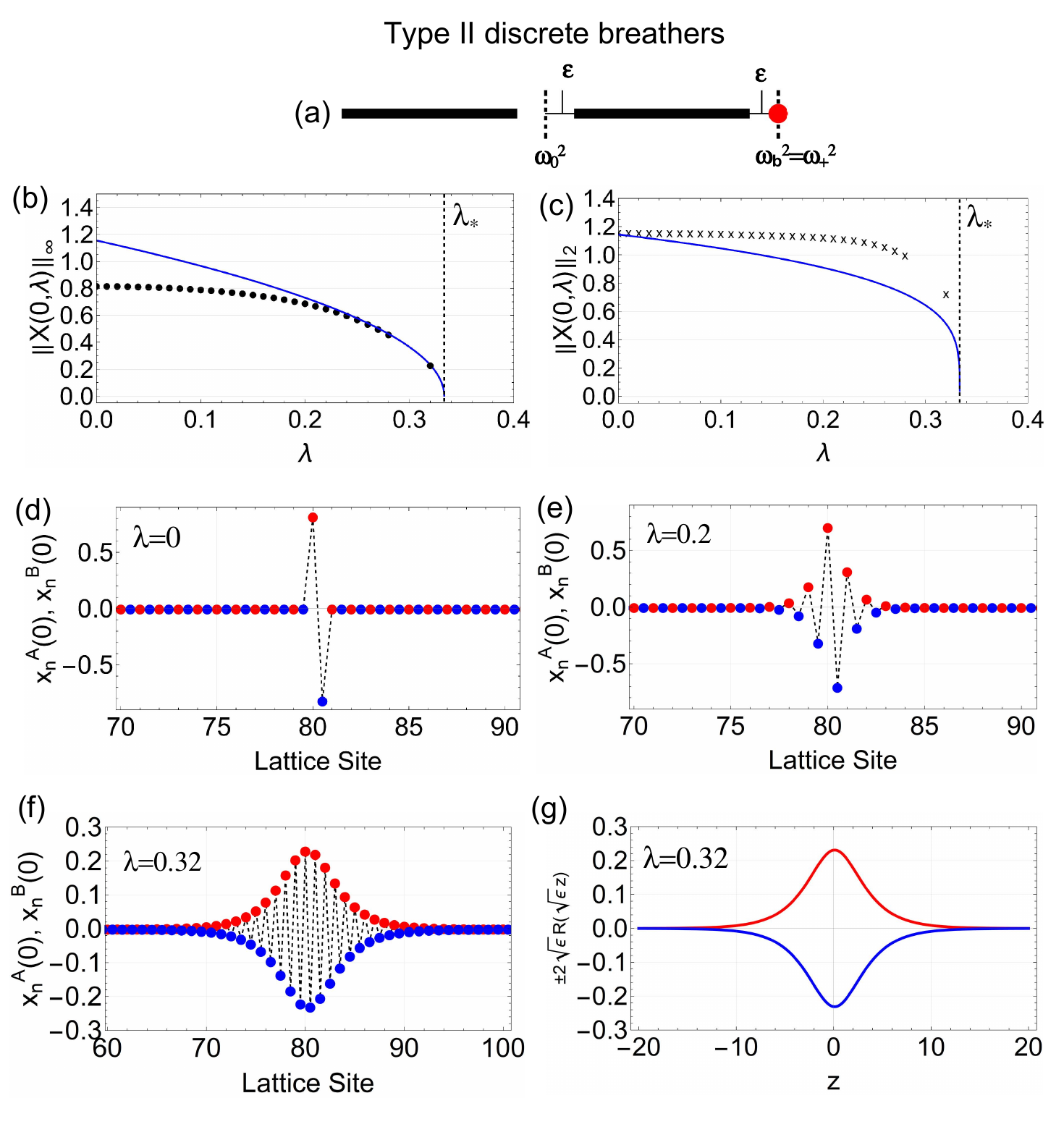}
   \caption{(a) phonon spectrum schematic (black) with breather frequency, $\omega_b$, (red); (b) $l^\infty$ norm of computed (dotted) discrete breather and its approximation from the weakly nonlinear long wave theory \eqref{ab}; (c) $l^2$ norm of computed (dotted) discrete breather and its approximation from the weakly nonlinear long wave theory \eqref{nls10}; d), e) and f) show numerically computed discrete breather profiles for $\lambda=0,.2,.32$, respectively; and (g) shows envelope obtained from analytical approximation $z\mapsto R(z)$, a homoclinic orbit of the system \eqref{nls10}. The continuation is initialized, for $\lambda=0$,
   with a anti-continuum out-of-phase periodic orbit of the nonlinear dimer \eqref{eq4} of frequency $\omega_b=\omega_+=2$, corresponding to the initial value parameter $a_*\sim 0.82$; see Section \ref{2classes}. Parameter values: $\Gamma=1$, $\gamma_{\text{in}}=0.5$, $\gamma_{\text{out}}=1.5$ ($\lambda_*=1/3$), $N=161$. }
\end{figure}

We remark that the spatial asymptotic descriptions in regimes (B) and (C) are identical, with the exception of the vector $\xi^{(\pm)}$ in \eqref{ab}. The vector $\xi^{(\pm)}$ determines the sign of the envelope solution on each sub-lattice site, either constant or alternating.  To leading-order we observe this behavior in the computed discrete breathers in regimes (B) and (C), respectively (see Figures \ref{acoustic} and \ref{optical}).  As before, the parameter $\nu$ in the asymptotic expression \eqref{ab} represents a modulation about the frequency $\omega_{\pm}$.  In the following figures we set $\nu=0$.

First, we consider regime (B). Figure \ref{acoustic} shows a family of continued breathers with frequency $\omega_-$, seeded by Type I data and with a softening potential, $\Gamma<0$. Panels (b) and (c) in Figure \ref{acoustic} again track the $\ell^{\infty}$ and $\ell^{2}$-norms as the $\mathcal{O}(\epsilon)$-distance between $\omega_-$ and the acoustic band approaches zero. The norms of the asymptotic expressions, \eqref{nls10} and \eqref{ab}, are again shown by the solid curves.  Figure \ref{acoustic} (b) and (c) show that both norms of $X^{\lambda}(0)$, seeded by Type I data, can be continued near $\lambda_*$ and closely approach the asymptotic curves, which in this case both limit to zero. Panels (d)-(f) in Figure \ref{acoustic} show the deformation of Type I discrete breather profiles at different values of $\lambda$ and panel (g) shows the corresponding NLS soliton from $\eqref{ab}$ at $\lambda=0.32$, again showing excellent agreement. 
Figure \ref{optical} is analogous to Figure \ref{acoustic}, but with a breather frequency fixed just above the optical band (regime (C)).  To respect the sub-lattice spatial symmetry of the envelope solution \eqref{ab}, we seed the breather family with Type II data. Figure \ref{optical} shows that out-of-phase states can indeed be continued near the band-edge, where our asymptotics again capture the breather's spatial outline.
  
\subsection{Dynamical stability}\label{sec:stability}
In this section we discuss the linear dynamical stability of discrete breathers. In particular we study the time evolution for the linearization of the dynamical system \eqref{eq1A} about representative numerically computed discrete breathers.
This dynamical system is an infinite system of coupled linear ordinary equations (ODEs) with $T_b=2\pi/\omega_b-$ periodic coefficients.
Its stability/instability properties are characterized  by the spectrum of the {\it monodromy operator}:  the operator which maps an initial state to the state at time $T_b$. The monodromy operator is the Jacobian of the Poincar\'e mapping, $\partial_u G(u(0);\lambda)$, where the Poincar\'e map:  $X(0)\mapsto G(u(0);\lambda)$ maps an initial data vector $X(0)$ for \eqref{eq1A} into the solution at time $T_b$. The spectrum of
$\partial_u G(u(0);\lambda)$ is called the {\it Floquet spectrum} and points in the Floquet spectrum are called {\it Floquet multipliers}.  A  discrete breather, $X^\lambda$, is spectrally stable if its associated Floquet spectrum lies in the closed  unit disc in $\mathbb{C}$.
Since our coupled dimer network is Hamiltonian, if 
$\mu$ is a Floquet multiplier then so are  $\bar{\mu}$, $1/\mu$ and $1/\bar{\mu}$; complex Floquet multipliers come in quartets and real Floquet multipliers in reciprocal pairs. It follows that a discrete breather of the system is spectrally stable only if the Floquet spectrum is a subset of the unit circle.  

If, for the purpose of numerical approximation, we  truncate the linearized dynamics to a finite system with $N-$ dimers, we have a system of $2N$ coupled linear ODEs of second order with $T_b=2\pi/\omega_b-$ periodic coefficients. Written as a system of first order ODEs, we have an equivalent system of $4N$ ODEs. The Poincar\'e mapping is then a mapping $u(0)\in \mathbb{R}^{4N}\mapsto G^{(N)}(u(0);\lambda)\in \mathbb{R}^{4N}$,
 and the monodromy operator is approximated by $\partial_uG^{(N)}(u(0);\lambda)$, a $4N\times4N$ matrix. We investigate the linear spectral stability of representive choices of numerically computed breathers by computing the eigenvalues of $\partial_uG^{(N)}(u(0);\lambda)$, for tractable and appropriately large values of $N$. These eigenvalues are taken as approximations to the exact Floquet spectrum associated with the $X^\lambda$. The choices of $N$ are provided in figures.

\begin{figure}[H]\label{floq}
  \centering
  \includegraphics[scale=0.6]{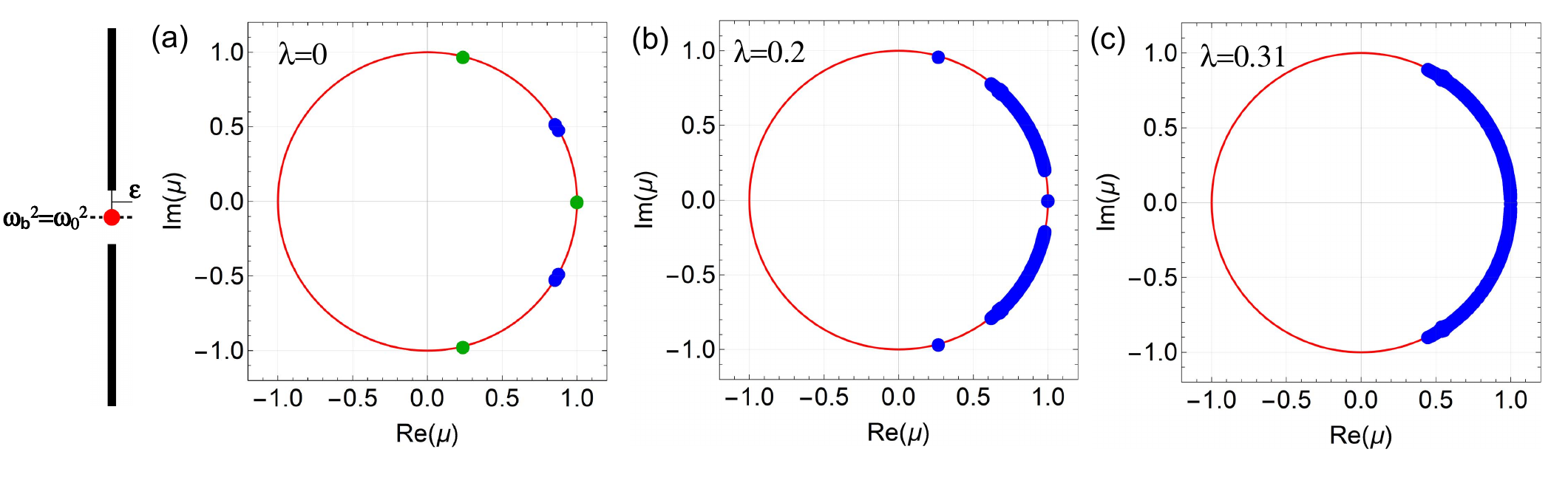}
   \vspace{-0.3cm}
   \caption{The corresponding numerically-obtained Floquet spectra (linear stability) for the converged breathers shown in Figure \ref{fig:continuation} with a frequency centered inside the phonon gap at various values of $\lambda$.  The panels indicate that the discrete breathers obtained in Figure \ref{fig:continuation} are \underline{\textbf{linearly spectrally stable}}. }
\end{figure}
Figure \ref{floq} shows the eigenvalues $\mu_j \ (1\leq j\leq 4N)$ of  $\partial_u G^{(N)}(u(0);\lambda)$ for numerically computed discrete breathers with values of $\lambda$ (and other parameters) corresponding to those for Figure \ref{fig:continuation}d-f.   

At $\lambda=0$,
$\partial_u G^{(N)}(u(0);\lambda)$ is block diagonal with each $4\times4$ block corresponding to one of the $N-$  non-interacting dimers. The $n=0$ block has four Floquet multipliers.  One Floquet multiplier is equal to $+1$. It has  algebraic multiplicity 2 and geometric multiplicity 1; this follows because $\dot{z}_*(t;E)$ is periodic solution of $L_{*-}Y=0$ and  $\partial_E z(t;E)$ solves this ODE and has linear growth in $t$;
see Section \ref{sec:existence}. The other two Floquet multipliers are associated with the linear time-periodic ODE $L_{*,+}Y=0$. Their product is equal to one.
These values are indicated with x's  on the unit circle in Figure \ref{floq}a.  

The $n\ne0$ blocks are identical (corresponding to a constant coefficient system of ODEs) and thus each $n$ contributes the same four Floquet multipliers.
They are of the form  $e^{i\omega T_b}$ where $\omega$ varies over the four roots of:  $\det(-\omega^2+V''(0)-\gamma_{\rm in}\sigma_1)=0$.
We indicate these multipliers with blue points in Figure \ref{floq}a.
In this particular simulation, there are two very nearby Floquet multipliers on the unit circle in the second and fourth quadrants. 
For $\lambda=0$, these four points on the unit circle are eigenvalues of the monodromy operator of infinite multiplicity; they lie in the essential spectrum.

As $\lambda$ is varied away from zero, these infinitely degenerate eigenvalues perturb into continuous arcs of spectrum along the unit circle. These arcs of Floquet multipliers, for $\lambda\ne0$, are clearly seen in the center panel of Figure \ref{floq}. $\mu=1$ remains a Floquet multiplier and the other two simple Floquet multipliers, associated with $n=0$ for $\lambda=0$, can move around and perhaps collide with these growing arcs of Floquet multiplier spectrum.  In  Figure \ref{floq}, the Floquet multipliers continue to lie along the unit circle as  $\lambda$ is varied till close to $\lambda_*$. We conclude that the  discrete breather discussed in  Figure \ref{fig:continuation} is linearly spectrally stable. For the other branches of discrete breathers discussed in Section \ref{sec:numerics}, we found that those corresponding to frequencies near the phonon band edge were linearly stable as well.
%We also computed the Floquet spectra of the other discrete breathers discussed in Section \ref{sec:numerics}.  We find that each of the breathers near the phonon bands, close to $\lambda_*$, are also linearly stable.

Finally,  in  Hamiltonian contexts similar to this, it has been shown that
 spectrally stable states are stable on exponentially long, but finite, time scales \cite{bambusi}.  
 
\subsection{More general nonlinearity}\label{sec:nonlin}
In this section we remark on extensions of our results to a general class of  anharmonic potentials: \begin{equation}\label{higher} 
V(x)=\dfrac{3}{2}x^2+\dfrac{\Gamma}{2\sigma+2}|x|^{2\sigma}x^2,\quad \sigma>0,
\end{equation}
which satisfies \eqref{pot} and agrees with special case considered above for $\sigma=1$.
Since the potential \eqref{higher} is $C^2$ and even, Theorems \ref{thm:breather} and \ref{thm:exp_decay} on  existence and exponential decay of discrete breathers, for $\lambda$ small, extend to the dimer network \eqref{nloc-case} with potential, $V(x)$, given by  \eqref{higher}.

Our asymptotic analysis of the weakly nonlinear long wave regime $\lambda\to\lambda_\star (\epsilon\to0)$ can be implemented for sufficiently smooth nonlinearities  and yields  envelope equations (effective nonlinear Dirac equations for $\omega_b$ in-gap; Case (A) of Section  \ref{sec:continuum}) and effective nonlinear Schroedinger equations for $\omega_b$ out-of-gap; Cases (B) or (C) of Section \ref{sec:continuum}). We next highlight characteristics of discrete breathers in this regime which depend strongly on the nonlinearity, in particular the parameter $\sigma$.

Consider the case where $\omega_b$ is in-gap; Case (A) of Section  \ref{sec:continuum} and  $\sigma\in\mathbb{N}$. In this case, our asymptotic analysis yields, at leading order, a discrete breather  $X^\lambda(t)=\{x_n(t,\epsilon)\}_{n\in\mathbb{Z}}$ of the form:
 \begin{equation}
x_n(t,\epsilon)\ \equiv \begin{pmatrix}
    x^A_n(t,\epsilon)\\  x^B_n(t,\epsilon) \end{pmatrix} \approx 2\epsilon^{1/2\sigma} (-1)^n 
    \begin{pmatrix} U(\epsilon n;\nu)\\ V(\epsilon n;\nu) 
    \end{pmatrix} \cos\left[\ \left(\omega_0-\epsilon\frac{ \nu}{2\omega_0}\right)t\right],
        \label{envelope2} \end{equation}
here $(U,V)$ is a decaying solution (homoclinic orbit to $(0,0)$)  of the system
\begin{align}\label{eps72}
&\gamma_{\text{in}}U'=U-\nu V-\binom{2\sigma+1}{\sigma} \Gamma V^{2\sigma+1}\\ \nonumber
&\gamma_{\text{in}}V'=-V+\nu U+\binom{2\sigma+1}{\sigma}\Gamma U^{2\sigma+1}.
\end{align}
A computation analogous to the one given in Section \ref{sec:continuum} yields
\begin{equation}\label{nonlin_l2}
\|X^\lambda\|^2_2\approx 4\epsilon^{\frac{1}{\sigma}-1}\left(\|U(z;\nu)\|_{L^{2}(\mathbb{R})}^2+\|V(z;\nu)\|_{L^{2}(\mathbb{R})}^2\right).
\end{equation}
The case $\sigma=\sigma_D=1$ is $L^2-$ scaling critical.
\begin{figure}
  \centering
  \includegraphics[scale=0.8]{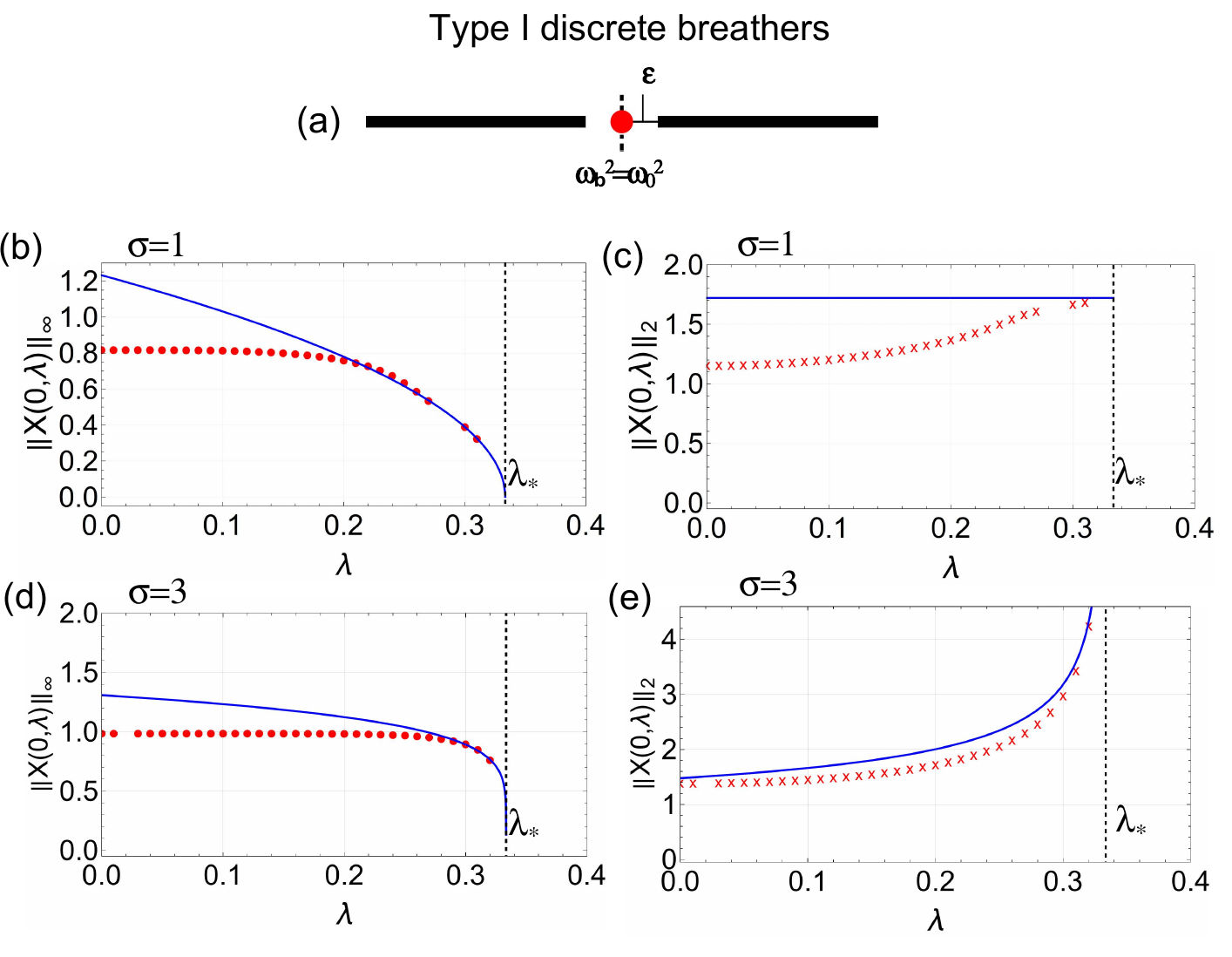}
  \vspace{-0.4cm}
  \caption{Bifurcation curves of mid-gap discrete breathers with $\sigma=1$ and $\sigma=3$: (a) phonon spectrum schematic (black) with breather frequency, $\omega_b$, (red); (b) and (d) show the $l^\infty$ norm of computed (dotted) discrete breather and its approximation from the weakly nonlinear long wave theory \eqref{envelope2} (blue); (c) and (e) show the corresponding $l^2$ norm of computed (x's) discrete breather and its approximation from the weakly nonlinear theory \eqref{envelope2} (blue).}
  \label{fig:nonlin_h}
\end{figure}
In Figure \ref{fig:nonlin_h} we contrast the $\ell^{\infty}$ and $\ell^{2}$-norms of mid-gap discrete breathers continued from the anti-continuum in the cases: $\sigma= 1$ and $\sigma= 3$. The behavior of the  $l^\infty$ and $l^2$ norms  $X^{\lambda}(0)$ agree well with the  asymptotic expressions, based on \eqref{envelope2} and \eqref{eps72}, as $\lambda\to\lambda_*$. 
\begin{remark}
We also considered the case $\sigma=1/2$, where the potential is $C^3$ but not smooth enough near $0$ to implement our multiple scale expansion. Our calculations are consistent with $l^2-$ subcritical scaling behavior; both the $l^\infty$ and $l^2$ norms appeared to approach zero as $\lambda\to\lambda_\star$. %Nevertheless, Figure \ref{fig:nonlin_h} (b) %and (c) show a behavior with the natural %dilation scalings given in equations %\eqref{envelope2} and \eqref{nonlin_l2}.
\end{remark} 

Here, we have assumed that $\epsilon$ is positive, however the emergent symmetry argument given in Section \ref{sec:continuum} can be used again to obtain discrete breather envelopes, from \eqref{envelope2} and \eqref{eps72}, in the topologically non-trivial regime ($\epsilon$ small and negative). 

As $\lambda\to\lambda_*$, the tails of the mid-gap breathers in Figure \ref{fig:nonlin_h} have asymmetric decay rates on A and B sites. For $\lambda\uparrow\lambda_*$: $x_n^B\gg x_n^A\sim (x_n^B)^{2\sigma+1}$ as $n\to+\infty$ and $x_n^A\gg x_n^B\sim (x_n^A)^{2\sigma+1}$ as $n\to-\infty$. Again, our long-wave asymptotics implies that past the topological transition point of the linear system for $\lambda\downarrow\lambda_*$: $x_n^A\gg x_n^B\sim (x_n^A)^{2\sigma+1}$ as $n\to+\infty$ and $x_n^B\gg x_n^A\sim (x_n^B)^{2\sigma+1}$ as $n\to-\infty$.

Finally, a short remark on the
out-of-gap cases (B) or (C) of Section \ref{sec:continuum}.  Here there is an effective nonlinear Schroedinger equation and we have that
\[\|X^\lambda\|^2_2\approx 4\epsilon^{\frac{2}{\sigma}-1}\|R(z;\nu)\|_{L^{2}(\mathbb{R})}^2,\]
where $R(Z,\nu)$ is the localized solution of \eqref{nls10} with a $2\sigma+1$-order nonlinearity.  Hence, for breathers in regimes (B) and (C) $\sigma=\sigma_S=2$ is $L^2-$ scaling critical.

\section{Discussion and final remarks}
\label{sec:conclusions}
We have studied - analytically and numerically - discrete breathers of a SSH-like network of nonlinear dimer oscillators. In particular, we have proved existence of discrete breathers (or DBs, solutions which are periodic in time and exponentially localized in space,) near the anti-continuous limit, in which our coupling parameter, $\lambda$ is small and non-zero. We numerically continue branches of discrete breathers till near the critical coupling value, $\lambda_\star$, at which the phonon gap closes, corresponding to the topological transition in linear SSH.
 In this latter limiting regime,  discrete breathers have a multi-scale wave-packet structure, where the envelope of this wave-packet is characterized by long-wave envelope nonlinear envelope PDEs of nonlinear Dirac type (for DB frequencies in the phonon gap) and nonlinear Schroedinger type  (for DB frequencies above or below the phonon spectrum). For discrete breather's whose frequencies are ``in-gap'', the continuum limit states are a type of gap-soliton.  Our envelope theory shows excellent agreement with simulations of the discrete breathers for $\lambda$ near $\lambda_\star$.
 Moreover,
 an emergent symmetry, present for the continuum (vanishing phonon gap) limit, but not 
 present in the discrete model enables us to construct --- from DBs for $\lambda$ in the topologically trivial linear regime of SSH ($\lambda-\lambda_\star<0$ and small) ---   DBs  in the 
 topologically nontrivial linear regime ($\lambda-\lambda_\star>0$ and small).  
A further consequence of our analysis is clarification of the chiral character of mid-gap discrete breathers.

 This work is influenced by the agenda (for photonics, phononics, and mechanical systems) of  exploring the interplay between nonlinearity and novel band structures for which, in the linear regime, there are topological phenomena; see references in Section \ref{prev-work} and those cited therein. While aspects of linear band structure topology arise in the bifurcation and continuation of nonlinear discrete breathers, a theory of topological states in nonlinear systems remains an open challenge.

\appendix

\section{Fourier Coefficient Method}
A numerical method for computing discrete breathers was originally described in \cite{ma96} based on solving for fixed points of a Poincar\'e mapping. Here we describe a Fourier-based numerical method for constructing time-reversible discrete breathers of system \eqref{eq3a}, in the setting of  \cref{thm:breather}.
%\footnote{\textcolor{magenta}{Did we review earlier why one expects 
%fixed $\omega_b$ breathers for all $\lambda$
% and relation to time-reversability due to %the potential's symmetry?}}

Given that both the non-resonance and non-degeneracy conditions in \cref{thm:breather} are satisfied, we know that there exists $\lambda_b>0$ and a unique time-periodic and reversible, exponentially localized solution to \eqref{eq3a} for $0\leq\lambda<\lambda_b$.  To construct such solutions, we will truncate the infinite-dimensional system in space and time and iteratively solve a large system of nonlinear equations, transforming between the time- and frequency-domains. 

We will employ the Newton-Raphson method to solve the system, which requires knowledge of the non-singular Jacobian of the mapping \eqref{eq5}.  To begin we define the Fourier series
\begin{equation*}
    x_n^{A,B}(t)=\sum_{m\in\mathbb{Z}}\hat{x}_{n,m}^{A,B}e^{i\omega_b mt},\qquad \hat{x}_{n,m}^{A,B}=\dfrac{1}{2\pi}\int_0^{2\pi}x_n^{A,B}(t)e^{-i\omega_b mt}dt
\end{equation*}
where, recall, $\omega_b=2\pi/T_b$ is the breather frequency.  Next we define the transformed mapping for $n,m\in\mathbb{Z}$
\begin{equation}\label{num1} 
\mathcal{F}\{F(X(t),\lambda)\}=
\left\{
\begin{aligned}
&-m^2\omega_b^2\hat{x}_{n,m}^A+\mathcal{F}\{V'(x_{n}^A)\}_m-\lambda\gamma_{\text{out}}\hat{x}_{n-1,m}^B-\gamma_{\text{in}}\hat{x}_{n,m}^B\\ 
&-m^2\omega_b^2\hat{x}_{n,m}^B+\mathcal{F}\{{V'(x_{n}^B)}\}_m-\lambda\gamma_{\text{out}}\hat{x}_{n+1,m}^A-\gamma_{\text{in}}\hat{x}_{n,m}^A.
\end{aligned}
\right.
\end{equation}
where $\mathcal{F}$ denotes the isometric Fourier mapping $\mathcal{F}:L^2_{\text{per}}[0,T_b]\to l^2(\mathbb{Z})$. Linearizing the above system about $X(t)$ in the time-domain and transforming leads to
\begin{equation}\label{num2}
\mathcal{F}\{F_X(X(t),\lambda)Y(t)\}=
\left\{
\begin{aligned}
&-m^2\omega_b^2\hat{y}_{n,m}^A+\mathcal{F}\{V''(x_{n}^A)y_n^A\}_m-\gamma_{\text{in}}\hat{y}_{n,m}^B-\lambda\gamma_{\text{out}}\hat{y}_{n-1,m}^B\\ 
&-m^2\omega_b^2\hat{y}_{n,m}^B+\mathcal{F}\{V''(x_{n}^B)y_n^B\}_m-\gamma_{\text{in}}\hat{y}_{n,m}^A-\lambda\gamma_{\text{out}}\hat{y}_{n+1,m}^A.
\end{aligned}
\right.
\end{equation}
where $Y=\{y_j\in H^2(\mathbb{R} / \mathbb{Z}T_b;\mathbb{R}^2)\}_ {j\in\mathbb{Z}}$. The expression above can be rewritten as
\begin{align}\label{num3}
\mathcal{F}\{F_X(X(t),\lambda)\}*\hat{Y}(t)=\qquad\qquad\qquad\qquad\qquad\qquad\qquad\qquad\qquad\qquad\qquad
\\ \nonumber
\left\{
\begin{aligned}
&-m^2\omega_b^2\hat{y}_{n,m}^A+\left(\mathcal{F}\{V''(x_{n}^A)\}*\hat{y}_n^A\right)_m-\gamma_{\text{in}}\hat{y}_{n,m}^B-\lambda\gamma_{\text{out}}\hat{y}_{n-1,m}^B\\ 
&-m^2\omega_b^2\hat{y}_{n,m}^B+\left(\mathcal{F}\{V''(x_{n}^B)\}*\hat{y}_n^B\right)_m-\gamma_{\text{in}}\hat{y}_{n,m}^A-\lambda\gamma_{\text{out}}\hat{y}_{n+1,m}^A,
\end{aligned}
\right.
\end{align}
where $*$ denotes the convolution operation: $l^2(\mathbb{Z})\times l^2(\mathbb{Z})\to l^2(\mathbb{Z})$.   

We now truncate the above exact expressions for $N$ lattice sites with zero boundary conditions at the lattice edges, as an approximation for exponentially decaying lattice breathers.  Taking $N_t$ equally-spaced points in $[0,T_b]$, we use the following definition of the discrete Fourier transform 
\begin{equation}
    \hat{X}_m=\sum_{p=0}^{N_t-1}X_pe^{-i\omega_bp\Delta t m}=\sum_{p=0}^{N_t-1}X_pe^{-2\pi ipm/N_t}
\end{equation}
and inverse
\begin{equation}
    X_p=\dfrac{1}{N_t}\sum_{m=0}^{N_t-1}\hat{X}_me^{2\pi ipm/N_t}.
\end{equation}

We can write the $2NN_t\times 2NN_t$ truncated block tri-diagonal Jacobian matrix of \eqref{num3} as
\begin{equation}\label{num4}
\hspace{-0.2cm}
\mathcal{F}\{F_X(X,\lambda)\}=
 \begin{pmatrix}
   M(x_0^A) & -\gamma_{\text{in}}I_{N_t,N_t}  & 0 & \cdots & 0 \\
   -\gamma_{\text{in}}I_{N_t,N_t} & M(x_0^B)  & -\lambda\gamma_{\text{out}}I_{N_t,N_t} & 0 & \\
   0 & -\lambda\gamma_{\text{out}}I_{N_t,N_t} & M(x_1^A) & -\gamma_{\text{in}}I_{N_t,N_t} & \\ 
   \vdots &  & \ddots & \ddots& \ddots\\ 
0 &  &   &  & M(x_{N}^B)
 \end{pmatrix}
\end{equation}
where $I_{N_t,N_t}$ is the $N_t\times N_t$ identity matrix. $M(x)$ is the $N_t\times N_t$ circulant matrix formed by incrementally circularly shifting the vector $\widehat{V''(x)}/N_t$ and then adding the diagonal matrix $\text{diag}\{\left(-0^2\omega_b^2, -1^2\omega_b^2,\cdots,-(N_t-1)^2\omega_b^2\right)\}$. 

We can now utilize fast-Fourier-transform FFT libraries and evaluate the nonlinear terms in the systems above by evaluating them in the time-domain via an inverse FFT and then transforming back into the frequency-domain. Also recall that a necessary condition to invoke \cref{thm:breather} is the restriction of the solution space to be time-reversible, i.e. $X(t)=X(-t)$.  Since $X(t)$ is real-valued, we have $\hat{X}_m=\hat{X}_{-m}^*$ and furthermore, since we require the solutions to be \textit{even} in time, we have $\hat{X}_m=\hat{X}_{-m}$. Thus the number of nonlinear equations to solve is essentially reduced by half, which is equivalent to using the discrete cosine transform.  

To solve $\mathcal{F}\{F(X(t),\lambda)\}=0$ for $X(t)$ with $\lambda\neq 0$, we use the following Newton-Raphson scheme
\begin{equation}
    \hat{X}^{(j+1)}=\hat{X}^{(j)}+\left[\mathcal{F}\{F_X(X^{(j)},\lambda)\}\right]^{-1}\mathcal{F}\{F(X^{(j)},\lambda)\}
\end{equation}
and iterate to a prescribed tolerance (here $j$ denotes the iteration counter). To initialize the iteration, we use the anti-continuum solution: $\hat{X}^{(0)}=\hat{X}_*$. Note, the inverse above is never explicitly computed and instead the system is solved via, for instance, $LU$-decomposition.  Finally, we remark that in order to analyze the stability of breathers, once the solution has converged, we take the initial values in the time-domain and numerically integrate the linearized equations independently out to $T_b$ to construct the monodromy matrix and compute its eigensystem (see section \ref{sec:stability}).

\bibliographystyle{siamplain}
\bibliography{references}
\end{document}